\documentclass[12pt]{article}
\usepackage{latexsym}
\usepackage{amsmath,amsfonts}
\usepackage{times}
\allowdisplaybreaks[4]

\catcode`\@=11

\newcount\hour
\newcount\minute
\newtoks\amorpm \hour=\time\divide\hour by 60\minute
=\time{\multiply\hour by 60 \global\advance\minute by-\hour}
\edef\standardtime{{\ifnum\hour<12 \global\amorpm={am}%
        \else\global\amorpm={pm}\advance\hour by-12 \fi
        \ifnum\hour=0 \hour=12 \fi
        \number\hour:\ifnum\minute<10
        0\fi\number\minute\the\amorpm}}
\edef\militarytime{\number\hour:\ifnum\minute<10
0\fi\number\minute}

\def\draftlabel#1{{\@bsphack\if@filesw {\let\thepage\relax
   \xdef\@gtempa{\write\@auxout{\string
      \newlabel{#1}{{\@currentlabel}{\thepage}}}}}\@gtempa
   \if@nobreak \ifvmode\nobreak\fi\fi\fi\@esphack}
        \gdef\@eqnlabel{#1}}
\def\@eqnlabel{}
\def\@vacuum{}
\def\marginnote#1{}
\def\draftmarginnote#1{\marginpar{\raggedright\scriptsize\tt#1}}
\def\draft{
        \pagestyle{plain}
        \overfullrule=2pt
        \oddsidemargin -.5truein
        \def\@oddhead{\sl \phantom{\today\quad\militarytime} \hfil
        \smash{\Large\sl DRAFT} \hfil \today\quad\militarytime}
        \let\@evenhead\@oddhead
        \let\label=\draftlabel
        \let\marginnote=\draftmarginnote
        \def\ps@empty{\let\@mkboth\@gobbletwo
        \def\@oddfoot{\hfil \smash{\Large\sl DRAFT} \hfil}
        \let\@evenfoot\@oddhead}
        \def\@eqnnum{(\theequation)\rlap{\kern\marginparsep\tt\@eqnlabel}%
        \global\let\@eqnlabel\@vacuum}  }

\newcommand{\rf}[1]{(\ref{#1})}
\renewcommand{\theequation}{\thesection.\arabic{equation}}
\renewcommand{\thefootnote}{\fnsymbol{footnote}}
\newcommand{\newsection}{    
\setcounter{equation}{0}\section}

\def\appendix#1{\addtocounter{section}{1}\setcounter{equation}{0}
\renewcommand{\thesection}{\Alph{section}}
\section*{Appendix \thesection\protect\indent \parbox[t]{11.15cm}{#1}}
\addcontentsline{toc}{section}{Appendix \thesection\ \ \ #1}}

\def\nline{\,\nabla\kern -0.7em\raise0.2ex\hbox{/}\,\,}
\def\yline{\,y\kern -0.47em /}
\def\aline{\,a\kern -0.49em /}
\def\parline{\,\partial\kern -0.55em /\,\,}

\newcommand{\Co}{\mathbb{C}}
\newcommand{\Eo}{\mathbb{E}}

\newcommand{\Mo}{\mathbb{M}}
\newcommand{\No}{\mathbb{N}}
\newcommand{\Po}{\mathbb{P}}
\newcommand{\Ro}{\mathbb{R}}

\def\be{\begin{equation}}
\def\ee{\end{equation}}
\def\beq{\begin{eqnarray}}
\def\eeq{\end{eqnarray}}
\def\beqq{\begin{eqnarray*}}
\def\eeqq{\end{eqnarray*}}

\def\CSFsm{{\scriptscriptstyle CSF}}

\def\smone{{\scriptscriptstyle (1)}}

\def\smpt{{\scriptscriptstyle [2]}}
\def\smp3{{\scriptscriptstyle [3]}}

\def\smpn{{\scriptscriptstyle [n]}}

\def\Thsm{{\scriptscriptstyle \rm Th}}
\def\Wsm{{\scriptscriptstyle \rm W}}

\def\Bbf{{\bf B}}
\def\Cbf{{\bf C}}
\def\Jbf{{\bf J}}

\def\Pbf{{\bf P}}

\def\Qbf{{\bf Q}}

\def\ibf{{\bf i}}
\def\iibf{{\bf ii}}
\def\iiibf{{\bf iii}}
\def\ivbf{{\bf iv}}

\def\NN{{\cal N}}
\def\PP{{\cal P}}
\def\SSc{{\cal S}}

\def\vph{{\vphantom{5pt}}}

\def\half{\frac{1}{2}}

\def\alphab{\bar{\alpha}}
\def\upsilonb{\bar{\upsilon}}
\def\zetab{\bar{\zeta}}

\def\fb{{\bar{f}}}

\def\mb{{\bar{m}}}

\def\irm{{\rm i}}
\def\frm{{\rm f}}
\def\msv{{\rm msv}}
\def\msl{{\rm msl}}

\def\dyn{{\rm dyn}}
\def\minrm{{\rm min}}

\def\even{{\rm even}}
\def\odd{{\rm odd}}

\def\osc{{\rm osc}}
\def\vecrm{{\rm vec}}
\def\frd{{\rm frd}}
\def\crt{{\rm crt}}
\def\Erm{{\rm E}}
\def\EPLrm{{\rm EPL}}
\def\PLrm{{\rm PL}}

\def\smBB{{\scriptscriptstyle BB}}

\def\betach{\check{\beta}}

\def\noinbf#1{\noindent {\bf #1}}

\begin{document}


\begin{flushright}
FIAN-TD-2025-8 \hspace{0.5cm} \,
\\
arXiv:2505.02817 V3\hspace{0cm} \,
\end{flushright}

\vspace{1cm}

\begin{center}

{\Large \bf Interacting massive/massless

\medskip
continuous-spin fields and integer-spin fields}

\vspace{2.5cm}

R.R. Metsaev\footnote{ E-mail: metsaev@lpi.ru }

\vspace{1cm}

{\it Department of Theoretical Physics, P.N. Lebedev Physical
Institute, \\ Leninsky prospect 53,  Moscow 119991, Russia }

\vspace{3cm}

{\bf Abstract}

\end{center}

In the framework of light-cone gauge approach, interacting continuous-spin fields and integer-spin fields propagating in flat space are studied. The continuous-spin fields are considered by using a light-cone gauge vector superspace formulation. Description of massive continuous-spin fields associated with the principal, complementary and discrete series is presented. For the massive continuous-spin fields of the principal and complementary series and massless continuous-spin fields, all parity-even cubic vertices realized as functions on the light-cone gauge vector superspace are obtained. Cubic vertices for a cross-interaction of massive/massless continuous spin fields and massive/massless integer-spin fields are also derived. These results for cubic vertices are complete for the dimensions of space-time greater than four. The use of the light-cone gauge vector superspace formulation considerably simplifies the cubic vertices as compared to the ones of oscillator formulation. Some cubic vertices realized as distributions are also found. Map between the oscillator formulation and the vector superspace formulation of the continuous-spin fields is explicitly described. An equivalence of the light-cone gauge and Lorentz covariant formulations of the free continuous-spin fields is also demonstrated.

\vspace{3cm}


\newpage
\renewcommand{\thefootnote}{\arabic{footnote}}
\setcounter{footnote}{0}

\section{ \large Introduction}

In view of various interesting features a continuous-spin field (CSF) has attracted permanent attention for a long period of time. For review and list of references in earlier literature, see Refs.\cite{Bekaert:2006py}-\cite{Brink:2002zx}.
Lagrangian formulation of a gauge bosonic CSF proposed in Ref.\cite{Schuster:2013pta} has triggered interest in this topic.
Lagrangian formulation of a gauge fermionic CSF was soon developed in Ref.\cite{Najafizadeh:2015uxa}, while interacting CSFs were studied in Refs.\cite{Metsaev:2017cuz}-\cite{Rivelles:2023hzo}.
A generalization of worldline formalism to CSF with applications to QED and gravitational physics is presented in Refs.\cite{Schuster:2023xqa}, while towards the $S$-matrix description of CSF may be found in Refs.\cite{Schuster:2013pxj,Bellazzini:2024dco}. Thermodynamics of CSF is considered in Ref.\cite{Schuster:2024wjc}, while Lagrangian formulation of a supersymmetric CSF was studied in Refs.\cite{Najafizadeh:2019mun}-\cite{Buchbinder:2022msd}.%
\footnote{CSF in AdS space was studied in Refs.\cite{Metsaev:2016lhs}-\cite{Buchbinder:2024hea}. See also discussion in Ref.\cite{Basile:2023vyg}. Study of mixed-symmetry CSF may be found in
Refs.\cite{Alkalaev:2017hvj,Metsaev:2021zdg}. Unfolded CSF was analysed in Refs.\cite{Ponomarev:2010st,Khabarov:2020glf}.}

In Refs.\cite{Metsaev:2017cuz,Metsaev:2018moa}, we developed the light-cone gauge formulation of  massive/massless CSFs propagating in $\Ro^{d-1,1}$, $d\geq 4$. In these references, to realize an infinite number of spin degrees of freedom of CSF we used oscillators. For this reason we shall refer the formulation in Refs.\cite{Metsaev:2017cuz,Metsaev:2018moa} to as oscillator formulation. In Ref.\cite{Metsaev:2017cuz,Metsaev:2018moa}, using our oscillator formulation, we obtained all parity-even cubic vertices which involve at least one massive/massless CSF.%
\footnote{Using the oscillator formulation, we found all parity-even cubic vertices expandable in transverse momenta. We have no proof and hence do not state that the oscillator formulation provides us all cubic vertices which could be found by other approaches. In view of the infinite spin degrees of freedom of CSF, an equivalence of different approaches is not obvious a priori.}

Unfortunately, the light-cone gauge oscillator formulation leads to complicated expressions for cubic vertices. Depending on mass values, the cubic vertices turn out to be expressible in terms of special functions: hypergeometric functions and Bessel functions. Besides such special functions the cubic vertices involve some complicated dressing operators acting on the just mentioned special functions. This motivates us to look for alternative light-cone gauge formulations of CSFs.

One of the alternative light-cone gauge formulations of a massless CSF is based on using a vector superspace in place of oscillators. For massless CSFs in $\Ro^{3,1}$ and $\Ro^{4,1}$, such alternative light-cone gauge formulation was studied in Ref.\cite{Brink:2002zx}. Starting with the formulation in Ref.\cite{Brink:2002zx}, we realize a vector superspace of a massless CSF propagating in $\Ro^{d-1,1}$ as a sphere $S^{d-3}$ embedded in Euclidean space $\Eo^{d-2}$. This leads us to the embedding space realization of a vector superspace. For a massive CSF in $\Ro^{d-1,1}$, we develop the light-cone gauge formulation by using a vector superspace which is also realized as $S^{d-3}$ embedded in $\Eo^{d-2}$. We refer such formulation of massive/massless CSF to as light-cone gauge vector formulation or simply to as vector formulation.

We apply the vector formulation of CSFs for the study of cubic vertices for CSFs and integer-spin fields. We show that the vector formulation, in contrast to the oscillator formulation, leads to simple solutions to cubic vertices (rational or exponential functions). Solutions to vertices realized as functions on the vector superspace are referred to as {\bf f-solutions}. We note then that besides the f-solutions we find a few particular distributional solutions.

To describe a general classification of cubic vertices we use shortcuts for fields under investigation. For a massive CSF, we use shortcut $(m,\SSc)_\CSFsm$, where $m$ and $\SSc$ stand for the respective mass and spin parameters, $m^2 < 0$, $\SSc \in \Co$, while, for a massless CSF, we use shortcut $(0,\kappa)_\CSFsm$, where $\kappa$ stands for spin parameter, $\kappa^2>0$. For a massive integer-spin field, we use shortcut $(m,s)$, where $m$ and $s$ stand for the respective mass and spin parameters, $m^2>0$, $s\in \No_0$, while, for a massless integer-spin field, we use shortcut $(0,s)$, where $s$ stands for spin parameter, $s\in \No_0$. We summarize our shortcuts as
{\small
\beqq
\begin{array}{ll}
(m,\SSc)_\CSFsm, \ \  m^2 < 0, \ \ \SSc\in \Co,   \ \ \hbox{massive CSF}; \hspace{0.5cm} & (m,s), \ \ m^2 > 0,  \ s\in \No_0, \ \hbox{massive integer-spin field};
\\[7pt]
(0,\kappa)_\CSFsm,  \ \ \ \kappa^2 > 0 \hspace{2cm} \hbox{massless CSF};  &  (0,s), \ \ s\in \No_0, \hspace{1.7cm} \hbox{massless integer-spin field};\qquad
\end{array}
\eeqq
}
Using such shortcuts, the general classification of cubic vertices which are non-trivial a priori and involve, among other fields, at least one massive/massless CSF can be presented as follows.
{\small
\beq
&& \hspace{-1.5cm} \hbox{\bf Cubic vertices:}
\nonumber\\
&& \hspace{-1cm} \hbox{\bf Three continuous-spin fields:}
\nonumber\\
\label{26032025-man01-01} && \hspace{-1cm} (m_1,\SSc_1)_\CSFsm\hbox{-}(m_2,\SSc_2)_\CSFsm\hbox{-}(m_3,\SSc_3)_\CSFsm\,,
\\
\label{26032025-man01-02} && \hspace{-1cm} (m_1,\SSc_1)_\CSFsm\hbox{-}(m_2,\SSc_2)_\CSFsm\hbox{-}(0,\kappa_3)_\CSFsm\,, \hspace{0.5cm} m_1^2 \ne m_2^2 \,;
\\
\label{26032025-man01-03} && \hspace{-1cm} (m_1,\SSc_1)_\CSFsm\hbox{-}(m_2,\SSc_2)_\CSFsm\hbox{-}(0,\kappa_3)_\CSFsm\,, \hspace{0.5cm}  m_1^2=m_2^2\,;
\\
\label{26032025-man01-04} && \hspace{-1cm} (m_3,\SSc_3)_\CSFsm\hbox{-}(0,\kappa_2)_\CSFsm\hbox{-}(0,\kappa_3)_\CSFsm\,,
\\
\label{26032025-man01-05} && \hspace{-1cm} (0,\kappa_1)_\CSFsm\hbox{-}(0,\kappa_2)_\CSFsm\hbox{-}(0,\kappa_3)_\CSFsm\,,  \hspace{2.7cm} \hspace{1cm}  \hbox{distributional solution}
\\[10pt]
&& \hspace{-1cm}  \hbox{\bf Two continuous-spin fields and one integer-spin field:}
\nonumber\\
\label{26032025-man01-06} && \hspace{-1cm} (m_1,\SSc_1)_\CSFsm\hbox{-}(m_2,\SSc_2)_\CSFsm\hbox{-}(m_3,s_3)\,,
\\
\label{26032025-man01-07}  && \hspace{-1cm} (m_1,\SSc_1)_\CSFsm\hbox{-}(m_2,\SSc_2)_\CSFsm\hbox{-}(0,s_3)\,, \hspace{1.5cm}  m_1^2\ne m_2^2\,;
\\
\label{26032025-man01-08}  && \hspace{-1cm} (m_1,\SSc_1)_\CSFsm\hbox{-}(m_2,\SSc_2)_\CSFsm\hbox{-}(0,s_3)\,,  \hspace{1.5cm}    m_1^2 =  m_2^2\,;
\\
\label{26032025-man01-09}  && \hspace{-1cm} (m_1,\SSc_1)_\CSFsm \hbox{-}(0,\kappa_2)_\CSFsm\hbox{-}(m_3,s_3)\,,
\\
\label{26032025-man01-10} && \hspace{-1cm} (m_1,\SSc_1)_\CSFsm\hbox{-}(0,\kappa_2)_\CSFsm\hbox{-}(0,s_3)\,,
\\
\label{26032025-man01-11} && \hspace{-1cm} (0,\kappa_1)_\CSFsm\hbox{-}(0,\kappa_2)_\CSFsm\hbox{-}(m_3,s_3)\,,
\\
\label{26032025-man01-12} &&  \hspace{-1cm} (0,\kappa_1)_\CSFsm\hbox{-}(0,\kappa_2)_\CSFsm\hbox{-}(0,s_3)\,,  \hspace{3.5cm}       \hspace{0.8cm} \hbox{distributional solution}
\\[10pt]
&& \hspace{-1cm}  \hbox{\bf One continuous-spin field and two integer-spin fields:}
\nonumber\\
\label{26032025-man01-13}  && \hspace{-1cm} (m_1,\SSc_1)_\CSFsm^\vph\hbox{-}(m_2,s_2)\hbox{-}(m_3,\kappa_3)\,,
\\
\label{26032025-man01-14} && \hspace{-1cm} (m_1,\SSc_1)_\CSFsm^\vph\hbox{-}(m_2,s_2)\hbox{-}(0,s_3)\,,
\\
\label{26032025-man01-15} &&  \hspace{-1cm} (m_1,\SSc_1)_\CSFsm^\vph\hbox{-}(0,s_2)\hbox{-}(0,s_3)\,,
\\
\label{26032025-man01-16} && \hspace{-1cm} (0,\kappa_1)_\CSFsm \hbox{-}(m_2,s_2)\hbox{-}(m_3,s_3)\,,  \hspace{0.6cm} m_2^2  \ne m_3^2\,;
\\
\label{26032025-man01-17} && \hspace{-1cm} (0,\kappa_1)_\CSFsm\hbox{-}(m_2,s_2)\hbox{-}(m_3,s_3)\,, \hspace{0.6cm} m_2^2 = m_3^2\,,\hspace{1cm}  \hspace{1cm} \hbox{distributional solution}
\\
\label{26032025-man01-18} && \hspace{-1cm} (0,\kappa_1)_\CSFsm\hbox{-}(m_2,s_2)\hbox{-}(0,s_3)\,,
\\
\label{26032025-man01-19} && \hspace{-1cm} (0,\kappa_1)_\CSFsm\hbox{-}(0,s_2)\hbox{-}(0,s_3)\,,  \hspace{4cm} -
\eeq
}

Let us briefly summarize our results in this paper. For cubic vertices  \rf{26032025-man01-05}, \rf{26032025-man01-12}, \rf{26032025-man01-17}, \rf{26032025-man01-19}, we conclude that f-solutions to our equations do not exist, while, for the remaining  cubic vertices \rf{26032025-man01-01}-\rf{26032025-man01-19}, we find all parity-even f-solutions to our equations.
Finding of all distributional solutions to equations for vertices  \rf{26032025-man01-01}-\rf{26032025-man01-19} is an open problem. Presently, we are not aware of a systematic method for finding distributional solutions. Therefore, discussion of full list of distributional vertices is beyond scope of the present paper. Nevertheless, for vertices \rf{26032025-man01-05}, \rf{26032025-man01-12}, \rf{26032025-man01-17}, we managed to find some particular distributional solutions, which we also present in this paper.
Our study provides all f-solutions for parity-even cubic vertices when $d>4$. For $d=4$, our analysis requires to be carried out separately. To avoid too long discussion and unnecessary technical complications in our presentation, the case $d=4$ will be considered elsewhere.

We now briefly describe our results obtained in the framework of the oscillator approach in Refs.\cite{Metsaev:2017cuz,Metsaev:2018moa}. For cubic vertices \rf{26032025-man01-12}, \rf{26032025-man01-19}, we shown that solution to our equations does not exist, while, for cubic vertices \rf{26032025-man01-01}-\rf{26032025-man01-04}, \rf{26032025-man01-06}-\rf{26032025-man01-11}, and \rf{26032025-man01-13}-\rf{26032025-man01-18} we found all parity-even solutions to our equations. Vertices \rf{26032025-man01-05} were not studied in Refs.\cite{Metsaev:2017cuz,Metsaev:2018moa}.%
\footnote{The oscillator approach hides the separation of vertices into the f-vertices and distributional vertices. Such separation becomes visible in the framework of the vector formulation.}
Opposite results for vertices \rf{26032025-man01-12} obtained in the oscillator and vector formulations is related to the fact that in this paper we impose more weak conditions on the cubic vertices (see our comment in Sec.\,\ref{sub-sec-5-7}).

Though our major interest is related to cubic vertices we investigate an equivalence of various light-cone gauge and Lorentz covariant formulations of {\it free} CSF. We show that our light-cone gauge vector formulation of massive CSF is derivable from the Lorentz covariant constraints proposed by Bekaert and Boulanger  (BB-constraints) in Ref.\cite{Bekaert:2006py}, while the light-cone gauge vector formulation of massless CSF is derivable from the Lorentz covariant Wigner constraints which may be found, e.g., in Ref.\cite{Bekaert:2006py}.%
\footnote{Interesting study of so called modified Wigner constraints may be found in Refs.\cite{Najafizadeh:2017tin}.
}
Also we demonstrate that our light-cone gauge vector formulation for massive/massless CSFs proposed in this paper is equivalent to the light-cone gauge oscillator formulation for massive/massless CSFs proposed in Refs.\cite{Metsaev:2017cuz,Metsaev:2018moa}.

This paper is organized as follows. In Sec.\,\ref{sec-02}, we present our light-cone gauge vector formulation of massive/massless CSFs and recall the textbook  light-cone gauge oscillator formulation of massive/massless integer-spin fields.
The general form of equations required to determine all cubic vertices uniquely has been discussed in Refs.\cite{Metsaev:2017cuz,Metsaev:2018moa}. In Sec.\,\ref{sec-03}, in order to make our paper self-contained,  we first discuss briefly the general form of our equations and then we present the particular form of the equations suitable for the study of parity-even cubic vertices of CSFs considered in the framework of the vector formulation.

In Sec.\,\ref{sec-3csf}, we present our result for parity-even cubic vertices for three massive/massless CSFs.
We study cubic vertices for cross-interactions of two massive CSFs and one massless CSF and the ones for cross-interactions of one massive CSF and two massless CSFs.  Cubic vertices for self-interacting massive/massless CSF are also investigated.
In Sec.\,\ref{sec-2csf}, we present our result for parity-even cubic vertices for cross interactions of two massive/massless CSFs and one massive/massless integer-spin field, while, in Sec.\,\ref{sec-1csf}, we study parity-even cubic vertices for cross-interactions of one massive/massless CSF and two massive/massless integer-spin fields.%
\footnote{For $d\geq 4$, light-cone gauge vertices for integer-spin fields were studied in Refs.\cite{Bengtsson:1983pd}-\cite{Metsaev:2022yvb}, while their Lorentz covariant cousins  were investigated, e.g., in Refs.\cite{Manvelyan:2010jr,Sagnotti:2010at}. For integer-spin fields in 3d,  light-cone gauge vertices for integer-spin fields were studied in Refs.\cite{Metsaev:2020gmb}, while Lorentz covariant cubic vertices were considered in Refs.\cite{Mkrtchyan:2017ixk}-\cite{Delplanque:2024enh}.
}

In Sec.\,\ref{concl}, we summarize our study. Notation and conventions are given in Appendix A. In Appendices B and C, we present the derivation of the light-cone gauge vector formulation for massive and massless CSFs starting with the respective covariant BB-constraints
and Wigner constraints. In Appendix D, we provide a definition and useful relations for $c$ -- distribution. In Appendix E, we describe interrelations between the oscillator and vector formulations of CSF. In Appendix F, we provide some technical details of the derivation of a distributional solution for cubic vertex.

\newsection{ \large Free light-cone gauge continuous-spin fields and integer-spin fields}\label{sec-02}

In order to make our presentation self-contained and fix our notation, we provide short review of the basic elements of the light-cone formulation of fields under our consideration. Light-cone gauge vector formulation of massive CSF we present in this Section has not been discussed earlier in the context of massive CSF. For massless CSF, we use our update of the light-cone gauge formulation in Ref.\cite{Brink:2002zx} in terms of embedding space approach.%
\footnote{ In Ref.\cite{Brink:2002zx}, the light-cone gauge formulation of massless CSF  was obtained by using a group-theoretical method. In Appendix C, we show that such formulation can be obtained by using the Lorentz covariant Wigner constraints.}
For the integer-spin fields, we still use the old-fashioned light-cone gauge oscillator formulation.

\noindent {\bf Notation and Poincar\'e algebra in light-cone frame}. We use approach proposed in Ref.\cite{Dirac:1949cp} which tells us that the problem of finding a dynamical system  amounts to a problem of finding a solution for commutators of a basic symmetry algebra. In our case, basic symmetries are governed by the Poincar\'e algebra.
We start therefore with the description how the Poincar\'e algebra is realized on a space of light-cone gauge fields.

First, the commutators of the Poincar\'e algebra $iso(d-1,1)$ are fixed to be%
\footnote{  For vector indices of the Lorentz algebra $so(d-1,1)$, we use the Greek indices $\mu,\nu,\rho,\sigma = 0,1,\ldots,d-1$.}
\be \label{27032025-man01-01}
{} [P^\mu,\,J^{\nu\rho}]=\eta^{\mu\nu} P^\rho - \eta^{\mu\rho} P^\nu\,,
\qquad {} [J^{\mu\nu},\,J^{\rho\sigma}] = \eta^{\nu\rho} J^{\mu\sigma} + 3\hbox{ terms}\,,
\ee
where $J^{\mu\nu}$ are generators of
the $so(d-1,1)$ Lorentz algebra, while $P^\mu$ are the translation generators.
The flat metric $\eta^{\mu\nu}$ is assumed to be mostly positive.

Second, in place of the Lorentz basis coordinates $x^\mu$ we introduce the light-cone basis
coordinates $x^\pm$, $x^I$, where vector indices of the $so(d-2)$ algebra take values $I,J=1,\ldots,d-2$, while the coordinates $x^\pm$ are fixed to be $x^\pm := (x^{d-1}  \pm x^0)/\sqrt{2}$. The $x^+$ is taken to be an evolution parameter. The $so(d-1,1)$ Lorentz algebra vector $X^\mu$ is then decomposed as $X^+,X^-,X^I$, and non vanishing elements of the flat metric are given by $\eta_{+-} = \eta_{-+}=1$, $\eta_{IJ} = \delta_{IJ}$.

Third, in the light-cone approach, the Poincar\'e algebra generators are
separated into kinematical generators $P^+$, $P^I$, $J^{+I}$, $J^{+-}$, $J^{IJ}$ and dynamical generators $P^-$, $J^{-I}$.
For $x^+=0$, the kinematical generators are quadratic in fields, while, for arbitrary $x^+$,
the dynamical generators involve quadratic and higher order terms in fields.%
\footnote{ The generators $P^I$, $P^+$, $J^{IJ}$ are independent of $x^+$, while the generators $J^{+I}$, $J^{+-}$  are representable as $G= G_1 + x^+ G_2$, where $G_1$ is quadratic in fields, while $G_2$ involves quadratic and higher order terms in fields.}
In order to provide a field realization of the Poincar\'e algebra generators we now proceed with the discussion of the light-cone gauge description of CSF and integer-spin fields.

\noindent {\bf Continuous-spin field}. Light-cone gauge massive/massless CSF propagating in $\Ro^{d-1,1}$ is described by the following set of fields defined in a momentum space
{\small
\beq \label{27032025-man01-08}
&& \bigoplus_{n=n_{\min}}^\infty   \,\,\phi^{I_1\ldots I_n}(p)\,, \hspace{3cm}  \phi^{III_3\ldots I_n}(p) = 0\,,
\nonumber\\
&& n_\minrm =
\left\{\begin{array}{ll}
0 & \hbox{for massive CSF of principal series};
\\
0 & \hbox{for massive CSF of complementary series};
\\
s+1\,, \ \ s \in \No_0\,, \hspace{0.3cm} & \hbox{for massive CSF of discrete series};
\\
0 &   \hbox{for massless CSF};
\end{array} \right.
\eeq
}
\!where the dependence of the fields on $x^+$ is implicit, while the argument $p$ stands for the momenta $p^I$, $\beta$, $\beta:=p^+$. The fields $\phi^{I_1\ldots I_n}(p)$ with $n=0$, $n=1$, and $n\geq 2$ are the respective scalar, vector, and rank-$n$ totally symmetric {\it traceless} tensor fields of the $so(d-2)$ algebra. All fields are assumed to be complex-valued. To simplify the presentation we use the unit vector $u^I$, $u^Iu^I=1$ which describes a sphere $S^{N-1}$ embedded in $\Eo^N$, $N=d-2$. The continuous-spin field $\phi(p,u)$ is defined then by the relation
{\small
\be \label{27032025-man01-10}
\phi(p,u) = \sum_{n=n_\minrm}^\infty \phi_n(p,u)\,,\quad \ \phi_n(p,u) = \frac{1}{ n!\sqrt{\mu_n\tau_n S_{N-1}^{\phantom{1}} } } u^{I_1} \ldots u^{I_n} \phi^{I_1\ldots I_n}(p)\,,\quad \ N=d-2\,,
\ee
}
\!where $n_\minrm$ is given in \rf{27032025-man01-08}, while $\mu_n$, $\tau_n$, $S_{N-1}$ are defined in \rf{17042025-appk-01}, \rf{17042025-appk-03} in Appendix E.%
\footnote{Interestingly, for $n_\minrm=0$, the field content in \rf{27032025-man01-10} is the same as in higher-spin field theory in Refs.\cite{Vasiliev:1990en}. Recent studies of higher-spin field theory and extensive list of references may be found, e.g., in Refs.\cite{Didenko:2021vdb}-\cite{DeFilippi:2019jqq}.
}
As side remark, the field $\phi_n(p,u)$ obeys the equation $(\PP^I \PP^I + n(n+N-2))\phi_n=0$\,, where $\PP^I$ is a derivative of the unit vector $u^I$ (see Appendix A).

\noindent {\bf Integer-spin fields}. Light-cone gauge integer-spin fields in $\Ro^{d-1,1}$ are described by the following set of fields defined in momentum space:
\beq
\label{27032025-man01-25}  && \bigoplus_{n=0}^s \,\,\phi^{I_1\ldots I_n}(p)\,, \hspace{4cm} \hbox{for massive spin-$s$ field};
\\[-2pt]
\label{27032025-man01-26} && \phi^{I_1\ldots I_s}(p)\,, \hspace{1cm} \phi^{III_3\ldots I_s}(p) = 0 \,, \hspace{1.2cm} \hbox{for massless spin-$s$ field};
\eeq
where, in \rf{27032025-man01-25}, fields $\phi^{I_1\ldots I_n}(p)$ with $n=0$, $n=1$, and $n\geq 2$ are the respective scalar, vector, and rank-$n$ totally symmetric tensor fields of the $so(d-2)$ algebra. Traceless constraint for fields \rf{27032025-man01-25} is given below in \rf{27032025-man01-39}. In \rf{27032025-man01-26}, fields $\phi^{I_1\ldots I_s}(p)$ with $s=0$, $s=1$, and $n\geq 2$ are the respective scalar, vector, and traceless rank-$s$ totally symmetric tensor fields of the $so(d-2)$ algebra.

To use index-free notation we introduce oscillators  $\alpha^I$, $\zeta$ and collect fields \rf{27032025-man01-25}, \rf{27032025-man01-26} into respective ket-vectors defined by the relations
{\small
\beq
\label{27032025-man01-30} && \hspace{-1cm} \phi_s(p,\alpha) = \sum_{n=0}^s  \frac{\zeta^{s-n}}{n!\sqrt{(s-n)!}} \alpha^{I_1} \ldots \alpha^{I_n}
\phi^{I_1\ldots I_n}(p)\,, \hspace{0.7cm} \hbox{for massive spin-$s$ field};
\\
\label{27032025-man01-31} && \hspace{-1cm} \phi_s(p,\alpha) = \frac{1}{s!} \alpha^{I_1} \ldots \alpha^{I_s} \phi^{I_1\ldots I_s}(p) \,, \hspace{3.1cm} \hbox{for massless spin-$s$ field};
\eeq
}
\!where $\alpha$ in $\phi_s(p,\alpha)$ stands for the respective set of oscillators $\alpha^I$, $\zeta$ and $\alpha^I$. From \rf{27032025-man01-30}, \rf{27032025-man01-31}, we see that the ket-vectors satisfy the following homogeneity conditions
\beq
\label{27032025-man01-36}  && (N_\alpha + N_\zeta -s) \phi_s  = 0\,, \hspace{0.9cm}   \hbox{for massive spin-$s$ field};
\nonumber\\
&& (N_\alpha - s) \phi_s  = 0\,,   \hspace{1.8cm} \hbox{for massless spin-$s$ field};
\eeq
which tell us that ket-vector \rf{27032025-man01-30} is a degree-$s$ homogeneous polynomial in the oscillators $\alpha^I$, $\zeta$, while ket-vector \rf{27032025-man01-31} is a degree-$s$ homogeneous polynomial in the oscillators $\alpha^I$.
Ket-vectors of massive/massless spin-$s$ fields \rf{27032025-man01-30}, \rf{27032025-man01-31} should obey the traceless constraints given by,
\beq
\label{27032025-man01-39} &&  (\alphab^2 + \zetab^2) \phi_s =0 \,, \hspace{1cm} \hbox{for massive spin-$s$ field};
\nonumber\\
&& \alphab^2  \phi_s =0 \,, \hspace{2.2cm} \hbox{for massless spin-$s$ field}.\qquad
\eeq
Field \rf{27032025-man01-30}/\rf{27032025-man01-31} not satisfying  constraints \rf{27032025-man01-39} will be referred to as massive/massless spin-$s$ triplet field. Lorentz covariant studies of the triplet field may be found, e.g., in Refs.\cite{Bengtsson:1986ys}-\cite{Campoleoni:2012th}.

In order to simplify the presentation of our  results we use the chain of massive/massless fields
\be  \label{27032025-man01-41}
\phi(p,\alpha) = \sum_{s=0}^\infty \phi_s(p,\alpha)\,.
\ee
Ignoring constraints \rf{27032025-man01-39}, we get ket-vector \rf{27032025-man01-41} which describes tower of massive/massless integer-spin triplet fields, while, using constraints  \rf{27032025-man01-39}, implies that ket-vector  \rf{27032025-man01-41} describes chain of massive/massless fields which consists of every spin-$s$ field just once. For the case of CSF, we never relax the traceless constraint \rf{27032025-man01-08}.

\noindent {\bf Field-theoretical realization of Poincar\'e algebra}. We start with the realization of the Poincar\'e algebra generators in term of differential operators acting on the fields under consideration,%
\footnote{ Note that, without loss of generality, the generators of the Poincar\'e algebra are considered for $x^+=0$.}
{\small
\beq
\label{27032025-man01-43} && P^I=p^I\,, \qquad \qquad\quad P^+=\beta\,, \qquad J^{+I}=\partial_{p^I} \beta\,, \qquad \quad \ J^{+-} = \partial_\beta \beta\,,\qquad
\nonumber\\
&& J^{IJ}=p^I \partial_{p^J} - p^J \partial_{p^I} + M^{IJ}\,,
\nonumber\\
&& P^- =  -\frac{p^I p^I + m^2}{2\beta}\,, \qquad J^{-I} = - \partial_\beta p^I + \partial_{p^I} P^- + \frac{1}{\beta}(M^{IJ} p^J + M^I)\,,
\\
\label{27032025-man01-46}  && \hspace{1cm} \beta := p^+\,,\qquad \partial_\beta := \partial/\partial \beta\,, \quad \partial_{p^I} :=  \partial/\partial p^I\,,
\eeq
}
where, in \rf{27032025-man01-46}, we present our shortcut notation.

Operator $M^{IJ}$ appearing \rf{27032025-man01-43} stands for a spin operator of the $so(d-2)$ algebra. The operators $M^{IJ}$ and $M^I$ satisfy the commutators and hermitian conjugation rules given by
{\small
\beq
\label{27032025-man01-49} && [M^{IJ},M^{KL}] =  \delta^{JK} M^{IL} + 3\hbox{ terms}\,, \qquad  [M^I,M^{JK}] = \delta^{IJ} M^K - \delta^{IK} M^J \,,
\nonumber\\
&& [M^I, M^J ] = - m^2 M^{IJ}\,,
\qquad M^{IJ\dagger} = - M^{IJ}\,, \hspace{2cm} M^{I\dagger} = - M^I\,,
\eeq
}
\!where $m^2$ is a square of mass,
\be \label{27032025-man01-53}
m^2 < 0\,, \quad  \hbox{for massive CSF}; \qquad   m^2> 0\,, \quad  \hbox{for massive integer-spin field}.
\ee
For CSF, the realization of the spin operators $M^{IJ}$, $M^I$ is then given by
{\small
\beq
\label{27032025-man01-55} && M^I =   - |m| \big( \PP^I +  \SSc u^I \big) \,,  \hspace{0.5cm} M^{IJ} = u^I \PP^J - u^J\PP^I\,, \qquad  \hbox{for massive CSF};\qquad
\nonumber\\[5pt]
&& M^I = - \irm \kappa u^I\,, \hspace{2.4cm} M^{IJ} = u^I \PP^J - u^J\PP^I\,, \hspace{0.5cm} \hbox{ for massless CSF};
\eeq
}
\!where $\PP^I$ is a derivative of the unit vector $u^I$ (see Appendix A).
For massless CSF, the parameter $\kappa$ in \rf{27032025-man01-55} is real-valued.
From \rf{27032025-man01-49}-\rf{27032025-man01-55}, we see that, for massive CSF, the operators $M^{IJ}$, $M^I$ realize commutators of $so(d-2,1)$ algebra. We note the following values of the label $\SSc$:
{\small
\beq
&& \hspace{-2cm} \SSc = \frac{3-d}{2} + q \,;
\nonumber\\
\label{27032025-man01-60} && \hspace{-2cm}\Re q = 0\,, \hspace{5cm} \hbox{for massive CSF of principal series};
\nonumber\\
&& \hspace{-2cm} \Im q =0\,,  \hspace{0.3cm} -\frac{d-3}{2} < q < \frac{d-3}{2}\,,  \hspace{1cm} \hbox{for massive CSF of complementary series};\qquad
\nonumber\\
&& \hspace{-2cm} q = \frac{3- d}{2} - s\,, \ \ s\in \No_0\,,  \hspace{2.4cm} \hbox{for massive CSF of discrete  series};\qquad
\eeq
}
\!where we introduce a new label $q$ which sometimes turns out to be convenient in our study. In this paper, we do not consider cubic vertices involving massive CSF of the discrete series.

For integer-spin fields, the spin operators $M^{IJ}$, $M^I$ are given by
{\small
\beq
\label{27032025-man01-68}  && M^I = m (\zeta\alphab^I - \alpha^I \zetab),\hspace{0.5cm} M^{IJ} = \alpha^I\alphab^J - \alpha^J\alphab^I\,, \hspace{0.5cm} \hbox{for massive integer-spin field};\qquad
\nonumber\\
&& M^I = 0\,,  \hspace{2.6cm} M^{IJ} = \alpha^I\alphab^J - \alpha^J\alphab^I\,, \hspace{0.5cm} \hbox{for massless integer-spin field}.
\eeq
}
From \rf{27032025-man01-49}, \rf{27032025-man01-53}, and \rf{27032025-man01-68}, we see that, for massive integer-spin field, the operators $M^{IJ}$, $M^I$ realize commutators of $so(d-1)$ algebra as it should be.

Eigenvalues of the 2nd- and 4th-order Casimirs of the Poincar\'e algebra $iso(d-1,1)$ are as follows: $C_2=m^2$, $C_4=m^2\SSc(\SSc+d-3)$ for massive CSF; $C_2=0$, $C_4=\kappa^2$ for massless CSF;  $C_2=m^2$, $C_4=m^2s(s+d-3)$ for massive spin-$s$ field and $C_2=0$, $C_4=0$ for massless spin-$s$ field. We note the expression for the 4th order Casimir operator, $\hat{C}_4 = - M^IM^I- \half m^2 M^{IJ}M^{IJ}$.

Now, using the notation $G_\smpt$ for the field realization of the Poincar\'e algebra generators \rf{27032025-man01-01} at quadratic order in fields, we note the relation
\be \label{27032025-man01-75}
G_\smpt  =\int \beta d^{d-1}p\, \phi^*(p,u,\alpha)\cdot G \phi(p,u,\alpha)\,, \qquad
d^{d-1}p :=  d\beta d^{d-2}p\,,
\ee
where $G$ stands for operators given in
\rf{27032025-man01-43}, while the dot $\cdot$ is used for the inner product defined in \rf{12042025-man01-08}, \rf{12042025-man01-08-u} in Appendix A. In \rf{27032025-man01-75} and below, $\phi(p,u,\alpha)$ stands for $\phi(p,u)$ and $\phi(p,\alpha)$. The light-cone gauge action takes the form
{\small
\be \label{27032025-man01-76}
S = \int dx^+  d^{d-1} p\,\, \phi^*(p,u,\alpha) \cdot \irm \beta
\partial^- \phi(p,u,\alpha) +\int dx^+ P^-\,,
\ee
}
\!where $\partial^-=\partial/\partial x^+$ and $P^-$ is the light-cone Hamiltonian. For the case of free fields, the $P_\smpt^-$ is obtained by plugging $P^-$ \rf{27032025-man01-43} into \rf{27032025-man01-75}.%
\footnote{Inclusion of internal symmetry algebras could be done, e.g., as in Refs.\cite{Konstein:1989ij}-\cite{Skvortsov:2020wtf}.
}

\noinbf{Massless CSF from massive CSF of principal series}. Consider the principal series \rf{27032025-man01-60} and take the limit
\be  \label{27032025-man01-80}
|m| \rightarrow 0\,, \qquad |q| \rightarrow \infty\,, \qquad |m| q = \irm \kappa\,.
\ee
From \rf{27032025-man01-55}, we then see that, in the limit \rf{27032025-man01-80}, the spin operator of massive CSF is contracted to the one of massless CSF. Note also that, in view of \rf{27032025-man01-08}, ket-vectors of massless CSF and the one of massive CSF of principal series match. Obviously, the massive CSF of the complementary and discrete series cannot be contracted to massless CSF.%
\footnote{Contraction of massive integer-spin field to massless CSF was discussed in Ref.\cite{Bekaert:2005in}. See also interesting discussion in Refs.\cite{Khan:2004nj}.
}

\vspace{-0.2cm}
\newsection{ \large Cubic interaction vertices and light-cone gauge dynamical principle} \label{sec-03}

For interacting fields, the dynamical generators $G^\dyn = P^-, J^{-I}$ are presented as
{\small
\be \label{29032025-man01-01}
G^\dyn  = \sum_{n=2}^\infty  G_\smpn^\dyn\,,
\ee
}
\!where $G_\smpn^\dyn$ is a functional that has $n$ powers of field $\phi$. Dynamical generators of the second order in field are given in  \rf{27032025-man01-75}. In this section, we describe the complete system of equations which allows us to determine all solutions for the dynamical generators $P_\smp3^-$, $J_\smp3^{-I}$.

Let us start with general expressions for the dynamical generators $P_\smp3^-$, $J_\smp3^{-I}$ given by
{\small
\be  \label{29032025-man01-05} P_\smp3^- =  \int\!\! d\Gamma_\smp3 \,\, \Phi_\smp3^* \cdot p_\smp3^-  \,, \hspace{1cm} J_\smp3^{-I} = \int \!\! d\Gamma_\smp3\,\,   \Phi_\smp3^* \cdot j_\smp3^{-I} -  \frac{1}{3} \bigl(\sum_{a=1,2,3} \partial_{p_a^I}\Phi_\smp3^* \bigr) \cdot p_\smp3^-\,,
\ee
}
\!where the following notation is used:%
\footnote{We use hermitian $P^-$ and anti-hermitian $J^{-I}$. If, in \rf{29032025-man01-05}, $P_\smp3^-$ is not hermitian, while $J_\smp3^{-I}$ is not anti-hermitian, then the expressions in \rf{29032025-man01-05} should be supplemented by suitable hermitian conjugated cousins.}
{\small
\beq
\label{29032025-man01-10}  &&  \Phi_\smp3^* := \prod_{a=1,2,3} \phi_a^*(p_a,u_a,\alpha_a)\,,
\nonumber\\
&& d\Gamma_\smp3 := (2\pi)^{d-1} \delta^{d-1} \bigl( \sum_{a=1,2,3} p_a \bigr)
\prod_{a=1,2,3} \frac{d^{d-1} p_a}{(2\pi)^{(d-1)/2}} \,, \qquad d^{d-1}p_a = d^{d-2}p_ad\beta_a\,,
\eeq
}
and the densities $p_\smp3^-$ and $j_\smp3^{-I}$ in \rf{29032025-man01-05} are presented as
\be \label{29032025-man01-15}
p_\smp3^- = p_\smp3^-(\Po,\beta_a, u_a,\alpha_a,\zeta_a)\,, \qquad
j_\smp3^{-I} = j_\smp3^{-I}(\Po,\beta_a, u_a, \alpha_a,\zeta_a)\,.
\ee
The density $p_\smp3^-$ is refereed to as cubic vertex. The index $a=1,2,3$ labels three fields entering dynamical generators \rf{29032025-man01-05}, while $\beta_a$ in \rf{29032025-man01-15} stand for three light-cone momenta \rf{27032025-man01-46}. The arguments $u_a$ in \rf{29032025-man01-10}, \rf{29032025-man01-15} stand for unit vectors entering CSFs \rf{27032025-man01-10}, while $\alpha_a$, $\zeta_a$ are oscillators entering integer-spin fields. A new momentum variable $\Po^I$ in \rf{29032025-man01-15} is expressed in terms of the standard momenta $p_a^I$ and $\beta_a$ as
\be \label{29032025-man01-18}
\Po^I := \frac{1}{3}\sum_{a=1,2,3} \betach_a p_a^I\,, \qquad
\betach_a:= \beta_{a+1}-\beta_{a+2}\,, \quad \beta_a:=
\beta_{a+3}\,.
\ee

\noinbf{System of equations for cubic vertex}.
Here we write down the general form of equations and restrictions for the cubic vertex $p_\smp3^-$ and the density $j_\smp3^{-I}$ obtained in Refs.\cite{Metsaev:2017cuz,Metsaev:2018moa},
{\small
\beq
\label{29032025-man01-20}  && \Jbf^{+-}  p_\smp3^-  = 0 \,, \hspace{5.5cm} \hbox{kinematical } \  J^{+-}-\hbox{ symmetry};
\\
\label{29032025-man01-21}  &&  \Jbf^{IJ} p_\smp3^-  = 0\,, \hspace{5.7cm} \hbox{kinematical } \  J^{IJ}-\hbox{ symmetries};
\\
\label{29032025-man01-22}  && j_\smp3^{-I} = - (\Pbf^-)^{-1} \Jbf^{-I} p_\smp3^- \,, \qquad \hspace{2.8cm} \hbox{ dynamical } P^-, J^{-I} \hbox{ symmetries };\qquad
\\
&& \hspace{1.8cm} \hbox{ \it Light-cone gauge dynamical principle}
\nonumber\\
\label{29032025-man01-25}  && p_\smp3^-\,, \ j_\smp3^{-I} \hspace{4mm} \hbox{are expandable in $\Pbf^-$ and well defined for} \ \Pbf^-=0 \,; \quad p_\smp3^-\ne 0 \ \hbox{for} \ \Pbf^-=0\,,
\\
\label{29032025-man01-23}  && p_\smp3^- \,, \ j_\smp3^{-I} \hspace{4mm} \hbox{are expandable in} \  \Po^I\,,
\\
\label{29032025-man01-24}  && p_\smp3^-  \ne \Pbf^- V_\frd\,,\quad  \hbox{if $V_\frd$ is expandable in $\Pbf^-$ and well defined for $\Pbf^-=0$} \,,
\eeq
}
\!
where we use the notation
{\small
\beq
\label{29032025-man01-30} && \hspace{-1.3cm} \Jbf^{+-} =  \Po^I\partial_{\Po^I} + \sum_{a=1,2,3} \beta_a\partial_{\beta_a} \,, \qquad  \Jbf^{IJ}  =  \Po^I \partial_{\Po^J}  -
\Po^J \partial_{\Po^I} + \sum_{a=1,2,3} M_a^{IJ}\,,
\nonumber\\
&& \hspace{-1.3cm} \Pbf^-  = \frac{\Po^I \Po^I}{2\beta} - \sum_{a=1,2,3} \frac{m_a^2}{2\beta_a}\,, \hspace{1.3cm}  \Jbf^{-I}  =   - \frac{\Po^I}{\beta} \No_\beta + \frac{1}{\beta} \Mo^{IJ} \Po^J + \sum_{a=1,2,3}  \frac{\check\beta_a }{6\beta_a} m_a^2 \partial_{\Po^I}  - \frac{1}{\beta_a} M_a^I\,,
\nonumber\\
&& \hspace{-1.3cm} \No_\beta    :=  \frac{1}{3}\sum_{a=1,2,3} \check\beta_a \beta_a \partial_{\beta_a}\,, \hspace{1.5cm}
\Mo^{IJ}   := \frac{1}{3}\sum_{a=1,2,3} \check\beta_a M_a^{IJ}\,, \qquad \beta  :=  \beta_1 \beta_2 \beta_3\,,  \qquad
\eeq
}
\!while the notion of expandable functions we use is defined as follows. If a power series expansion of a function $f(x)$ in the $x$ consists only non-negative integer powers of $x$ then the function $f$ is said to be expandable in $x$,
\be \label{30032025-man01-50}
f(x) = \sum_{n\in \No_0} f_n x^n\,, \qquad \hbox{for function expandable in} \ x\,.
\ee
\!The following comments are in order.

\noinbf{i}) The restriction $\Pbf^-=0$ amounts to the energy conservation law and this restriction is realized either on a physical sheet or on an unphysical sheet. Matrix elements of the vertex $p_\smp3^-$, $j_\smp3^{-I}$ are computed on the surface $\Pbf^-=0$ and we recall that the matrix elements of $p_\smp3^-$ are realized as 3-point scattering amplitudes. In our study, we never ignore the restrictions in \rf{29032025-man01-25}.

\noinbf{ii}) If $V_\frd$ satisfies the restriction in \rf{29032025-man01-24}, then the vertex $p_\smp3^- = \Pbf^- V_\frd$ does not contribute to 3-point scattering amplitudes and can be removed by using field redefinitions (see, e.g., Appendix B in Ref.\cite{Metsaev:2005ar}). Therefore, such vertex is eliminated from our study.
Note however, if $V_\frd$ does not satisfy the restriction in \rf{29032025-man01-24}, while the vertex $p_\smp3^- = \Pbf^- V_\frd$ satisfies restrictions \rf{29032025-man01-25}, then such vertex is allowed in our study.

\noinbf{iii}) In this paper, in order to get most general solutions, we ignore requirement \rf{29032025-man01-23}. This leads to new solutions in Secs. \ref{sec-4-5} and  \ref{sub-sec-5-7}. All our remaining solutions automatically obey \rf{29032025-man01-23}.

\subsection{ Equations for parity-even cubic interaction vertices}\label{sec-3-1}

In general, the cubic vertex $p_\smp3^-$ depends on the following variables
\be \label{30032025-man01-01}
\Po^I\,,\quad  \beta_a\,, \quad u_a^I\,,\quad \alpha_a^I\,, \quad \zeta_a\,, \qquad a=1,2,3\,.
\ee
The $so(d-2)$ algebra symmetries
\rf{29032025-man01-21} tell that cubic vertex $p_\smp3^-$ depends on invariants constructed out of $\Po^I$, $u_a^I$, $\alpha_a^I$, the delta-Kroneker $\delta^{IJ}$, and the Levi-Civita symbol $\epsilon^{ I_1\ldots I_{d-2} }$ of the $so(d-2)$ algebra. The variables $\beta_a$ and $\zeta_a$ are invariants of the $so(d-2)$ algebra. In this paper, to avoid our study being too long, we ignore invariants involving the Levi-Civita symbol.%
\footnote{ For massless integer-spin fields in $\Ro^{4,1}$, the discussion of parity-odd light-cone gauge cubic vertices may be found in Sec.\,8.1 in Ref.\cite{Metsaev:2005ar}, while, for the ones in $\Ro^{2,1}$, the discussion of parity-odd Lorentz covariant vertices may be found in Ref.\cite{Kessel:2018ugi}. An attractive feature of the parity-even vertices is that such vertices admit relatively straightforward generalization to their BRST cousins. BRST studies of interacting integers-spin fields may be found, e.g., in Refs.\cite{Bengtsson:1987jt}-\cite{Vasiliev:2025hfh}, while BRST studies of free CSF may be found, e.g., in Refs.\cite{Bengtsson:2013vra}-\cite{Burdik:2020ror}.
}
So we consider cubic vertices depending on the following invariants of the $so(d-2)$ algebra symmetries:
\beq
\label{30032025-man01-05} && B_a^\alpha, \quad  B_a^u\,,\quad q_{ab}^{\alpha\alpha}\,, \quad q_{ab}^{\alpha u}\,,  \quad q_{ab}^{u\alpha}\,, \quad q_{ab}^{uu}\,,\quad \beta_a\,, \quad \zeta_a\,,
\\
\label{30032025-man01-06} && \Pbf^- \,,
\eeq
where we use the notation%
\footnote{ In \rf{30032025-man01-05}-\rf{30032025-man01-10}, in place of the invariants $\Po^I\alpha_a^I$, $\Po^I u_a^I$, and $\Po^I\Po^I$, we find it convenient to use the respective invariants $B_a^\alpha$, $B_a^u$, and $\Pbf^-$, where $\Pbf^-$ is defined in \rf{29032025-man01-30}.}
\beq
\label{30032025-man01-10} && B_a^\alpha  =  \frac{\Po^I\alpha_a^I}{\beta_a}\,,\qquad  B_a^u  =  \frac{\Po^Iu_a^I}{\beta_a}\,,
\nonumber\\
&& q_{ab}^{\alpha\alpha}  =  \alpha_a^I \alpha_b^I\,, \qquad q_{ab}^{\alpha u}  =  \alpha_a^I u_b^I\,, \qquad q_{ab}^{u\alpha}  =  u_a^I \alpha_b^I\,,\qquad  q_{ab}^{uu}  =  u_a^I u_b^I\,.
\eeq

Cubic vertices depending only on variables \rf{30032025-man01-05}, \rf{30032025-man01-06} are referred to as parity-even cubic vertices or simply to as cubic vertices.
Using field redefinitions, we can remove $\Pbf^-$-terms in the cubic vertex.
So we deal with cubic vertex $p_\smp3^-$ which depends only on variables \rf{30032025-man01-05}.%
\footnote{ Only in Sec.\,\ref{subsec-6-5}, upon considering distributional vertices, we prefer to use cubic vertex depending on $\Pbf^-$-terms.
}

Using \rf{29032025-man01-30}, we find the realization of $\Jbf^{-I}$ on $p_\smp3^-$ in terms of variables \rf{30032025-man01-05},

\be \label{30032025-man01-15}
\Jbf^{-I}  =    \Pbf^- \sum_{a=1,2,3} \frac{2\betach_a}{3\beta_a}\big(\alpha_a^I \partial_{B_a^\alpha} + u_a^I \partial_{B_a^u} \big) + \Po^I G_\beta + \sum_{a=1,2,3} \frac{1}{\beta_a} \big( \alpha_a^I G_a^\alpha +  u_a^I G_a^u\big)\,,
\ee
where $G_\beta$, $G_a^\alpha$, $G_a^u$ are presented in Appendix A. Using the explicit expressions of $G_\beta$, $G_a^\alpha$, $G_a^u$, we learn that action of the operators $G_\beta$, $G_a^\alpha$, $G_a^u$ on the vertex $p_\smp3^-$ does not produce terms proportional to $\Pbf^-$. Using \rf{29032025-man01-22}, \rf{29032025-man01-25}, we conclude then that the vertex $p_\smp3^-$ should satisfy the equations

\beq
\label{30032025-man01-16} && G_a^\alpha p_\smp3^- =0\,,\qquad  a=1,2,3;\qquad G_b^u p_\smp3^- =0\,,\qquad  b=1,2,3;\qquad a\ne b\,,
\\[5pt]
\label{30032025-man01-17} && G_\beta p_\smp3^- = 0\,,
\eeq
where restriction $a\ne b$ \rf{30032025-man01-16} reflects the fact that the $p_\smp3^-$ cannot depend on the variables $B_a^\alpha$ and $B_a^u$ having one and same index $a$.
The remaining equation to be taken into account is given in \rf{29032025-man01-20}. In terms of the vertex $p_\smp3^-$ that depends on the variables in \rf{30032025-man01-05}, equation \rf{29032025-man01-20} is represented as

\be \label{30032025-man01-25}
\sum_{a=1,2,3} \beta_a \partial_{\beta_a} p_\smp3^-=0\,.
\ee
Use of \rf{29032025-man01-22}, \rf{29032025-man01-25} and equations \rf{30032025-man01-16}, \rf{30032025-man01-17} provides us the expression for the density $j_\smp3^{-I}$,

\be \label{30032025-man01-30}
j_\smp3^{-I} = -\sum_{a=1,2,3} \frac{2\betach_a}{3\beta_a}\big(\alpha_a^I \partial_{B_a^\alpha} + u_a^I \partial_{B_a^u} \big) p_\smp3^-\,.
\ee

The following remarks are in order.

\medskip
\noinbf{i)} Equations \rf{30032025-man01-16}-\rf{30032025-man01-25} constitute the complete system of equations which allows us determine all f-solutions for cubic vertices in the framework of the vector formulation we use in this paper. The corresponding expression for the density $j_\smp3^{-I}$ is given in \rf{30032025-man01-30}.

\medskip
\noinbf{ii)} As we demonstrate below, equations \rf{30032025-man01-16}-\rf{30032025-man01-25} allow us also to find some {\it particular distributional solutions}. Note however it is unlikely that these equations are  applicable for finding {\it all  distributional solutions} for cubic vertices. This is to say that, for example, equations \rf{30032025-man01-16} for the vertices involving three CSFs given by $G_a^u p_\smp3^- =0$, $a=1,2,3$, are obtained  under the assumption that the three vectors $u_a^I$, $a=1,2,3$ are independent. Obviously this is not the case for the distributional cubic vertices like $p_\smp3^- \sim \delta(u_1-u_2)\delta(u_2-u_3)$. All in all we have no proof and hence do not state that equations \rf{30032025-man01-16}-\rf{30032025-man01-25} are  applicable for finding all distributional cubic vertices. In our view, a systematic method for finding all distributional cubic vertices remains to be understood.

\medskip
\noinbf{iii)} Equations \rf{30032025-man01-16}-\rf{30032025-man01-17} are first-order differential equations with respect to variables \rf{30032025-man01-05}. They can systematically be solved by applying the standard method of characteristics. The procedure for finding solutions to equations \rf{30032025-man01-16}-\rf{30032025-man01-25} is the same as the one in Appendix D in Ref.\cite{Metsaev:2005ar}. Therefore, in the interest of the brevity, we skip the details of the derivation and present only our final results for f-solutions. We will make a few comments concerning details of the derivation of distributional solutions.

\medskip
\noinbf{Local and non-local vertices}. Before to proceed we explain our terminology. If a cubic vertex $p_\smp3^-$ is a finite-order polynomial in the momentum $\Po^I$, then such vertex is referred to as local vertex. Otherwise the vertex is referred to as non-local vertex. As we demonstrate below, all solutions for our cubic vertices involving at least one massive/massless continuous-spin field turn out to be non-local.

Dependence of cubic vertices on the momentum $\Po^I$ is realized only trough the variables $B_a^\alpha$ and $B_a^u$ defined in \rf{30032025-man01-10}. The variables $B_a^\alpha$ are relevant for cubic vertices involving at least one integer-spin or triplet field, while the variables $B_a^u$ are relevant for cubic vertices involving at least one  CSF. If the index $a$ labels the tower of integer-spin or triplet fields shown in \rf{27032025-man01-41}, then the cubic vertex is expandable in $B_a^\alpha$, while, if the index $a$ labels integer-spin or triplet field, then the cubic vertex is realized as finite order polynomial in $B_a^\alpha$. If the index $a$ labels CSF, then the dependence of the cubic vertex on  variables $B_a^u$ turns out more complicated. It is the dependence on the variables $B_a^u$ that leads to appearance of non-locality. Depending on the way in which the non-locality manifests itself in our f-solutions for cubic vertices, we use the following terminology:

{\small
\beq
\label{30032025-man01-35} && \hspace{-1.5cm} \begin{array}{ll}
e^{W_\Erm},  \hspace{7mm}   W_\Erm = X_{(1)} \,, \hspace{2cm}   & \hbox{exponential non-locality ($E$-non-locality)};
\\[5pt]
X_\PLrm^\NN, \hspace{7mm}  X_\PLrm = X_{(1)}\,, \hspace{0.5cm} \NN \in \Co /\No_0\,,  & \hbox{power-law non-locality ($PL$-non-locality)};
\\[5pt]
e^{W_\EPLrm},  \hspace{3mm}  W_\EPLrm = X_\smone / Y_\smone   \qquad &\hbox{exponential-power-law non-locality ($EPL$-non-locality)};
\end{array}
\nonumber\\
&& \hspace{3cm} X_\smone\,, \ Y_\smone -\hbox{ are degree-1 polynomials in $\Po^I$}.
\eeq
}
The explicit light-cone gauge representation for the degree-1 polynomials $X_{(1)}$, $Y_{(1)}$ entering non-localities \rf{30032025-man01-35}  may be found below in Secs.\ref{sec-3csf}-\ref{sec-1csf}. Here we would like to discuss briefly a Lorentz covariant representation of the non-localities shown in \rf{30032025-man01-35}. To this end we use the Lorentz covariant description of massive/massless CSFs discussed in Appendices B, C.  This implies that Lorentz covariant cousins of our light-cone gauge vertices depend on three momenta $p_a^\mu$ and three Lorentz $so(d-1,1)$ algebra vectors $\xi_a^\mu$, where the index $a=1,2,3$ label three fields entering the vertices, while $\mu=0,1,\ldots,d-1$ is a vector index of the Lorentz algebra. By using the following shortcuts $\xi_a\xi_b:=\xi_a^\mu \xi_b^\mu$, $p_a \xi_b: = p_a^\mu \xi_b^\mu$, we now discuss the Lorentz covariant representation of non-localities \rf{30032025-man01-35} in turn.

\medskip
\noinbf{E-non-locality}. Such non-locality is realized for the vertices given in \rf{26032025-man01-02}, \rf{26032025-man01-04}, \rf{26032025-man01-09}-\rf{26032025-man01-11}, \rf{26032025-man01-16}, \rf{26032025-man01-18}.
These vertices involve at least one massless CSFs and at least one massive CSF (or one massive integer-spin field). The Lorentz covariant representation of the quantity $W_\Erm$ \rf{30032025-man01-35} takes the following form:
for \rf{26032025-man01-02}, $W_\Erm = \frac{2\irm}{m_2^2-m_1^2} p_2\xi_3$; for \rf{26032025-man01-04}, $W_\Erm  = \frac{2\irm}{m_1^2}\big(p_1\xi_2 + p_1\xi_3\big)$; for \rf{26032025-man01-09}, $W_\Erm = \frac{2\irm}{m_1^2-m_3^2} p_1\xi_2$; for \rf{26032025-man01-10}, $W_\Erm = \frac{2\irm}{m_1^2} p_1\xi_2$;  for \rf{26032025-man01-11}, $W_\Erm =\frac{2\irm}{m_3^2}\big( p_3\xi_1 + p_3\xi_2\big)$;
for \rf{26032025-man01-16}, $W_\Erm = \frac{2\irm}{m_3^2-m_2^2} p_3\xi_1$;
for \rf{26032025-man01-18}, $W_\Erm = \frac{2\irm}{m_2^2} p_2\xi_1$.

\medskip
\noinbf{PL-non-locality}. Such non-locality is realized for the vertices involving massive CSFs. These vertices are given in \rf{26032025-man01-01}-\rf{26032025-man01-04}, \rf{26032025-man01-06}-\rf{26032025-man01-10}, and \rf{26032025-man01-13}-\rf{26032025-man01-15}. The Lorentz covariant representation for the PL-non-locality takes the form $\prod_a ( p_{a-1} \xi_a)^{\NN_a}$, $\NN_a \in \Co /\No_0$, where the product is performed for the indices $a$ corresponding to the massive CSFs.

\medskip
\noinbf{EPL-non-locality}. Such non-locality is realized only for the vertex given in \rf{26032025-man01-03}. The corresponding Lorentz covariant representation of the quantity $W_\EPLrm$ \rf{30032025-man01-35} takes the following form: $W_\EPLrm = \frac{\irm}{2}\big( \frac{\xi_1\xi_3}{\xi_1 p_3 } + \frac{\xi_2\xi_3}{\xi_2 p_3 }\big)$\,.

\newsection{ \large Cubic vertices for three continuous-spin fields}\label{sec-3csf}

In this Section, we discuss cubic vertices which involve three massive/massless CSFs. As seen from our classification in \rf{26032025-man01-01}-\rf{26032025-man01-05}, cubic vertices for such fields can be separated into the five particular cases. We now present our results for these particular cases in turn.

\vspace{-0.2cm}
\subsection{   Three massive CSFs}

Using the shortcut $(m,\SSc)_\CSFsm$ for a massive continuous-spin field, we consider the cubic vertex that involves the following fields:
\beq
\label{31032025-man01-1-01}&& \hspace{-1cm} (m_1,\SSc_1)_\CSFsm\hbox{-}(m_2,\SSc_2)_\CSFsm\hbox{-}(m_3,\SSc_3)_\CSFsm\,, \qquad    m_1^2 < 0\,,\qquad    m_2^2 < 0\,,\qquad m_3^2 < 0 \,,
\nonumber\\
&& \hspace{-1cm} \hbox{\small three massive continuous-spin fields.}\quad
\eeq
Our general f-solution to equations for cubic vertex $p_\smp3^-$  \rf{30032025-man01-16}-\rf{30032025-man01-25} can be presented as
\beq
\label{31032025-man01-1-05} && \hspace{-1cm} p_\smp3^- = (L_1^u)^{ \SSc_1}(L_2^u)^{ \SSc_2}(L_3^u)^{ \SSc_3} V(\Qbf_{12}^{uu}\,, \Qbf_{23}^{uu}\,, \Qbf_{32}^{uu})\,,
\\
\label{31032025-man01-1-06} &&  L_a^u =  B_a^u + \frac{\betach_a}{2\beta_a} |m_a| - \frac{m_{a+1}^2 - m_{a+2}^2}{2|m_a|}\,,
\nonumber\\
&& Q_{aa+1}^{uu}  =  q_{aa+1}^{uu} + \frac{1}{|m_a|} L_{a+1}^u  -  \frac{1}{ |m_{a+1}| } L_a^u  + \frac{1}{ 2|m_a| |m_{a+1}|} (m_{a+2}^2 - m_a^2 - m_{a+1}^2  )\,,
\nonumber\\
&& \Qbf_{aa+1}^{uu} = \frac{Q_{aa+1}^{uu} }{L_a^u L_{a+1}^u }\,,
\eeq
where, in \rf{31032025-man01-1-05}, we introduce vertex $V$ which depends on three variables shown explicitly in \rf{31032025-man01-1-05} and defined in \rf{31032025-man01-1-06}, \rf{30032025-man01-10}. The following remarks are in order.

\noinbf{i)} The vertex $V$ depending on the variables $\Qbf_{12}^{uu}$, $\Qbf_{23}^{uu}$, $\Qbf_{31}^{uu}$ is not fixed by equations \rf{30032025-man01-16}-\rf{30032025-man01-25}. The vertex $V$ is a freedom of our solution to cubic vertex $p_\smp3^-$. So there are many cubic vertices.

\noinbf{ii)} The momentum $\Po^I$ enters the variables $B_a^u$ \rf{30032025-man01-10}. From  \rf{31032025-man01-1-05}, \rf{31032025-man01-1-06}, we then see that
the dependence of $p_\smp3^-$ on $\Po^I$ is governed by the variables $L_1^u$, $L_2^u$, $L_3^u$,  and  $\Qbf_{12}^{uu}$, $\Qbf_{23}^{uu}$, $\Qbf_{31}^{uu}$ which turn out to be  expandable in $\Po^I$ (see definition in \rf{30032025-man01-50}).

\noinbf{iii)} Vertex $p_\smp3^-$ \rf{31032025-man01-1-05} is non-polynomial in $\Po^I$ and hence non-local. For example, taking into account the  values of $\SSc$ given in \rf{27032025-man01-60} and the pre-factor $(L_1^u)^{ \SSc_1}(L_2^u)^{ \SSc_2}(L_3^u)^{ \SSc_3}$ \rf{31032025-man01-1-05}, we see that vertex $p_\smp3^-$ \rf{31032025-man01-1-05} exhibits $PL$-non-locality defined in \rf{30032025-man01-35}. This $PL$-non-locality is unavoidable feature of the vertex $p_\smp3^-$ as either choice of the vertex $V$ leads to the $PL$-non-local vertex $p_\smp3^-$.

\vspace{-0.2cm}
\subsection{  Two massive CSFs (non-equal masses) and one massless CSF}

Using the shortcuts $(m,\SSc)_\CSFsm$ and $(0,\kappa)_\CSFsm$ for the respective  massive and massless continuous-spin  fields, we consider the cubic vertex that involves the following fields:
\beq
\label{31032025-man01-2-01} && \hspace{-1.2cm} (m_1,\SSc_1)_\CSFsm\hbox{-}(m_2,\SSc_2)_\CSFsm\hbox{-}(0,\kappa_3)_\CSFsm\,, \qquad     m_1^2<0\,, \qquad m_2^2<0\,, \qquad m_1^2 \ne m_2^2 \,,
\nonumber\\
&& \hspace{-1.2cm} \hbox{\small two massive CSFs with non-equal masses and one massless CSF.}\quad
\eeq
Our general f-solution to equations for cubic vertex \rf{30032025-man01-16}-\rf{30032025-man01-25}  can be presented as
\beq
\label{31032025-man01-2-05} && \hspace{-1cm} p_\smp3^- =  e^W   (L_1^u)^{\SSc_1}(L_2^u)^{\SSc_2} V(
\Qbf_{12}^{uu},\Qbf_{23}^{uu}, \Qbf_{31}^{uu})\,,
\\
&& W = - \frac{2\irm \kappa_3 B_3^u}{m_1^2-m_2^2}\,,
\nonumber\\
\label{31032025-man01-2-06} &&  L_1^u =  B_1^u + \frac{\betach_1}{2\beta_1} |m_1| - \frac{m_2^2}{2|m_1|}\,,\qquad  L_2^u =  B_1^u + \frac{\betach_2}{2\beta_2} |m_2| + \frac{m_1^2}{2|m_2|}\,,
\nonumber\\
&& Q_{12}^{uu}  =  q_{12}^{uu} + \frac{1}{|m_1|} L_2^u  -  \frac{1}{ |m_2| } L_1^u  - \frac{1}{ 2|m_1| |m_2| } ( m_1^2 + m_2^2  )\,,
\nonumber\\
&& Q_{23}^{uu}  =  q_{23}^{uu} + \frac{1}{|m_2|} B_3^u  -  \frac{2}{ m_1^2-m_2^2 } L_2^u B_3^u\,,
\nonumber\\
&& Q_{31}^{uu}  =  q_{31}^{uu} - \frac{1}{|m_1|} B_3^u  +  \frac{2}{ m_1^2-m_2^2 } B_3^u L_1^u\,,
\nonumber\\
&& \Qbf_{12}^{uu} = \frac{ Q_{12}^{uu} }{ L_1^u L_2^u }\,, \qquad
\Qbf_{23}^{uu} = \frac{ Q_{23}^{uu} }{L_2^u}\,,\qquad \Qbf_{31}^{uu} = \frac{ Q_{31}^{uu} }{L_1^u}\,,
\eeq
where, in \rf{31032025-man01-2-05}, we introduce vertex $V$ which depends on three variables shown explicitly in \rf{31032025-man01-2-05} and defined in \rf{31032025-man01-2-06}, \rf{30032025-man01-10}. The following remarks are in order.

\noinbf{i)} Vertex $V$ \rf{31032025-man01-2-05} depending on the variables $\Qbf_{12}^{uu}$, $\Qbf_{23}^{uu}$, $\Qbf_{31}^{uu}$ is not fixed by equations \rf{30032025-man01-16}-\rf{30032025-man01-25}. The vertex $V$ is a freedom of our solution to cubic vertex $p_\smp3^-$. So there are many cubic vertices.

\noinbf{ii)} The momentum $\Po^I$ enters the variables $B_a^u$ \rf{30032025-man01-10}. From  \rf{31032025-man01-2-05},\rf{31032025-man01-2-06}, we see then that
the dependence of the vertex $p_\smp3^-$ on the momentum $\Po^I$ is governed by the variables $W$, $L_1^u$, $L_2^u$,  and  $\Qbf_{12}^{uu}$, $\Qbf_{23}^{uu}$, $\Qbf_{31}^{uu}$ which turn out to be  expandable in $\Po^I$ (see definition \rf{30032025-man01-50}).

\noinbf{iii)} Vertex $p_\smp3^-$ \rf{31032025-man01-2-05} is non-polynomial in the momentum $\Po^I$ and hence non-local. For example, taking into account the  values of the spin parameter $\SSc$ given in \rf{27032025-man01-60} and the pre-factor $e^W (L_1^u)^{ \SSc_1}(L_2^u)^{ \SSc_2}$ in \rf{31032025-man01-2-05}, we see that vertex $p_\smp3^-$ \rf{31032025-man01-2-05} exhibits both $E$- and $PL$-non-localities defined in \rf{30032025-man01-35}. Note that these  non-localities are unavoidable feature of the vertex $p_\smp3^-$ as either choice of the vertex $V$ leads to the $E$- and $PL$-non-local vertex $p_\smp3^-$.

\vspace{-0.2cm}
\subsection{  Two massive CSFs (equal masses) and one massless CSF}

Using the shortcuts $(m,\SSc)_\CSFsm$ and $(0,\kappa)_\CSFsm$ for the respective  massive and massless continuous-spin fields, we consider the cubic vertex that involves the following fields:
\beq
\label{31032025-man01-3-01} && \hspace{-1cm} (m_1,\SSc_1)_\CSFsm\hbox{-}(m_2,\SSc_2)_\CSFsm\hbox{-}(0,\kappa_3)_\CSFsm\,, \qquad     m_1^2=m^2\,, \qquad m_2^2 = m^2\,, \qquad m^2 < 0 \,,
\nonumber\\
&& \hspace{-1cm} \hbox{\small two massive CSFs with equal masses and one massless CSF.}\quad
\eeq
General solution to equations for cubic vertex \rf{30032025-man01-16}-\rf{30032025-man01-25} can be presented as
\beq
\label{31032025-man01-3-05} && \hspace{-1cm} p_\smp3^- = e^W   (L_1^u)^{\SSc_1}(L_2^u)^{\SSc_2} V(B_3^u\,,
\Qbf_{12}^{uu},\Qbf^{uu})\,,
\\
\label{31032025-man01-3-06} &&  L_1^u =  B_1^u - \frac{\beta_3}{\beta_1}|m|\,, \qquad L_2^u =  B_2^u + \frac{\beta_3}{\beta_2} |m| \,, \qquad
\nonumber\\
&& Q_{12}^{uu} = q_{12}^{uu} + \frac{1}{|m|} L_2^u - \frac{1}{|m|} L_1^u + 1\,,
\nonumber\\
&& Q_{23}^{uu} = q_{23}^{uu} + \frac{1}{|m|} B_3^u \,, \hspace{1cm}  Q_{31}^{uu} = q_{31}^{uu} - \frac{1}{|m|} B_3^u\,,
\nonumber\\
&& \Qbf_{12}^{uu} = \frac{ Q_{12}^{uu} }{ L_1^u L_2^u }\,, \hspace{2.3cm}
\Qbf_{23}^{uu} = \frac{ Q_{23}^{uu} }{L_2^u}\,,\qquad \Qbf_{31}^{uu} = \frac{ Q_{31}^{uu} }{L_1^u}\,,
\nonumber\\
&& W = \frac{\irm \kappa_3}{2} (  \Qbf_{31}^{uu} - \Qbf_{23}^{uu})\,, \qquad \Qbf^{uu} = \Qbf_{23}^{uu} + \Qbf_{31}^{uu}\,,
\eeq
where, in \rf{31032025-man01-3-05}, we introduce vertex $V$ which depends on three variables shown explicitly in \rf{31032025-man01-3-05} and defined in \rf{31032025-man01-3-06}, \rf{30032025-man01-10}. The following remarks are in order.

\noinbf{i)} Vertex $V$ \rf{31032025-man01-3-05} is not fixed by equations \rf{30032025-man01-16}-\rf{30032025-man01-25}. The vertex $V$ is a freedom of our solution to cubic vertex $p_\smp3^-$. This implies that there are many cubic vertices.

\noinbf{ii)} The momentum $\Po^I$ enters the variables $B_a^u$ \rf{30032025-man01-10}. From  \rf{31032025-man01-3-05}, \rf{31032025-man01-3-06}, we then see that
the dependence of the vertex $p_\smp3^-$ on the momentum $\Po^I$ is governed by the variables $W$, $L_1^u$, $L_2^u$,  and $B_3^u$, $\Qbf_{12}^{uu}$, $\Qbf^{uu}$ which turn out to be  expandable in $\Po^I$ (see definition in \rf{30032025-man01-50}).

\noinbf{iii)} Vertex $p_\smp3^-$ \rf{31032025-man01-3-05} is non-polynomial in the momentum $\Po^I$ and hence non-local. For example, taking into account the values of $\SSc$ given in \rf{27032025-man01-60} and the pre-factor $e^W (L_1^u)^{ \SSc_1} (L_2^u)^{ \SSc_2}$ in \rf{31032025-man01-3-05}, we see that vertex $p_\smp3^-$ \rf{31032025-man01-3-05} exhibits both $PL$- and $EPL$- non-localities defined in \rf{30032025-man01-35}. These non-localities are unavoidable feature of the vertex $p_\smp3^-$ as either choice of the vertex $V$ leads to the $PL$- and $EPL$-non-local vertex $p_\smp3^-$. We note that, among all f-solutions  considered in this paper, the solution given in \rf{31032025-man01-3-05} is the only f-solutions which exhibits $EPL$-non-locality.

\vspace{-0.2cm}
\subsection{ One massive CSF and two massless CSFs }

Using the shortcuts $(m,\SSc)_\CSFsm$ and $(0,\kappa)_\CSFsm$ for the respective  massive and massless continuous-spin fields, we consider the cubic vertex that involves the following fields:
\beq
\label{31032025-man01-4-01} && \hspace{-1cm} (m_1,\SSc_1)_\CSFsm\hbox{-}(0,\kappa_2)_\CSFsm\hbox{-}(0,\kappa_3)_\CSFsm  \,, \qquad     m_1^2 < 0 \,,
\nonumber\\
&& \hspace{-1cm} \hbox{\small one massive CSF and two massless CSFs.}\quad
\eeq
We find the following general solution to equations for cubic vertex \rf{30032025-man01-16}-\rf{30032025-man01-25}:
\beq
\label{31032025-man01-4-05} && \hspace{-1cm} p_\smp3^- = e^W (L_1^u)^{\SSc_1} V(\Qbf_{12}^{uu}\,, Q_{23}^{uu}\,, \Qbf_{31}^{uu})\,,
\\
\label{31032025-man01-4-06} && W = \frac{2\irm}{m_1^2} \big(\kappa_2 B_2^u - \kappa_3 B_3^u\big)\,,  \hspace{1.5cm} L_1^u =  B_1^u   + \frac{\betach_1}{2\beta_1} |m_1|\,,
\nonumber\\
&& Q_{23}^{uu}  = q_{23}^{uu} -  \frac{2}{m_1^2} B_2^u B_3^u\,,
\hspace{2cm} Q_{31}^{uu}  = q_{31}^{uu} - \frac{1}{|m_1|} B_3^u +  \frac{2}{m_1^2} B_3^u L_1^u\,,
\nonumber\\
&& Q_{12}^{uu}  = q_{12}^{uu} + \frac{1}{|m_1|} B_2^u + \frac{2}{m_1^2} L_1^u B_2^u \,,
\nonumber\\
&& \Qbf_{12}^{uu} = \frac{ Q_{12}^{uu} }{L_1^u}\,,\qquad \Qbf_{31}^{uu} = \frac{ Q_{31}^{uu} }{L_1^u}\,,
\eeq
where, in \rf{31032025-man01-4-05}, we introduce vertex $V$ which depends on three variables shown explicitly in \rf{31032025-man01-4-05} and defined in \rf{31032025-man01-4-06}, \rf{30032025-man01-10}. The following remarks are in order.

\noinbf{i)} Vertex $V$ \rf{31032025-man01-4-05} is not fixed by equations \rf{30032025-man01-16}-\rf{30032025-man01-25}. The vertex $V$ is a freedom of our solution to cubic vertex $p_\smp3^-$. In other words, there are many cubic vertices.

\noinbf{ii)} The momentum $\Po^I$ enters the variables $B_a^u$ \rf{30032025-man01-10}. From  \rf{31032025-man01-4-05}, \rf{31032025-man01-4-06}, we then see that
the dependence of the vertex $p_\smp3^-$ on the momentum $\Po^I$ is governed by the variables $W$, $L_1^u$, and  $\Qbf_{12}^{uu}$, $Q_{23}^{uu}$, $\Qbf_{31}^{uu}$ which turn out to be  expandable in $\Po^I$ (see definition in \rf{30032025-man01-50}).

\noinbf{iii)} Vertex $p_\smp3^-$ \rf{31032025-man01-4-05} is non-polynomial in the momentum $\Po^I$ and hence non-local. For example, taking into account the  values of the spin parameter $\SSc$ given in \rf{27032025-man01-60} and the pre-factor $e^W (L_1^u)^{ \SSc_1}$ in \rf{31032025-man01-4-05}, we see that cubic vertex \rf{31032025-man01-4-05} exhibits both $E$- and $PL$-non-localities defined in \rf{30032025-man01-35}. Note that these  non-localities are unavoidable feature of the vertex $p_\smp3^-$ as either choice of the vertex $V$ leads to the $E$- and $PL$-non-local vertex $p_\smp3^-$.

\vspace{-0.2cm}
\subsection{ Three massless CSFs }\label{sec-4-5}

Using the shortcut $(0,\kappa)_\CSFsm$ for a massless continuous spin field, we consider the cubic vertex that involves the following fields:
\beq
\label{31032025-man01-5-01} && \hspace{-1cm} (0,\kappa_1)_\CSFsm\hbox{-}(0,\kappa_2)_\CSFsm\hbox{-}(0,\kappa_3)_\CSFsm\,,
\qquad \hbox{\small three massless CSFs.}\quad
\eeq
We formulate our 2 statements.

\noinbf{Statement 1}. There are no f-solutions to equations for cubic vertex \rf{30032025-man01-16}-\rf{30032025-man01-25}.

\noinbf{Statement 2}. The particular distributional solution to equations for cubic vertex \rf{30032025-man01-16}-\rf{30032025-man01-25} takes the following form:
\beq
\label{31032025-man01-5-05} && \hspace{-1cm} p_\smp3^- =  e^W V(B_1^u,B_2^u,B_3^u, C_{123}^{uuu})\delta( \sum_{a=1,2,3} \kappa_a B_a^u)\,,
\\
\label{31032025-man01-5-06} && W = \irm\sum_{a=1,2,3} \frac{ \kappa_a B_a^u - \kappa_{a+1} B_{a+1}^u }{ 3B_a^u B_{a+1}^u } q_{aa+1}^{uu}\,,
\hspace{1cm} C_{123}^{uuu} = B_1^u q_{23}^{uu} + B_2^u q_{31}^{uu} + B_3^u q_{12}^{uu}\,, \qquad
\eeq
where, in \rf{31032025-man01-5-05}, a new vertex $V$ depends formally on four variables $B_1^u$, $B_2^u$, $B_3^u$, $C_{123}^{uuu}$ defined in \rf{31032025-man01-5-06}, \rf{30032025-man01-10}, while $\delta$ stands for the $1$-dimensional Dirac-delta function. Obviously, in view of $\delta$-function, the vertex $V$ actually depends on three variables. The following remarks are in order.

\noinbf{i)} Vertex $V$ \rf{31032025-man01-5-05} is not fixed by equations \rf{30032025-man01-16}-\rf{30032025-man01-25}. The vertex $V$ is a freedom of our solution to cubic vertex $p_\smp3^-$. In other words, there are many cubic vertices.

\noinbf{ii)} The momentum $\Po^I$ enters the variables $B_a^u$ \rf{30032025-man01-10}. From  \rf{31032025-man01-5-05}, \rf{31032025-man01-5-06}, we see that
the dependence of $p_\smp3^-$ on $\Po^I$ is governed by $B_1^u$, $B_2^u$, $B_3^u$, $C_{123}^{uuu}$, and $W$. The variables $B_1^u$, $B_2^u$, $B_3^u$, $C_{123}^{uuu}$ are expandable in $\Po^I$, while the variable $W$ does not (see definition in \rf{30032025-man01-50}).

\noinbf{iii)} Vertex $p_\smp3^-$ \rf{31032025-man01-5-05} is non-polynomial in $\Po^I$ and hence non-local. In view of the variable $W$, the vertex $p_\smp3^-$ is non-expandable in $\Po^I$ and hence does not meet restriction \rf{29032025-man01-23}.

\noinbf{iv)} Using $\Pbf^-$ \rf{29032025-man01-30}, we note that, for massless fields, the restriction $\Pbf^-=0$ amounts to the restriction $\Po^I\Po^I=0$, which in turn implies $\Po^I=0$ and hence, on the physical sheet, the vertex $p_\smp3^-$ does not meet restriction \rf{29032025-man01-25}. In order to respect restriction \rf{29032025-man01-25} at least formally, one can use the unphysical sheet, $\Po^I \in \Co$.

\noinbf{Proof of Statements 1, 2}. From \rf{30032025-man01-17}, \rf{30032025-man01-25}, we learn that $p_\smp3^-$ is independent of the momenta $\beta_1$, $\beta_2$, $\beta_3$. This implies that cubic vertex depends on the variables $B_a^u$, $q_{aa+1}^{uu}$, $a=1,2,3$ and the remaining equations \rf{30032025-man01-16} take the form
\be \label{31032025-man01-5-10}
G_a^u p_\smp3^- = 0 \,, \qquad G_a^u =   B_{a+2}^u  \partial_{ q_{a+2 a}^{uu} } -  B_{a+1}^u  \partial_{ q_{aa+1}^{uu} }  + \irm \kappa_a\,, \qquad a=1,2,3\,.
\ee
In view of the relation for the operators $G_a^u$,
\be \label{31032025-man01-5-12}
\sum_{a=1,2,3} B_a^u G_a^u = \irm \Bbf_\kappa\,, \qquad \Bbf_\kappa= \kappa_1 B_1^u + \kappa_2 B_2^u + \kappa_3 B_3^u\,,
\ee
we see then that equations \rf{31032025-man01-5-10} lead to the relation $\Bbf_\kappa p_\smp3^-=0$ which implies that f-solution to equations \rf{31032025-man01-5-10} does not exist and this is the content of our {\bf Statement 1}. The only possibility for a non-trivial solution is to consider the distributional cubic vertex $p_\smp3^-$ given by
\be \label{31032025-man01-5-15}
p_\smp3^- = V_\frm \delta(\Bbf_\kappa)\,,\qquad  V_\frm  = V_\frm(B_1^u,B_2^u,B_3^u,q_{12}^{uu},q_{23}^{uu},q_{31}^{uu})\,,
\ee
where $V_\frm$ is new vertex. Plugging $p_\smp3^-$ \rf{31032025-man01-5-15} in equations \rf{31032025-man01-5-10}, we find solution \rf{31032025-man01-5-05}.

\vspace{-0.2cm}
\newsection{ \large Cubic vertices for two continuous-spin fields and one integer-spin field}\label{sec-2csf}

In this Section, we discuss cubic vertices which involve two massive/massless CSFs and one massive/massless integers-spin field. From our classification in \rf{26032025-man01-06}-\rf{26032025-man01-12}, we note the seven particular cases. We now present our results for these particular cases in turn.

\vspace{-0.2cm}
\subsection{ \large Two massive CSFs and one massive integer-spin field}\label{sec-5-1}

Using the shortcuts $(m,\SSc)_\CSFsm$ and $(m,s)$ for the respective  massive continuous-spin field and massive integer-spin field, we are going to find the cubic vertex that involves the following fields:
\beq
\label{31032025-man01-6-01} && (m_1,\SSc_1)_\CSFsm\hbox{-}(m_2,\SSc_2)_\CSFsm \hbox{-}(m_3,s_3) \,,\qquad m_1^2 < 0\,, \qquad m_2^2 < 0\,, \qquad m_3^2 > 0\,,
\nonumber\\
&&  \hbox{\small two massive CSFs and one massive integer-spin field.}\quad
\eeq

\noinbf{Two massive CSFs and one tower of massive integer-spin triplet fields}.
Before discussing cubic vertex for fields in \rf{31032025-man01-6-01}, we consider cubic vertex for two CSFs shown in \rf{31032025-man01-6-01} and one tower of massive triplet fields \rf{27032025-man01-41} having mass $m_3$. For this case, our general solution to equations \rf{30032025-man01-16}-\rf{30032025-man01-25} can be presented as
\beq
\label{31032025-man01-6-05} && \hspace{-1cm} p_\smp3^- = (L_1^u)^{ \SSc_1} (L_2^u)^{ \SSc_2} V(L_3^\alpha\,, \Qbf_{12}^{uu}\,, \Qbf_{23}^{u\alpha}\,, \Qbf_{31}^{\alpha u}\,;\, Q_{33}^{\alpha\alpha})\,,
\\
\label{31032025-man01-6-06} &&  L_a^u = B_a^u + \frac{\betach_a}{2\beta_a} |m_a| - \frac{m_{a+1}^2 - m_{a+2}^2}{2|m_a|}\,,\qquad a =1,2\,,
\nonumber\\
&&  L_3^\alpha =  B_3^\alpha - \frac{\betach_3}{2\beta_3} m_3\zeta_3 - \frac{m_1^2 - m_2^2}{2m_3}\zeta_3\,,
\nonumber\\
&& Q_{12}^{uu}  =  q_{12}^{uu} + \frac{1}{|m_1|} L_2^u  -  \frac{1}{|m_2|} L_1^u  + \frac{1}{ 2|m_1| |m_2|} (m_3^2 - m_1^2 - m_2^2 )\,,
\nonumber\\
&& Q_{23}^{u\alpha}  =  q_{23}^{u\alpha} + \frac{1}{|m_2|} L_3^\alpha  -  \frac{\zeta_3}{ m_3 } L_2^u  + \frac{ \zeta_3  }{ 2 |m_2| m_3 } (m_1^2 - m_2^2 - m_3^2  )\,,
\nonumber\\
&& Q_{31}^{\alpha u}  =  q_{31}^{\alpha u} + \frac{\zeta_3}{m_3} L_1^u  -  \frac{1}{ |m_1| } L_3^\alpha  + \frac{ \zeta_3  }{ 2 m_3 |m_1| } (m_2^2 - m_3^2 - m_1^2 )\,,
\nonumber\\
&& \Qbf_{12}^{uu}  = \frac{Q_{12}^{uu}}{L_1^uL_2^u}\,, \qquad \Qbf_{23}^{u\alpha}  = \frac{Q_{23}^{u\alpha}}{L_2^u}\,,\qquad \Qbf_{31}^{\alpha u}  = \frac{Q_{31}^{\alpha u}}{L_1^u}\,,
\nonumber\\
&& Q_{33}^{\alpha\alpha} = q_{33}^{\alpha\alpha} + \zeta_3\zeta_3\,,
\eeq
where, in \rf{31032025-man01-6-05}, we introduce vertex $V$ which depends on five variables shown explicitly in \rf{31032025-man01-6-05} and defined in \rf{31032025-man01-6-06}, \rf{30032025-man01-10}. The following remarks are in order.

\noinbf{i)} Explicit form of vertex $V$ \rf{31032025-man01-6-05} is not fixed by equations \rf{30032025-man01-16}-\rf{30032025-man01-25}. The vertex $V$ is a freedom of our solution to cubic vertex $p_\smp3^-$. In other words, there are many cubic vertices.

\noinbf{ii)} The momentum $\Po^I$ enters the variables $B_a^u$, $B_a^\alpha$ \rf{30032025-man01-10}. From  \rf{31032025-man01-6-05}, \rf{31032025-man01-6-06}, we then see that
the dependence of $p_\smp3^-$ on $\Po^I$ is governed by the variables $L_1^u$, $L_2^u$, and  $L_3^\alpha$, $\Qbf_{12}^{uu}$, $\Qbf_{23}^{u\alpha}$, $\Qbf_{31}^{\alpha u}$.

\noinbf{iii)} Cubic vertex \rf{31032025-man01-6-05} is non-polynomial in $\Po^I$ and hence non-local. For example, taking into account the  values of $\SSc$ given in \rf{27032025-man01-60} and the pre-factor $(L_1^u)^{ \SSc_1}(L_2^u)^{ \SSc_2}$ in \rf{31032025-man01-6-05}, we see that cubic vertex \rf{31032025-man01-6-05} exhibits the $PL$-non-locality defined in \rf{30032025-man01-35}. This $PL$-non-locality is unavoidable feature of the vertex $p_\smp3^-$ as either choice of the vertex $V$ leads to the $PL$-non-local vertex $p_\smp3^-$.

\noinbf{iv}) The variables $L_3^\alpha$, $\Qbf_{23}^{u\alpha}$, $\Qbf_{31}^{\alpha u}$ are linear forms of the oscillators $\alpha_3^I$, $\zeta_3$,  while the variable $Q_{33}^{\alpha\alpha}$ is a quadratical form of the oscillators $\alpha_3^I$, $\zeta_3$. In order for the vertex $p_\smp3^-$ \rf{31032025-man01-6-05} to be sensible, this vertex should be expandable in the just mentioned variables (see definition \rf{30032025-man01-50}).

\noinbf{Two massive CSFs and one massive integer-spin triplet field}. On the one hand, as the massive spin-$s_3$ triplet field obeys constraint \rf{27032025-man01-36}, the vertex should also obey the same constraint,
\be \label{31032025-man01-6-09}
(N_{\alpha_3} + N_{\zeta_3} - s_3) p_\smp3^- = 0 \,,
\ee
which tells us that $p_\smp3^-$ is the degree-$s_3$ homogeneous polynomial in $\alpha_3^I$, $\zeta_3$.
On the other hand, the operator $N_{\alpha_3} + N_{\zeta_3} - s_3$ in  \rf{31032025-man01-6-09} {\it commutes strongly} with all operators \rf{29032025-man01-30} entering basic equations \rf{29032025-man01-20}-\rf{29032025-man01-24}. This implies that vertex of our interest is obtainable from solution \rf{31032025-man01-6-05} by plugging $p_\smp3^-$ \rf{31032025-man01-6-05} into  \rf{31032025-man01-6-09} and finding solution to constraint \rf{31032025-man01-6-09}. Doing so, we find that constraint \rf{31032025-man01-6-09} amounts to the  constraint for the vertex $V$,
\be \label{31032025-man01-6-12}
(N_{\alpha_3} + N_{\zeta_3} - s_3) V = 0 \,.
\ee
The general solution to \rf{31032025-man01-6-12} is easily found to be
\beq
\label{31032025-man01-6-15} &&  V =
(L_3^\alpha)^k (\Qbf_{23}^{u\alpha})^{n_1} (\Qbf_{31}^{\alpha u})^{n_2} (Q_{33}^{\alpha\alpha})^l V_{n_1,n_2,l}(\Qbf_{12}^{uu})\,,\qquad k = s_3 -n_1 - n_2 - 2 l\,,\qquad
\nonumber\\
&& \hspace{1cm}  k \geq 0\,, \quad n_1 \geq 0 \,, \quad  n_2 \geq 0 \,, \quad l \geq 0\,,
\eeq
where a dependence of new vertex $V_{n_1,n_2,l}$ on $\Qbf_{12}^{uu}$ is not fixed. Three integers $n_1$, $n_2$, $l$ and the vertex $V_{n_1,n_2,l}$ express the freedom of our solution, while the inequalities \rf{31032025-man01-6-15} amount to requirement that the powers of all oscillator variables be non–negative. Plugging $V$ \rf{31032025-man01-6-15} into \rf{31032025-man01-6-05}, we get the cubic vertex for two CSFs shown in \rf{31032025-man01-6-01} and one massive spin-$s_3$ triplet field.

\noinbf{Two massive CSFs and one massive integer-spin field}.  On the one hand, as the massive spin-$s_3$ field obeys constraints \rf{27032025-man01-36}, \rf{27032025-man01-39}, the vertex should also obey the same constraints,
\be \label{31032025-man01-6-20}
\big( N_{\alpha_3} + N_{\zeta_3} - s_3 \big) p_\smp3^- = 0 \,, \qquad \big( \alphab_3^2 + \zetab_3^2 \big) p_\smp3^- = 0 \,.
\ee
On the other hand, the operators $N_{\alpha_3} + N_{\zeta_3} - s_3$ and $\alphab_3^2 + \zetab_3^2$ in  \rf{31032025-man01-6-20} {\it commute strongly} with all operators \rf{29032025-man01-30} entering basic equations \rf{29032025-man01-20}-\rf{29032025-man01-24}. This implies that vertex of our interest is obtainable from general solution \rf{31032025-man01-6-05} by plugging $p_\smp3^-$ \rf{31032025-man01-6-05} into  \rf{31032025-man01-6-20} and finding solution to constraints \rf{31032025-man01-6-20}. Plugging $p_\smp3^-$ \rf{31032025-man01-6-05} into \rf{31032025-man01-6-20} leads to the respective constraints for the vertex $V$,
\be \label{31032025-man01-6-25}
(N_{\alpha_3} + N_{\zeta_3} - s_3) V = 0 \,, \qquad (\alphab_3^2 + \zetab_3^2) V = 0 \,.
\ee
Solution to the first constraint in \rf{31032025-man01-6-25} has already been presented in \rf{31032025-man01-6-15}. It is easy to understand then that the most general solution to the 2nd constraint in \rf{31032025-man01-6-25} is given by
\beq
\label{31032025-man01-6-30} &&  V = (L_3^\alpha)^k (\Qbf_{23}^{u\alpha})^{n_1} (\Qbf_{31}^{\alpha u})^{n_2} V_{n_1,n_2}(\Qbf_{12}^{uu})\big|_\Thsm \,,\qquad k = s_3 -n_1 - n_2\,,
\nonumber\\
&& \hspace{1cm} k \geq 0\,, \qquad n_1\geq 0 \,, \qquad  n_2 \geq 0 \,,
\eeq
where a dependence of new vertex $V_{n_1,n_2}$ on $\Qbf_{12}^{uu}$ is not fixed by our equations. The integers $n_1$, $n_2$ and the vertex $V_{n_1,n_2}$ express the freedom of our solution, while the inequalities \rf{31032025-man01-6-30} amount to requirement that the powers of all oscillator variables be non–negative. The notation $|_\Thsm$ implies that, in \rf{31032025-man01-6-30} and \rf{30032025-man01-30}, the oscillators $\alpha_3^I$ and $\zeta_3$  should be replaced by the respective new oscillators $\alpha_{\Thsm\,3}^I$ and $\zeta_{\Thsm\,3}$ that respect the second constraint in \rf{31032025-man01-6-25},
\be \label{31032025-man01-6-35}
\alpha_3^I\rightarrow \alpha_{\Thsm\,3}^I \,, \qquad \zeta_3\rightarrow \zeta_{\Thsm\,3}\,,
\ee
where the oscillators $\alpha_\Thsm^I$, $\zeta_\Thsm$ are given in \rf{12042025-man01-05} in Appendix A. Briefly speaking, vertex \rf{31032025-man01-6-30} is obtainable from \rf{31032025-man01-6-15}  by setting $l=0$ and making replacement \rf{31032025-man01-6-35} in  \rf{31032025-man01-6-15}. Plugging vertex $V$ \rf{31032025-man01-6-30} into \rf{31032025-man01-6-05}, we get the cubic vertex $p_\smp3^-$ for fields in  \rf{31032025-man01-6-01}. Note that replacement \rf{31032025-man01-6-35} is immaterial in \rf{29032025-man01-05} because the generators $P_\smp3^-$ and  $J_\smp3^{-I}$ \rf{29032025-man01-05} are expressed in terms of the starred massive integer-spin field subjected to the constraint $\langle 0|\phi^*(\alpha)(\alpha^2+\zeta^2)=0$.

\vspace{-0.2cm}
\subsection{\large Two massive CSFs (non-equal masses) and one massless integer-spin field}\label{sec-5-2}

Using shortcuts $(m,\SSc)_\CSFsm$ and $(0,s)$ for the respective  massive continuous-spin field and massless integer-spin field, we are going to consider the cubic vertex that involves the following fields:
\beq
\label{31032025-man01-7-01} && (m_1,\SSc_1)_\CSFsm\hbox{-}(m_2,\SSc_2)_\CSFsm \hbox{-}(0,s_3) \,, \qquad m_1^2 < 0\,, \qquad m_2^2 < 0 \qquad m_1^2 \ne m_2^2\,,
\nonumber\\
&&  \hbox{\small two massive CSFs with non-equal masses and one massless integer-spin field.}\quad
\eeq

\noinbf{Two massive CSFs and one tower of massless integer-spin triplet fields}. Before discussing cubic vertex for fields in \rf{31032025-man01-7-01}, we consider cubic vertex for two CSFs in \rf{31032025-man01-7-01} and one tower of massless triplet fields \rf{27032025-man01-41}. For this case, our general solution to equations for cubic vertex \rf{30032025-man01-16}-\rf{30032025-man01-25} can be presented as
\beq
\label{31032025-man01-7-05} && \hspace{-1cm} p_\smp3^- = (L_1^u)^{ \SSc_1} (L_2^u)^{ \SSc_2} V(\Qbf_{12}^{uu}\,, \Qbf_{23}^{u\alpha}\,, \Qbf_{31}^{\alpha u}; q_{33}^{\alpha\alpha})\,,
\\
\label{31032025-man01-7-06} &&  L_1^u = B_1^u + \frac{\betach_1}{2\beta_1} |m_1| - \frac{m_2^2 }{2|m_1|}\,, \qquad L_2^u =  B_2^u + \frac{\betach_2}{2\beta_2} |m_2| + \frac{m_1^2 }{2|m_2|}\,,
\nonumber\\
&& Q_{12}^{uu}  =  q_{12}^{uu} + \frac{1}{|m_1|} L_2^u  -  \frac{1}{ |m_2| } L_1^u  - \frac{1}{ 2|m_1| |m_2|} (m_1^2 + m_2^2  )\,,
\nonumber\\
&& Q_{23}^{u\alpha}  =  q_{23}^{u\alpha} + \frac{1}{|m_2|} B_3^\alpha  -  \frac{2}{ m_1^2-m_2^2} L_2^u B_3^\alpha\,,
\nonumber\\
&& Q_{31}^{\alpha u}  =  q_{31}^{\alpha u} - \frac{1}{|m_1|} B_3^\alpha  + \frac{2}{ m_1^2-m_2^2} B_3^\alpha L_1^u \,,
\nonumber\\
&& \Qbf_{12}^{uu}  = \frac{Q_{12}^{uu}}{L_1^uL_2^u}\,, \qquad \Qbf_{23}^{u\alpha}  = \frac{Q_{23}^{u\alpha}}{L_2^u}\,,\qquad \Qbf_{31}^{\alpha u}  = \frac{Q_{31}^{\alpha u}}{L_1^u}\,,
\eeq
where, in \rf{31032025-man01-7-05}, we introduce vertex $V$ which depends on four variables shown explicitly in \rf{31032025-man01-7-05} and defined in \rf{31032025-man01-7-06}, \rf{30032025-man01-10}. The following remarks are in order.

\noinbf{i)} Explicit form of vertex $V$ \rf{31032025-man01-7-05} is not fixed by equations \rf{30032025-man01-16}-\rf{30032025-man01-25}. The vertex $V$ is a freedom of our solution to cubic vertex $p_\smp3^-$. This is to say that there are many cubic vertices.

\noinbf{ii)} The momentum $\Po^I$ enters the variables $B_a^u$, $B_a^\alpha$ \rf{30032025-man01-10}. From  \rf{31032025-man01-7-05}, \rf{31032025-man01-7-06}, we then see that the dependence of $p_\smp3^-$ on $\Po^I$ is governed by the variables $L_1^u$, $L_2^u$, and  $\Qbf_{12}^{uu}$, $\Qbf_{23}^{u\alpha}$, $\Qbf_{31}^{\alpha u}$.

\noinbf{iii)} Vertex $p_\smp3^-$ \rf{31032025-man01-7-05} is non-polynomial in $\Po^I$ and hence non-local. For example, taking into account the  values of $\SSc$ in \rf{27032025-man01-60} and the pre-factor $(L_1^u)^{ \SSc_1}(L_2^u)^{ \SSc_2}$ in \rf{31032025-man01-7-05}, we see that vertex $p_\smp3^-$ \rf{31032025-man01-7-05} exhibits the $PL$-non-locality defined in \rf{30032025-man01-35}. This $PL$-non-locality is unavoidable feature of the vertex $p_\smp3^-$ as either choice of the vertex $V$ leads to the $PL$-non-local vertex $p_\smp3^-$.

\noinbf{iv}) The variables $\Qbf_{23}^{u\alpha}$, $\Qbf_{31}^{\alpha u}$ are linear forms of the oscillators $\alpha_3^I$,  while the variable $q_{33}^{\alpha\alpha}$ is a quadratical form of the oscillators $\alpha_3^I$. Therefore in order for vertex $p_\smp3^-$ \rf{31032025-man01-7-05} to be sensible, this vertex should be expandable in the just mentioned variables (see definition \rf{30032025-man01-50}).

We now proceed with discussion of cubic vertex for two CSFs in \rf{31032025-man01-7-01} and one massless spin-$s_3$ (triplet) field. Derivation of such vertex is similar the one described in Sec.\,\ref{sec-5-1}. Therefore, in the interest of the brevity, we skip the details of the derivation and present only our results.

\noinbf{Two massive CSFs and one massless integer-spin triplet field}. Cubic vertex $p_\smp3^-$ for two CSFs in \rf{31032025-man01-7-05} and one massless spin-$s_3$ triplet field is given by \rf{31032025-man01-7-05}, \rf{31032025-man01-7-06}, where $V$ takes the form
\beq
\label{31032025-man01-7-15} && V =  (\Qbf_{23}^{u\alpha})^{n_1} (\Qbf_{31}^{\alpha u})^{n_2} (q_{33}^{\alpha\alpha})^l V_{n_1,n_2,l}(\Qbf_{12}^{uu})\,,
\nonumber\\
&& \hspace{1cm}   n_1 + n_2 + 2 l = s_3\,, \qquad \quad n_1 \geq 0 \,, \quad  n_2 \geq 0 \,, \quad l \geq 0\,,
\eeq
while new vertex $V_{n_1,n_2,l}$ depends only on $\Qbf_{12}^{uu}$. The integers $n_1$, $n_2$, $l$ subjected to the conditions in \rf{31032025-man01-7-15} and the vertex $V_{n_1,n_2,l}$ express the freedom of our solution for the vertex $V$, while the inequalities  amount to requirement that the powers of all oscillator variables be non-negative.

\noinbf{Two massive CSFs and one massless integer-spin field}. Cubic vertex $p_\smp3^-$ for fields in \rf{31032025-man01-7-01} is given by \rf{31032025-man01-7-05}, \rf{31032025-man01-7-06}, where $V$ takes the form
\be \label{31032025-man01-7-30}
V =   (\Qbf_{23}^{u\alpha})^{n_1} (\Qbf_{31}^{\alpha u})^{n_2} \, V_{n_1,n_2}(\Qbf_{12}^{uu})\big|_\Thsm \,,
\hspace{1cm} n_1 + n_2 = s_3 \,,\qquad n_1\,, n_2 \in \No_0\,,
\ee
while new vertex $V_{n_1,n_2}$ depends only on $\Qbf_{12}^{uu}$. The integers $n_1$, $n_2$ subjected to the conditions in \rf{31032025-man01-7-30} and the vertex $V_{n_1,n_2}$ express the freedom of our solution. Notation $|_\Thsm$ implies that, in \rf{31032025-man01-7-30} and \rf{30032025-man01-30}, we should make the replacement $\alpha_{\Thsm\,3}^I$, $\alpha_3^I\rightarrow \alpha_{\Thsm\,3}^I$, where $\alpha_\Thsm^I$ is given in \rf{12042025-man01-05x} in Appendix A.  This replacement is immaterial in \rf{29032025-man01-05} because generators $P_\smp3^-$ and  $J_\smp3^{-I}$ \rf{29032025-man01-05} are expressed in terms of the starred massless integer-spin field subjected to the constraint $\langle 0|\phi^*(\alpha) \alpha^2=0$.

\vspace{-0.2cm}
\subsection{\large Two massive CSFs (equal masses) and one massless integer-spin field}

Using the shortcuts $(m,\SSc)_\CSFsm$ and $(0,s)$ for the respective massive continuous-spin field and massless integer-spin field, we consider the cubic vertex that involves the following fields:
\beq
\label{31032025-man01-8-01} && (m_1,\SSc_1)_\CSFsm\hbox{-}(m_2,\SSc_2)_\CSFsm \hbox{-}(0,s_3) \,, \qquad m_1^2 = m^2\,, \qquad m_2^2 = m^2\,, \qquad m^2 < 0\,,
\nonumber\\
&&  \hbox{\small two massive CSFs with equal masses and one massless integer-spin field.}\quad
\eeq
\noinbf{Two massive CSFs and one tower of massless integer-spin triplet fields}.
We start with presentation of cubic vertex for two CSFs in \rf{31032025-man01-8-01} and one tower of massless integer-spin triplet fields \rf{27032025-man01-41}. For this case, our general solution to equations \rf{30032025-man01-16}-\rf{30032025-man01-25} can be presented as
\beq
\label{31032025-man01-8-05} && \hspace{-1cm} p_\smp3^- = (L_1^u)^{\SSc_1} (L_2^u)^{\SSc_2} V(B_3^\alpha,\Qbf_{12}^{uu}\,, \Cbf_{123}^{uu\alpha}; q_{33}^{\alpha\alpha})\,,
\\
\label{31032025-man01-8-06} && L_1^u =  B_1^u - \frac{\beta_3}{\beta_1} |m| \,,
\qquad  L_2^u =  B_2^u + \frac{\beta_3}{\beta_2} |m| \,,
\nonumber\\
&& Q_{12}^{uu}  =  q_{12}^{uu} + \frac{1}{|m|} L_2^u  -  \frac{1}{ |m| } L_1^u  + 1 \,,
\nonumber\\
&& C_{123}^{uu\alpha}  =  L_1^u \big(q_{23}^{u\alpha} + \frac{1}{|m|} B_3^\alpha\big)  + L_2^u \big( q_{31}^{\alpha u} - \frac{1}{|m|} B_3^\alpha\big)\,,
\nonumber\\
&& \Qbf_{12}^{uu}  = \frac{Q_{12}^{uu}}{L_1^uL_2^u}\,, \qquad \Cbf_{123}^{uu\alpha}  = \frac{C_{123}^{uu\alpha}}{L_1^uL_2^u}\,,
\eeq
where, in \rf{31032025-man01-8-05}, we introduce vertex $V$ which depends on four variables shown explicitly in \rf{31032025-man01-8-05} and defined in \rf{31032025-man01-8-06}, \rf{30032025-man01-10}. The following remarks are in order.

\noinbf{i)} Explicit form of vertex $V$ \rf{31032025-man01-8-05} is not fixed by equations \rf{30032025-man01-16}-\rf{30032025-man01-25}. The vertex $V$ is a freedom of our solution for cubic vertex $p_\smp3^-$. This is to say that there are many cubic vertices.

\noinbf{ii)} The momentum $\Po^I$ enters variables $B_a^u$, $B_a^\alpha$ \rf{30032025-man01-10}. From  \rf{31032025-man01-8-05}, \rf{31032025-man01-8-06}, we then see that the dependence of vertex $p_\smp3^-$ on $\Po^I$ is governed by the variables $L_1^u$, $L_2^u$, and $B_3^\alpha$, $\Qbf_{12}^{uu}$, $\Cbf_{123}^{uu\alpha}$.

\noinbf{iii)} Vertex $p_\smp3^-$ \rf{31032025-man01-8-05} is non-polynomial in $\Po^I$ and hence non-local. For example, taking into account the  values of $\SSc$ in \rf{27032025-man01-60} and the pre-factor $(L_1^u)^{ \SSc_1}(L_2^u)^{ \SSc_2}$ in \rf{31032025-man01-8-05}, we see that vertex $p_\smp3^-$ \rf{31032025-man01-8-05} exhibits the $PL$-non-locality defined in \rf{30032025-man01-35}. The $PL$-non-locality is unavoidable feature of the vertex $p_\smp3^-$ as either choice of the vertex $V$ leads to the $PL$-non-local vertex $p_\smp3^-$.

\noinbf{iv}) The variables $B_3^\alpha$, $\Cbf_{123}^{uu\alpha}$ are linear forms of the oscillators $\alpha_3^I$,  while the variable $q_{33}^{\alpha\alpha}$ is a quadratical form of the oscillators $\alpha_3^I$. In order for vertex $p_\smp3^-$ \rf{31032025-man01-8-05} to be sensible, this vertex should be expandable in the just mentioned variables (see definition \rf{30032025-man01-50}).

We now discuss cubic vertex for two CSFs shown in \rf{31032025-man01-8-01} and one massless spin-$s_3$ (triplet) field. As the derivation of such cubic vertex follows the procedure described in Sec.\,\ref{sec-5-1}, we skip the details of the derivation and present only our results.

\noinbf{Two massive CSFs and one massless integer-spin triplet field}. Cubic vertex $p_\smp3^-$ for two CSFs as in \rf{31032025-man01-8-01} and one massless spin-$s_3$ triplet field is given by \rf{31032025-man01-8-05}, \rf{31032025-man01-8-06}, where $V$ takes the form
\be
\label{31032025-man01-8-15}  V = (B_3^\alpha)^k (\Cbf_{123}^{uu\alpha})^n (q_{33}^{\alpha\alpha})^l V_{n,l}(\Qbf_{12}^{uu})\,, \hspace{1cm}   k = s_3 - n - 2 l\,, \qquad k\,, n \,, l \in \No_0\,,\qquad
\ee
while new vertex $V_{n,l}$ depends only on $\Qbf_{12}^{uu}$. The integers $n$, $l$ subjected to the conditions in \rf{31032025-man01-8-15} and the vertex $V_{n,l}$ express the freedom of our solution.

\noinbf{Two massive CSFs and one massless integer-spin field}. Cubic vertex $p_\smp3^-$ for fields in \rf{31032025-man01-8-01} is given by \rf{31032025-man01-8-05}, \rf{31032025-man01-8-06}, where $V$ takes the form
\be \label{31032025-man01-8-30}
V =   (B_3^\alpha)^k (\Cbf_{123}^{uu\alpha})^n \, V_n(\Qbf_{12}^{uu})\big|_\Thsm \,,\qquad k = s_3 -n\,,\qquad k\,, n \in \No_0 \,,
\ee
while new vertex $V_n$ depends only on $\Qbf_{12}^{uu}$. The integer $n$ subjected to the conditions in \rf{31032025-man01-8-30} and the vertex $V_n$ express the freedom of our solution. The notation $|_\Thsm$ implies that, in \rf{31032025-man01-8-30} and \rf{30032025-man01-30}, one just needs to make the replacement $\alpha_3^I\rightarrow \alpha_{\Thsm\,3}^I$, where $\alpha_\Thsm^I$  is given in \rf{12042025-man01-05x} in Appendix A. Such replacement  is immaterial in \rf{29032025-man01-05} (see our comment at the end of Sec.\,\ref{sec-5-2}).

\vspace{-0.2cm}
\subsection{\large One massive CSF, one massless CSF and one massive integer-spin field}

Using the shortcuts $(m,\SSc)_\CSFsm$, $(0,\kappa)_\CSFsm$, and $(m,s)$ for the respective  massive CSF, massless CSF, and massive integer-spin field, we consider the cubic vertex that involves the following fields:
\beq
\label{31032025-man01-9-01} && (m_1,\SSc_1)_\CSFsm\hbox{-}(0,\kappa_2)_\CSFsm \hbox{-}(m_3,s_3) \,,\qquad m_1^2 < 0 \,, \qquad m_3^2 > 0 \,,
\nonumber\\
&&  \hbox{\small one massive CSF, one massless CSF, and one massive integer-spin field.}\quad
\eeq
\noinbf{One massive CSF, one massless CSF, and one tower of massive integer-spin triplet fields}.
We start with presentation of cubic vertex for CSFs shown in \rf{31032025-man01-9-01} and tower of massive integer-spin triplet fields \rf{27032025-man01-41} having mass $m_3$. For this case, our general solution to equations for cubic vertex \rf{30032025-man01-16}-\rf{30032025-man01-25} can be presented as
\beq
\label{31032025-man01-9-05} &&  \hspace{-1cm} p_\smp3^- = e^W (L_1^u)^{\SSc_1}   V(L_3^\alpha,\Qbf_{12}^{uu}\,, Q_{23}^{u\alpha}\,,\Qbf_{31}^{\alpha u}; Q_{33}^{\alpha\alpha})\,,
\\
&& W = \frac{2\irm \kappa_2 B_2^u }{m_1^2-m_3^2}\,,
\nonumber\\
\label{31032025-man01-9-06} &&  L_1^u =  B_1^u + \frac{\betach_1}{2\beta_1} |m_1| + \frac{m_3^2}{2|m_1|}\,, \qquad L_3^\alpha =  B_3^\alpha - \frac{\betach_3}{2\beta_3} m_3 \zeta_3 - \frac{m_1^2}{2m_3}\zeta_3\,,
\nonumber\\
&& Q_{12}^{uu}  =  q_{12}^{uu} + \frac{1}{|m_1|} B_2^u  -  \frac{2}{m_3^2-m_1^2} L_1^u B_2^u \,,
\nonumber\\
&& Q_{23}^{u\alpha}  =  q_{23}^{u\alpha} - \frac{\zeta_3}{m_3} B_2^u  +  \frac{2}{m_3^2-m_1^2} B_2^u  L_3^\alpha \,,
\nonumber\\
&& Q_{31}^{\alpha u}  =  q_{31}^{\alpha u} + \frac{\zeta_3}{m_3} L_1^u - \frac{1}{|m_1|} L_3^\alpha  -  \frac{\zeta_3}{2m_3|m_1|}(m_3^2 + m_1^2) \,,
\nonumber\\
&& \Qbf_{12}^{uu}  = \frac{Q_{12}^{uu}}{L_1^u}\,, \qquad
\Qbf_{31}^{\alpha u}  = \frac{Q_{31}^{\alpha u}}{L_1^u}\,, \hspace{1cm}  Q_{33}^{\alpha\alpha} = q_{33}^{\alpha\alpha} + \zeta_3\zeta_3\,,
\eeq
where, in \rf{31032025-man01-9-05}, we introduce vertex $V$ which depends on the five variables shown explicitly in \rf{31032025-man01-9-05} and defined in \rf{31032025-man01-9-06}, \rf{30032025-man01-10}. The following remarks are in order.

\noinbf{i)} Vertex $V$ \rf{31032025-man01-9-05} is not fixed by equations \rf{30032025-man01-16}-\rf{30032025-man01-25}. The vertex $V$ is a freedom of our solution to cubic vertex $p_\smp3^-$. So, there are many cubic vertices.

\noinbf{ii)} The momentum $\Po^I$ enters variables $B_a^\alpha$, $B_a^u$ \rf{30032025-man01-10}. From  \rf{31032025-man01-9-05}, \rf{31032025-man01-9-06}, we see that the vertex $p_\smp3^-$ depends on the momentum $\Po^I$ through the variables $W$, $L_1^u$ and $L_3^\alpha$, $\Qbf_{12}^{uu}$, $Q_{23}^{u\alpha}$, $\Qbf_{31}^{\alpha u}$.

\noinbf{iii)} Vertex \rf{31032025-man01-9-05} is non-polynomial in $\Po^I$ and hence non-local. For example, taking into account the values of $\SSc$ given in \rf{27032025-man01-60} and the pre-factor $e^W (L_1^u)^{ \SSc_1}$ in \rf{31032025-man01-9-05}, we see that vertex \rf{31032025-man01-9-05} exhibits $E$- and $PL$-non-localities \rf{30032025-man01-35}. These $E$- and $PL$-non-localities are unavoidable feature of the vertex $p_\smp3^-$ as either choice of the vertex $V$ leads to $E$- and $PL$-non-local $p_\smp3^-$.

\noinbf{iv}) The variables $L_3^\alpha$, $Q_{23}^{u\alpha}$, $\Qbf_{31}^{\alpha u}$ are linear forms of the oscillators $\alpha_3^I$, $\zeta_3$,  while the variable $Q_{33}^{\alpha\alpha}$ is a quadratical form of the oscillators $\alpha_3^I$, $\zeta_3$. Therefore in order for vertex $p_\smp3^-$ \rf{31032025-man01-9-05} to be sensible, this vertex should be expandable in the just mentioned variables (see definition \rf{30032025-man01-50}).

\noinbf{One massive CSF, one massless CSF, and one massive integer-spin triplet field}. Cubic vertex $p_\smp3^-$ for two CSFs as in \rf{31032025-man01-9-01} and one massive spin-$s_3$ triplet field having mass $m_3$ is given by \rf{31032025-man01-9-05}, \rf{31032025-man01-9-06}, where $V$ takes the form
\beq
\label{31032025-man01-9-15} && V = (L_3^\alpha)^k (Q_{23}^{u\alpha})^{n_1} (\Qbf_{31}^{\alpha u})^{n_2} (Q_{33}^{\alpha\alpha})^l V_{n_1,n_2,l}(\Qbf_{12}^{uu})\,,
\nonumber\\
&& \hspace{1cm}   k = s_3 -n_1 - n_2 - 2 l\,, \qquad k \geq 0\,, \quad n_1 \geq 0 \,, \quad  n_2 \geq 0 \,, \quad l \geq 0\,,
\eeq
while new vertex $V_{n_1,n_2,l}$ depends only on $\Qbf_{12}^{uu}$. The integers $n_1$, $n_2$, $l$ subjected to the conditions in \rf{31032025-man01-9-15} and the vertex $V_{n_1,n_2,l}$ express the freedom of our solution, while the inequalities \rf{31032025-man01-9-15} amount to requirement that the powers of all oscillator variables be non-negative.

\noinbf{One massive CSF, one massless CSF, and one massive integer-spin  field}. The cubic vertex $p_\smp3^-$ for fields in \rf{31032025-man01-9-01} is given by \rf{31032025-man01-9-05}, \rf{31032025-man01-9-06}, where $V$ takes the form
\be \label{31032025-man01-9-30}
V =   (L_3^\alpha)^k (Q_{23}^{u\alpha})^{n_1} (\Qbf_{31}^{\alpha u})^{n_2} \, V_{n_1,n_2}(\Qbf_{12}^{uu})\big|_\Thsm \,,
\qquad k = s_3 -n_1 - n_2\,,\quad k\,, n_1 \,, n_2 \in \No_0\,,
\ee
while new vertex $V_{n_1,n_2}$ depends on $\Qbf_{12}^{uu}$. The integers $n_1$, $n_2$ subjected to the conditions in \rf{31032025-man01-9-30} and the vertex $V_{n_1,n_2}$ express the freedom of our solution. The notation $|_\Thsm$ implies that, in \rf{31032025-man01-9-30} and \rf{30032025-man01-30}, one needs to make the replacements $\alpha_3^I\rightarrow \alpha_{\Thsm\,3}^I$, $\zeta_3\rightarrow \zeta_{\Thsm\,3}$,
where $\alpha_\Thsm^I$, $\zeta_\Thsm$  are given in \rf{12042025-man01-05} in Appendix A.
Such replacements are immaterial in \rf{29032025-man01-05} (see comment at the end of Sec.\, \ref{sec-5-1}).

\vspace{-0.2cm}
\subsection{ One massive CSF, one massless CSF and one massless integer-spin field}

Using the shortcuts $(m,\SSc)_\CSFsm$, $(0,\kappa)_\CSFsm$, and $(0,s)$ for the respective  massive CSF, massless CSF and massless integer-spin field, we consider the cubic vertex that involves the following fields:
\beq
\label{31032025-man01-10-01} && (m_1,\SSc_1)_\CSFsm\hbox{-}(0,\kappa_2)_\CSFsm \hbox{-}(0,s_3) \,, \qquad m_1^2 < 0 \,,
\nonumber\\
&&  \hbox{\small one massive CSF, one massless CSF, and one massless integer-spin field.}\quad
\eeq

\noinbf{One massive CSF, one massless CSF, and one tower of massless triplet fields}.
We start with presentation of cubic vertex for CSFs shown in \rf{31032025-man01-10-01} and tower of massless integer-spin triplet fields \rf{27032025-man01-41}. For this case, our general solution to equations \rf{30032025-man01-16}-\rf{30032025-man01-25} can be presented as
\beq
\label{31032025-man01-10-05} &&\hspace{-1cm}  p_\smp3^- = e^W  (L_1^u)^{\SSc_1}   V(\Qbf_{12}^{uu}\,,  Q_{23}^{u\alpha}\,,\Qbf_{31}^{\alpha u};q_{33}^{\alpha\alpha})\,,
\\
\label{31032025-man01-10-06} && W = \frac{2\irm \kappa_2 B_2^u }{m_1^2}\,,  \hspace{1cm}   L_1^u =  B_1^u + \frac{\betach_1}{2\beta_1} |m_1| \,,
\nonumber\\
&& Q_{12}^{uu}  =  q_{12}^{uu} + \frac{1}{|m_1|} B_2^u  +  \frac{2}{m_1^2} L_1^u B_2^u \,,
\nonumber\\
&& Q_{23}^{u\alpha}  =  q_{23}^{u\alpha} -   \frac{2}{m_1^2} B_2^u  B_3^\alpha \,,
\hspace{2cm}  Q_{31}^{\alpha u}  =  q_{31}^{\alpha u} - \frac{1}{|m_1|} B_3^\alpha  + \frac{2}{m_1^2} B_3^\alpha L_1^u \,,
\nonumber\\
&& \Qbf_{12}^{uu}  = \frac{Q_{12}^{uu}}{L_1^u}\,, \qquad
\Qbf_{31}^{\alpha u}  = \frac{Q_{31}^{\alpha u}}{L_1^u}\,,
\eeq
where, in \rf{31032025-man01-10-05}, we introduce vertex $V$ which depends on four variables shown explicitly in \rf{31032025-man01-10-05} and defined in \rf{31032025-man01-10-06}, \rf{30032025-man01-10}. The following remarks are in order.

\noinbf{i)} Dependence of vertex $V$ \rf{31032025-man01-10-05} on its four arguments is not fixed by equations \rf{30032025-man01-16}-\rf{30032025-man01-25}. The vertex $V$ is a freedom of our solution to cubic vertex $p_\smp3^-$. So, there are many cubic vertices.

\noinbf{ii)} The momentum $\Po^I$ enters variables $B_a^u$, $B_a^\alpha$ \rf{30032025-man01-10}. From  \rf{31032025-man01-10-05}, \rf{31032025-man01-10-06}, we then see that the dependence of vertex $p_\smp3^-$ on the momentum $\Po^I$ is governed by $W$, $L_1^u$ and  $\Qbf_{12}^{uu}$, $Q_{23}^{u\alpha}$, $\Qbf_{31}^{\alpha u}$.

\noinbf{iii)} Vertex $p_\smp3^-$ \rf{31032025-man01-10-05} is non-polynomial in $\Po^I$ and hence non-local. For example, taking into account the  values of $\SSc$  \rf{27032025-man01-60} and the pre-factor $e^W (L_1^u)^{ \SSc_1}$ in \rf{31032025-man01-10-05}, we see that vertex $p_\smp3^-$ \rf{31032025-man01-10-05} exhibits $E$- and $PL$-non-localities \rf{30032025-man01-35}. These $E$- and $PL$-non-localities are unavoidable feature of the vertex $p_\smp3^-$ as either choice of the vertex $V$ leads to $E$- and $PL$-non-local $p_\smp3^-$.

\noinbf{iv}) The variables $Q_{23}^{u\alpha}$, $\Qbf_{31}^{\alpha u}$ are linear forms of the oscillators $\alpha_3^I$,  while the variable $q_{33}^{\alpha\alpha}$ is a quadratical form of the oscillators $\alpha_3^I$. In order for vertex $p_\smp3^-$ \rf{31032025-man01-10-05} to be sensible, this vertex should be expandable in the just mentioned variables (see definition \rf{30032025-man01-50}).

\noinbf{One massive CSF, one massless CSF, and one massless integer-spin triplet field}. Cubic vertex $p_\smp3^-$  for CSFs shown in \rf{31032025-man01-10-01} and one massless spin-$s_3$ triplet field is given by \rf{31032025-man01-10-05}, \rf{31032025-man01-10-06}, where the vertex $V$ takes the form
\be \label{31032025-man01-10-15}
V =  (Q_{23}^{u\alpha})^{n_1} (\Qbf_{31}^{\alpha u})^{n_2} (q_{33}^{\alpha\alpha})^l V_{n_1,n_2,l}(\Qbf_{12}^{uu})\,,\qquad n_1 + n_2 + 2 l = s_3 \,,   \quad n_1\,, n_2 \,,l \in \No_0\,,
\ee
while new vertex $V_{n_1,n_2,l}$ depends only on $\Qbf_{12}^{uu}$. The integers $n_1$, $n_2$, $l$ subjected to the conditions in \rf{31032025-man01-10-15} and the vertex $V_{n_1,n_2,l}$  express the freedom of our solution for the vertex $V$.

\noinbf{One massive CSF, one massless CSF, and one massless integer-spin field}. The cubic vertex $p_\smp3^-$ for fields in \rf{31032025-man01-10-01} is given by \rf{31032025-man01-10-05}, \rf{31032025-man01-10-06}, where $V$ takes the form
\be \label{31032025-man01-10-30}
V =   (Q_{23}^{u\alpha})^{n_1} (\Qbf_{31}^{\alpha u})^{n_2} \, V_{n_1,n_2}(\Qbf_{12}^{uu})\big|_\Thsm \,, \qquad  n_1 + n_2 = s_3\,, \quad n_1\,, n_2 \in \No_0\,,\qquad
\ee
while new vertex $V_{n_1,n_2}$ depends only on $\Qbf_{12}^{uu}$. The integers $n_1$, $n_2$  subjected to the conditions in \rf{31032025-man01-10-30} and the vertex $V_{n_1,n_2}$  express the freedom of our solution for the vertex $V$.
The notation $|_\Thsm$ implies that, in \rf{31032025-man01-10-30} and \rf{30032025-man01-30}, one just needs to make the replacement $\alpha_3^I\rightarrow \alpha_{\Thsm\,3}^I$, where $\alpha_\Thsm^I$  is given in \rf{12042025-man01-05x} in Appendix A. Such replacement is immaterial in \rf{29032025-man01-05} (see comment at the end of Sec.\,\ref{sec-5-2}).

\vspace{-0.2cm}
\subsection{ Two massless CSFs and one massive integer-spin field}

Using the shortcuts $(0,\kappa)_\CSFsm$, and $(m,s)$ for the respective  massless continuous-spin field and massive integer-spin field, we consider the cubic vertex that involves the following fields:
\beq
\label{31032025-man01-11-01} && (0,\kappa_1)_\CSFsm\hbox{-}(0,\kappa_2)_\CSFsm \hbox{-}(m_3,s_3) \,, \qquad m_3^2 > 0\,,
\nonumber\\
&&  \hbox{\small two massless CSFs and one massive integer-spin field.}\quad
\eeq
\noinbf{Two massless CSFs and tower of massive integer-spin triplet fields}.
We start with presentation of cubic vertex for CSFs shown in \rf{31032025-man01-11-01} and one tower of massive integer-spin triplet fields \rf{27032025-man01-41} having mass $m_3$. For this case, general solution to equations \rf{30032025-man01-16}-\rf{30032025-man01-25} can be presented as%
\beq
\label{31032025-man01-11-05} && \hspace{-1cm} p_\smp3^- = e^W  V(L_3^\alpha\,,  Q_{12}^{uu}\,,  Q_{23}^{u\alpha}\,,Q_{31}^{\alpha u}; Q_{33}^{\alpha\alpha})\,,
\\
&& W =  \frac{2\irm}{m_3^2} (\kappa_1 B_1^u- \kappa_2 B_2^u) \,, \hspace{2cm} L_3^\alpha =  B_3^\alpha - \frac{\betach_3}{2\beta_3} m_3 \zeta_3 \,,
\nonumber\\
\label{31032025-man01-11-06} && Q_{12}^{uu}  =  q_{12}^{uu} -  \frac{2}{m_3^2} B_1^u B_2^u \,, \hspace{2.5cm} Q_{23}^{u\alpha}  =  q_{23}^{u\alpha} - \frac{\zeta_3}{m_3} B_2^u +    \frac{2}{m_3^2} B_2^u  L_3^\alpha \,,
\nonumber\\
&& Q_{31}^{\alpha u}  =  q_{31}^{\alpha u} + \frac{\zeta_3}{m_3} B_1^u + \frac{2}{m_3^2} L_3^\alpha B_1^u \,, \hspace{1cm}
 Q_{33}^{\alpha\alpha} = q_{33}^{\alpha\alpha} + \zeta_3\zeta_3\,,
\eeq
where, in \rf{31032025-man01-11-05}, we introduce vertex $V$ which depends on five variables shown explicitly in \rf{31032025-man01-11-05} and defined in \rf{31032025-man01-11-06}, \rf{30032025-man01-10}. The following remarks are in order.

\noinbf{i)} Dependence of vertex $V$ \rf{31032025-man01-11-05} on its five arguments is not fixed by equations \rf{30032025-man01-16}-\rf{30032025-man01-25}. The vertex $V$ is a freedom of our solution to cubic vertex $p_\smp3^-$. So, there are many cubic vertices.

\noinbf{ii)} The momentum $\Po^I$ enters variables $B_a^u$, $B_a^\alpha$ \rf{30032025-man01-10}. From  \rf{31032025-man01-11-05}, \rf{31032025-man01-11-06}, we see that the vertex $p_\smp3^-$ depends on the momentum $\Po^I$ through the variables $W$ and $L_3^\alpha$, $Q_{12}^{uu}$, $Q_{23}^{u\alpha}$, $Q_{31}^{\alpha u}$.

\noinbf{iii)} Vertex \rf{31032025-man01-11-05} is non-polynomial in $\Po^I$ and hence non-local. For example, taking into account the pre-factor $e^W$, we see that vertex $p_\smp3^-$ \rf{31032025-man01-11-05} exhibits $E$-non-locality \rf{30032025-man01-35}. The $E$-non-locality is unavoidable feature of the vertex $p_\smp3^-$ as either choice of the vertex $V$ leads to $E$-non-local $p_\smp3^-$.

\noinbf{iv}) The variables $L_3^\alpha$, $Q_{23}^{u\alpha}$, $Q_{31}^{\alpha u}$ are linear forms of the oscillators $\alpha_3^I$, $\zeta_3$,  while the variable $Q_{33}^{\alpha\alpha}$ is a quadratical form of the oscillators $\alpha_3^I$, $\zeta_3$. In order for vertex $p_\smp3^-$ \rf{31032025-man01-11-05} to be sensible, this vertex should be expandable in the just mentioned variables (see definition \rf{30032025-man01-50}).

\noinbf{Two massless CSFs and one massive integer-spin triplet field}. Cubic vertex $p_\smp3^-$ for two CSFs \rf{31032025-man01-11-01} and one massive spin-$s_3$ triplet field having mass $m_3$ is given by \rf{31032025-man01-11-05}, \rf{31032025-man01-11-06}, where $V$ takes the form
\beq
\label{31032025-man01-11-15} && V = (L_3^\alpha)^k (Q_{23}^{u\alpha})^{n_1} (Q_{31}^{\alpha u})^{n_2} (Q_{33}^{\alpha\alpha})^l V_{n_1,n_2,l}(Q_{12}^{uu})\,,
\nonumber\\
&& \hspace{1cm}   k = s_3 -n_1 - n_2 - 2 l\,, \qquad k\,, n_1\,, n_2\,, l\in \No_0\,,
\eeq
while new vertex $V_{n_1,n_2,l}$ depends only on $\Qbf_{12}^{uu}$. The integers $n_1$, $n_2$, $l$ subjected to the conditions in \rf{31032025-man01-11-15} and the vertex $V_{n_1,n_2,l}$  express the freedom of our solution for the vertex $V$.

\noinbf{Two massless CSFs and one massive integer-spin field}. The cubic vertex $p_\smp3^-$ for fields in \rf{31032025-man01-11-01} is given by \rf{31032025-man01-11-05}, \rf{31032025-man01-11-06}, where $V$ takes the form
\be
\label{31032025-man01-11-30} V =   (L_3^\alpha)^k (Q_{23}^{u\alpha})^{n_1} (Q_{31}^{\alpha u})^{n_2} \, V_{n_1,n_2}(Q_{12}^{uu})\big|_\Thsm \,, \qquad k = s_3 -n_1 - n_2\,,\quad k\,,n_1 \,,n_2  \in \No_0\,,
\ee
while new vertex $V_{n_1,n_2}$ depends on $Q_{12}^{uu}$. The integers $n_1$, $n_2$  subjected to conditions \rf{31032025-man01-11-30} and the vertex $V_{n_1,n_2}$  express the freedom of our solution for the vertex $V$.
The notation $|_\Thsm$ implies that, in \rf{31032025-man01-11-30} and \rf{30032025-man01-30} one just needs to make the replacements $\alpha_3^I\rightarrow \alpha_{\Thsm\,3}^I$, $\zeta_3\rightarrow \zeta_{\Thsm\,3}$, where $\alpha_\Thsm^I$, $\zeta_\Thsm$  are given in \rf{12042025-man01-05} in Appendix A. Such replacements are immaterial in \rf{29032025-man01-05} (see our comment at the end of Sec.\,\ref{sec-5-1}).

\vspace{-0.2cm}
\subsection{ Two massless CSFs and one massless integer-spin field }\label{sub-sec-5-7}

Using the shortcuts $(0,\kappa)_\CSFsm$ and $(0,s)$ for the respective  massless continuous-spin field and massless integer-spin field, we consider the cubic vertex that involves the following fields:
\beq
\label{31032025-man01-12-01} && (0,\kappa_1)_\CSFsm\hbox{-}(0,\kappa_2)_\CSFsm \hbox{-}(0,s_3) \,,
\nonumber\\
&&  \hbox{\small two massless CSFs and one massless integer-spin field.}\quad
\eeq

\noinbf{Two massless CSFs and one tower of massless integer-spin triplet fields}.
We start with the presentation of our result about solutions to equations \rf{30032025-man01-16}-\rf{30032025-man01-25}. We formulate two Statements.

\noinbf{Statement 1}. There are no f-solutions to equations    \rf{30032025-man01-16}-\rf{30032025-man01-25}.

\noinbf{Statement 2}. Particular distributional solution to equations  \rf{30032025-man01-16}-\rf{30032025-man01-25} takes the form
\beq
\label{31032025-man01-12-05} && \hspace{-1cm} p_\smp3^- =  e^W V(B_1^u,B_2^u,B_3^\alpha, C_{123}^{uu\alpha};q_{33}^{\alpha\alpha})\delta(  \kappa_1 B_1^u  + \kappa_2 B_2^u)\,,
\\
\label{31032025-man01-12-06} && W = \irm \frac{ \kappa_1 B_1^u - \kappa_2 B_2^u }{ 2B_1^u B_2^u } q_{12}^{uu}\,,
\qquad  C_{123}^{uu\alpha} = B_1^u q_{23}^{u\alpha} + B_2^u q_{31}^{\alpha u} + B_3^\alpha q_{12}^{uu}\,,
\eeq
where, in \rf{31032025-man01-12-05}, new vertex $V$ depends formally on five variables   defined in \rf{31032025-man01-12-06} and \rf{30032025-man01-10}. The symbol $\delta$ stands for the  standard 1--dimensional Dirac-delta function.
In view of the $\delta$-function, the vertex $V$ actually depends on four variables.
The following remarks are in order.

\noinbf{i)} Vertex $V$ \rf{31032025-man01-12-05} is not fixed by equations \rf{30032025-man01-16}-\rf{30032025-man01-25}. So there are many cubic vertices.

\noinbf{ii)} The momentum $\Po^I$ enters variables $B_a^u$, $B_a^\alpha$ \rf{30032025-man01-10}. From  \rf{31032025-man01-12-05}, \rf{31032025-man01-12-06}, we then see that
the dependence of $p_\smp3^-$ on $\Po^I$ is governed by $B_1^u$, $B_2^u$, $B_3^\alpha$, $C_{123}^{uu\alpha}$, $q_{33}^{\alpha\alpha}$, and $W$. The variables $B_1^u$, $B_2^u$, $B_3^\alpha$, $C_{123}^{uu\alpha}$, $q_{33}^{\alpha\alpha}$ are expandable in $\Po^I$, while the variable $W$ does not (see definition in \rf{30032025-man01-50}).

\noinbf{iii)} Vertex $p_\smp3^-$ \rf{31032025-man01-12-05} is non-polynomial in $\Po^I$ and hence non-local. Note that the variable $W$ and the $\delta$-function in \rf{31032025-man01-12-05}, \rf{31032025-man01-12-06}  are non-expandable in  $\Po^I$. So, vertex \rf{31032025-man01-12-05} does not meet restriction \rf{29032025-man01-23}. In the oscillator approach in Refs.\cite{Metsaev:2017cuz,Metsaev:2018moa}, we considered only those vertices which are expandable in $\Po^I$ and hence we did not find the oscillator cousin of vertex \rf{31032025-man01-12-05}. Note also that all we said in our comment {\bf iv}) in Sec.\, \ref{sec-4-5} in this paper goes without changes for the vertex \rf{31032025-man01-12-05}.

\noinbf{iv}) The variables $B_3^\alpha$, $C_{123}^{uu\alpha}$ are linear forms of the oscillators $\alpha_3^I$,  while the variable $q_{33}^{\alpha\alpha}$ is a quadratical form of the oscillators $\alpha_3^I$. In order for vertex $p_\smp3^-$ \rf{31032025-man01-12-05} to be sensible, this vertex should be expandable in the just mentioned variables (see definition \rf{30032025-man01-50}).

As the derivation of $p_\smp3^-$ \rf{31032025-man01-12-05} is similar to the one in Sec.\, \ref{sec-4-5}, we omit the derivation here.

\noinbf{Two massless CSFs and one massless integer-spin triplet field. Distributional solution}. Distributional cubic vertex $p_\smp3^-$ for two massless CSFs shown in \rf{31032025-man01-12-01} and one massless spin-$s_3$ triplet field is given by \rf{31032025-man01-12-05}, \rf{31032025-man01-12-06}, where $V$ takes the form
\be
\label{31032025-man01-12-15} V = (B_3^\alpha)^k (C_{123}^{uu\alpha})^n  (q_{33}^{\alpha\alpha})^l V_{n,l}(B_1^u,B_2^u)\,,\qquad
 k = s_3 - n - 2 l\,, \quad k\,, n \,, l \in \No_0\,,
\ee
while new vertex $V_{n,l}$ depends only on $B_1^u$, $B_2$. The integers $n$, $l$ subjected to the conditions in \rf{31032025-man01-12-15} and the vertex $V_{n_1,n_2,l}$  express the freedom of our solution for the vertex $V$.

\noinbf{Two massless CSFs and one massless integer-spin field. Distributional solution}. Distributional cubic vertex $p_\smp3^-$ for fields in \rf{31032025-man01-12-01} is given by \rf{31032025-man01-12-05}, \rf{31032025-man01-12-06}, where $V$ takes the form
\be \label{31032025-man01-12-30}
V = (B_3^\alpha)^k (C_{123}^{uu\alpha})^n   V_n(B_1^u,B_2^u)\big|_\Thsm\,, \qquad   k = s_3 - n\,, \qquad k\,, n \in \No_0\,,
\ee
while new vertex $V_n$ depends only on the variables $B_1^u$, $B_2^u$. In view of $\delta$-function \rf{31032025-man01-12-05} only one of the variables is relevant. The integer $n$ subjected to the conditions in \rf{31032025-man01-12-30} and the vertex $V_n$  express the freedom of our solution for the vertex $V$.
The notation $|_\Thsm$ implies that, in \rf{31032025-man01-12-30} and \rf{30032025-man01-30}, one just needs ro make the replacement $\alpha_3^I\rightarrow \alpha_{\Thsm\,3}^I$, where $\alpha_\Thsm^I$  is given in \rf{12042025-man01-05x} in Appendix A.  Such replacement is immaterial in \rf{29032025-man01-05} (see our comment at the end of Sec.\,\ref{sec-5-2}).

\vspace{-0.2cm}
\newsection{ \large Cubic vertices for one continuous-spin fields and two integer-spin fields}\label{sec-1csf}

We now discuss cubic vertices involving one massive/massless CSF and two massive/massless integers-spin fields. From classification \rf{26032025-man01-12}-\rf{26032025-man01-19}, we see that such cubic vertices can be separated into seven particular cases. We now present our results for these particular cases in turn.

\vspace{-0.2cm}
\subsection{\large One massive CSF and two massive integer-spin fields}

Using the shortcuts $(m,\SSc)_\CSFsm$, $(m,s)$ for the respective  massive CSF and massive integer-spin field, we consider the cubic vertex that involves the following fields:
\beq
\label{31032025-man01-13-01} && (m_1,\SSc_1)_\CSFsm\hbox{-}(m_2,s_2) \hbox{-}(m_3,s_3) \,,\qquad m_1^2 < 0\,, \qquad m_2^2 > 0 \,, \qquad m_3^2 > 0\,,
\nonumber\\
&&  \hbox{\small one massive CSF and two massive integer-spin fields.}\quad
\eeq
\noinbf{One massive CSF and two towers of massive integer-spin triplet fields}.
First, we present cubic vertex for CSF shown in \rf{31032025-man01-13-01} and two towers of massive integer-spin triplet fields \rf{27032025-man01-41} having masses $m_2$ and $m_3$.
General solution to equations \rf{30032025-man01-16}-\rf{30032025-man01-25} can be presented as
\beq
\label{31032025-man01-13-05} && \hspace{-1cm} p_\smp3^- =  (L_1^u)^{\SSc_1} V(L_2^\alpha\,, L_3^\alpha\,, \Qbf_{12}^{u\alpha}\,, Q_{23}^{\alpha\alpha}\,, \Qbf_{31}^{\alpha u}; Q_{22}^{\alpha\alpha}, Q_{33}^{\alpha\alpha})\,,
\\
\label{31032025-man01-13-06} &&  L_1^u = B_1^u  + \frac{\betach_1}{2\beta_1} |m_1| - \frac{m_2^2 - m_3^2}{2|m_1|}\,,
\nonumber\\
&&  L_2^\alpha =  B_2^\alpha - \frac{\betach_2}{2\beta_2} m_2\zeta_2 - \frac{m_3^2 - m_1^2}{2m_2}\zeta_2\,, \qquad L_3^\alpha =  B_3^\alpha - \frac{\betach_3}{2\beta_3} m_3\zeta_3 - \frac{m_1^2 - m_2^2}{2m_3}\zeta_3\,,
\nonumber\\
&& Q_{12}^{u\alpha}  =  q_{12}^{u\alpha} + \frac{1}{|m_1|} L_2^\alpha  -  \frac{\zeta_2}{m_2} L_1^u  + \frac{m_3^2 - m_1^2 - m_2^2}{ 2|m_1| m_2} \zeta_2\,,
\nonumber\\
&& Q_{23}^{\alpha\alpha}  =  q_{23}^{\alpha\alpha} + \frac{\zeta_2}{m_2} L_3^\alpha  -  \frac{\zeta_3}{ m_3 } L_2^\alpha  + \frac{ m_1^2 - m_2^2 - m_3^2  }{ 2 m_2 m_3 } \zeta_2\zeta_3\,,
\nonumber\\
&& Q_{31}^{\alpha u}  =  q_{31}^{\alpha u} + \frac{\zeta_3}{m_3} L_1^u  -  \frac{1}{ |m_1| } L_3^\alpha  + \frac{ \zeta_3  }{ 2 m_3 |m_1| } (m_2^2 - m_3^2 - m_1^2 )\,,
\nonumber\\
&& \Qbf_{12}^{u\alpha}  = \frac{Q_{12}^{u\alpha}}{L_1^u}\,, \qquad \Qbf_{31}^{\alpha u}  = \frac{Q_{31}^{\alpha u}}{L_1^u}\,,
\hspace{1cm} Q_{aa}^{\alpha\alpha} = q_{aa}^{\alpha\alpha} + \zeta_a\zeta_a\,, \qquad a=1,2\,,
\eeq
where, in \rf{31032025-man01-13-05}, we introduce vertex $V$ which depends on seven variables  shown explicitly in \rf{31032025-man01-13-05} and defined in \rf{31032025-man01-13-06}, \rf{30032025-man01-10}. The following remarks are in order.

\noinbf{i)} Dependence of vertex $V$ \rf{31032025-man01-13-05} on the seven variables is not fixed by equations \rf{30032025-man01-16}-\rf{30032025-man01-25}. The vertex $V$ is a freedom of our solution to cubic vertex $p_\smp3^-$. So, there are many cubic vertices.

\noinbf{ii)} The momentum $\Po^I$ enters variables $B_a^u$, $B_a^\alpha$ \rf{30032025-man01-10}. From  \rf{31032025-man01-13-05}, \rf{31032025-man01-13-06}, we then see that the dependence of vertex $p_\smp3^-$ on $\Po^I$ is governed by the variables $L_1^u$ and $L_2^\alpha$, $L_3^\alpha$, $\Qbf_{12}^{u\alpha}$, $Q_{23}^{\alpha\alpha}$, $\Qbf_{31}^{\alpha u}$.

\noinbf{iii)} Vertex $p_\smp3^-$ \rf{31032025-man01-13-05} is non-polynomial in $\Po^I$ and hence non-local. For example, taking into account the  values of the spin parameter $\SSc$ given in \rf{27032025-man01-60} and the pre-factor $(L_1^u)^{\SSc_1}$ in \rf{31032025-man01-13-05}, we see that cubic vertex \rf{31032025-man01-13-05} exhibits $PL$-non-locality  \rf{30032025-man01-35}. Note that the $PL$-non-locality is unavoidable feature of the vertex $p_\smp3^-$ as either choice of the vertex $V$ leads to $PL$-non-local $p_\smp3^-$.

\noinbf{iv}) The variables $L_2^\alpha$, $L_3^\alpha$, $\Qbf_{12}^{u\alpha}$, $\Qbf_{31}^{\alpha u}$ and $Q_{22}^{\alpha\alpha}$, $Q_{23}^{\alpha\alpha}$, $Q_{33}^{\alpha\alpha}$  are the respective linear and quadratical forms of the oscillators. Therefore, in order for our solution to be sensible, vertex $p_\smp3^-$ \rf{31032025-man01-13-05} should be expandable in the just mentioned variables.

\noinbf{One massive CSF and two massive integer-spin triplet fields}. The cubic vertex $p_\smp3^-$ for CSF shown in \rf{31032025-man01-13-01} and two massive spin-$s_2$ and spin-$s_3$ triplet fields having the respective masses $m_2$ and $m_3$ is given by \rf{31032025-man01-13-05}, \rf{31032025-man01-13-06}, where $V$ takes the form
{\small
\beq
\label{31032025-man01-13-15} && \hspace{-1.2cm} V = (L_2^\alpha)^{k_2} (L_3^\alpha)^{k_3} (\Qbf_{12}^{u\alpha})^{n_3} (Q_{23}^{\alpha\alpha})^{n_1} (\Qbf_{31}^{\alpha u})^{n_2} (Q_{22}^{\alpha\alpha})^{l_2} (Q_{33}^{\alpha\alpha})^{l_3}\,,
\nonumber\\
&& \hspace{-0.3cm}   k_2 = s_2 - n_1 - n_3 - 2 l_2\,, \quad k_3  = s_3 - n_1 - n_2 - 2 l_3\,,\quad k_2\,, k_3\,, n_1\,, n_2\,, n_3\,, l_2\,, l_3 \in \No_0\,.
\eeq
}
The integers $n_1$, $n_2$, $n_3$, $l_2$, $l_3$ subjected to the conditions in \rf{31032025-man01-13-15} express the freedom of our solution for the vertex $V$.

\noinbf{One massive CSF and two massive integer-spin fields}. The cubic vertex $p_\smp3^-$ for fields in \rf{31032025-man01-13-01} is given by \rf{31032025-man01-13-05}, \rf{31032025-man01-13-06}, where $V$ takes the form
{\small
\beq
\label{31032025-man01-13-30} && \hspace{-1.2cm} V = (L_2^\alpha)^{k_2} (L_3^\alpha)^{k_3} (\Qbf_{12}^{u\alpha})^{n_3} (Q_{23}^{\alpha\alpha})^{n_1} (\Qbf_{31}^{\alpha u})^{n_2}\big|_\Thsm\,,
\nonumber\\
&& \hspace{-0.3cm}   k_2 = s_2 - n_1 - n_3 \,, \quad k_3  = s_3 - n_1 - n_2\,,\quad k_2\,, k_3\,, n_1\,, n_2\,, n_3 \in \No_0\,.
\eeq
}
The integers $n_1$, $n_2$, $n_3$  subjected to conditions in \rf{31032025-man01-13-30} express the freedom of our solution for the vertex $V$, while the notation $|_\Thsm$ implies that, in \rf{31032025-man01-13-30} and \rf{30032025-man01-30}, the oscillators $\alpha_a^I$, $\zeta_a$  should be replaced by the new oscillators $\alpha_{\Thsm\,a}^I$,  $\zeta_{\Thsm\,a}$,
\be \label{31032025-man01-13-35}
\alpha_a^I\rightarrow \alpha_{\Thsm\,a}^I\,, \qquad \zeta_a\rightarrow \zeta_{\Thsm\,a}\,,\qquad a=1,2\,,
\ee
where $\alpha_\Thsm^I$, $\zeta_\Thsm$  are given in \rf{12042025-man01-05} in Appendix A.
Replacements \rf{31032025-man01-13-35} are immaterial in \rf{29032025-man01-05} (see our comment at the end of Sec.\,\ref{sec-5-1}).

\vspace{-0.2cm}
\subsection{ \large One massive CSF, one massive integer-spin field and one massless integer-spin field}

Using the shortcuts $(m,\SSc)_\CSFsm$, $(m,s)$, and  $(0,s)$ for the respective  massive CSF,  massive integer-spin field and massless integer-spin field, we consider the cubic vertex for the following fields:
\beq
\label{31032025-man01-14-01} && (m_1,\SSc_1)_\CSFsm\hbox{-}(m_2,s_2) \hbox{-}(0,s_3) \,,\qquad m_1^2 < 0\,, \qquad m_2^2 > 0\,,
\nonumber\\
&&  \hbox{\small one massive CSF and one massive integer-spin field, and one massless integer-spin field.}\quad
\eeq
\noinbf{One massive CSF, one tower of massive integer-spin triplet fields and one tower of massless integer-spin triplet fields}. First, we present cubic vertex for CSF in \rf{31032025-man01-14-01}, one tower of massive integer-spin triplet fields having mass $m_2$ and one tower of massless integer-spin triplet fields \rf{27032025-man01-41}.
General solution to equations \rf{30032025-man01-16}-\rf{30032025-man01-25} can be presented as
\beq
\label{31032025-man01-14-05} && \hspace{-1cm} p_\smp3^- =  (L_1^u)^{\SSc_1} V(L_2^\alpha\,, \Qbf_{12}^{u\alpha}\,, Q_{23}^{\alpha\alpha}\,, \Qbf_{31}^{\alpha u}; Q_{22}^{\alpha\alpha},q_{33}^{\alpha\alpha})\,,
\\
\label{31032025-man01-14-06} &&  L_1^u =  B_1^u + \frac{\betach_1}{2\beta_1} |m_1| - \frac{m_2^2}{2|m_1|}\,, \qquad L_2^\alpha =  B_2^\alpha - \frac{\betach_2}{2\beta_2} m_2\zeta_2 + \frac{m_1^2}{2m_2}\zeta_2\,,
\nonumber\\
&& Q_{12}^{u\alpha}  =  q_{12}^{u\alpha} + \frac{1}{|m_1|} L_2^\alpha  -  \frac{\zeta_2}{m_2} L_1^u  - \frac{m_1^2 + m_2^2}{ 2|m_1| m_2} \zeta_2 \,,
\nonumber\\
&& Q_{23}^{\alpha\alpha}  =  q_{23}^{\alpha\alpha} + \frac{\zeta_2}{m_2} B_3^\alpha   -  \frac{ 2 }{ m_1^2 - m_2^2 } L_2^\alpha B_3^\alpha\,,
\nonumber\\
&& Q_{31}^{\alpha u}  =  q_{31}^{\alpha u} -  \frac{1}{ |m_1| } B_3^\alpha  + \frac{ 2 }{ m_1^2 - m_2^2 }  B_3^\alpha L_1^u\,,
\nonumber\\
&& \Qbf_{12}^{u\alpha}  = \frac{Q_{12}^{u\alpha}}{L_1^u}\,, \qquad \Qbf_{31}^{\alpha u}  = \frac{Q_{31}^{\alpha u}}{L_1^u}\,, \hspace{1cm} Q_{22}^{\alpha\alpha} = q_{22}^{\alpha\alpha} + \zeta_2\zeta_2\,,
\eeq
where, in \rf{31032025-man01-14-05}, we introduce vertex $V$ which depends on six variables  shown explicitly in \rf{31032025-man01-14-05} and defined in \rf{31032025-man01-14-06}, \rf{30032025-man01-10}. The following remarks are in order.

\noinbf{i)} Dependence of vertex $V$ \rf{31032025-man01-14-05} on the six variables is not fixed by equations \rf{30032025-man01-16}-\rf{30032025-man01-25}. The vertex $V$ is a freedom of our solution to cubic vertex $p_\smp3^-$. So, there are many cubic vertices.

\noinbf{ii)} The momentum $\Po^I$ enters variables $B_a^u$, $B_a^\alpha$  \rf{30032025-man01-10}. From  \rf{31032025-man01-14-05}, \rf{31032025-man01-14-06}, we then see that the dependence of $p_\smp3^-$ on $\Po^I$ is governed by the variables $L_1^u$ and $L_2^\alpha$, $\Qbf_{12}^{u\alpha}$, $Q_{23}^{\alpha\alpha}$, $\Qbf_{31}^{\alpha u}$.

\noinbf{iii)} Vertex $p_\smp3^-$ \rf{31032025-man01-14-05} is non-polynomial in $\Po^I$ and hence non-local. For example, taking into account the  values of $\SSc$ given in \rf{27032025-man01-60} and the pre-factor $(L_1^u)^{\SSc_1}$ in \rf{31032025-man01-14-05}, we see that vertex $p_\smp3^-$ \rf{31032025-man01-14-05} exhibits $PL$-non-locality defined in \rf{30032025-man01-35}. Note that the $PL$-non-locality is unavoidable feature of the vertex $p_\smp3^-$ as either choice of the vertex $V$ leads to $PL$-non-local $p_\smp3^-$.

\noinbf{iv}) The variables $L_2^\alpha$, $\Qbf_{12}^{u\alpha}$, $\Qbf_{31}^{\alpha u}$ and $Q_{22}^{\alpha\alpha}$,  $Q_{23}^{\alpha\alpha}$, $q_{33}^{\alpha\alpha}$ are the respective linear and quadratical forms of the oscillators. Therefore, in order for our solution to be sensible,  vertex $p_\smp3^-$ \rf{31032025-man01-14-05} should be expandable in the just mentioned variables (see definition \rf{30032025-man01-50}).

\noinbf{One massive CSF, one massive integer-spin triplet field and one massless integer-spin triplet field}. Cubic vertex $p_\smp3^-$ for CSF in \rf{31032025-man01-14-01}, one massive spin-$s_2$ triplet field having mass $m_2$ and one massless spin-$s_3$ triplet field is given by \rf{31032025-man01-14-05}, \rf{31032025-man01-14-06}, where $V$ takes the form
{\small
\beq
\label{31032025-man01-14-15} && \hspace{-1.2cm} V = (L_2^\alpha)^{k_2}  (\Qbf_{12}^{u\alpha})^{n_3} (Q_{23}^{\alpha\alpha})^{n_1} (\Qbf_{31}^{\alpha u})^{n_2} (Q_{22}^{\alpha\alpha})^{l_2} (q_{33}^{\alpha\alpha})^{l_3}\,,
\nonumber\\
&& \hspace{-0.3cm}   k_2 = s_2 - n_1 - n_3 - 2 l_2\,, \quad   n_1 +  n_2 + 2 l_3 = s_3\,,\quad k_2\,, n_1\,, n_2\,, n_3\,, l_2\,, l_3 \in \No_0\,.
\eeq
}
\!The integers $n_1$, $n_2$, $n_3$, $l_2$ in \rf{31032025-man01-14-15} express the freedom of our solution for the vertex $V$.

\noinbf{One massive CSF, one massive integer-spin field, and one massless integer-spin field}. Cubic vertex $p_\smp3^-$ for fields in \rf{31032025-man01-14-01} is given by \rf{31032025-man01-14-05}, \rf{31032025-man01-14-06}, where $V$ takes the form
{\small
\beq
\label{31032025-man01-14-30} && \hspace{-1.2cm} V = (L_2^\alpha)^{k_2}  (\Qbf_{12}^{u\alpha})^{n_3} (Q_{23}^{\alpha\alpha})^{n_1} (\Qbf_{31}^{\alpha u})^{n_2}\big|_\Thsm\,,
\nonumber\\
&& \hspace{-0.3cm}   k_2 = s_2 - n_1 - n_3 \,, \quad  n_1 +  n_2 = s_3\,,\quad k_2\,, n_1\,, n_2\,, n_3 \in \No_0\,.
\eeq
}
\!The integers $n_1$, $n_2$, $n_3$  subjected to the conditions in \rf{31032025-man01-14-30} express the freedom of our solution for the vertex $V$, while the notation $|_\Thsm$ implies that, in \rf{31032025-man01-14-30} and \rf{30032025-man01-30}, the oscillators $\alpha_2^I$, $\zeta_2$ and $\alpha_3^I$ should be replaced by the respective oscillators $\alpha_{\Thsm\,2}^I$,  $\zeta_{\Thsm\,2}$ and $\alpha_{\Thsm\,3}^I$,
\be \label{31032025-man01-14-35}
\alpha_2^I\rightarrow \alpha_{\Thsm\,2}^I\,, \qquad \zeta_2\rightarrow \zeta_{\Thsm\,2}\,, \qquad \alpha_3^I\rightarrow \alpha_{\Thsm\,3}^I\,,
\ee
where we use the notation given in \rf{12042025-man01-05} \rf{12042025-man01-05x} in Appendix A. Replacements \rf{31032025-man01-14-35} are immaterial in \rf{29032025-man01-05} (see our comment at the end of Secs. \ref{sec-5-1} and \ref{sec-5-2}).

\vspace{-0.2cm}
\subsection{\large One massive CSF and two massless integer-spin fields}

Using the shortcuts $(m,\SSc)_\CSFsm$ and $(0,s)$ for the respective  massive continuous-spin field and massless integer-spin field, we consider the cubic vertex that involves the following fields:
\beq
\label{31032025-man01-15-01} && (m_1,\SSc_1)_\CSFsm\hbox{-}(0,s_2) \hbox{-}(0,s_3) \,,\qquad m_1^2 < 0 \,,
\nonumber\\
&&  \hbox{\small one massive CSF and two massless integer-spin fields.}\quad
\eeq
\noinbf{One massive CSF and two towers of massless integer-spin triplet fields}. First, we present cubic vertex for CSF shown in \rf{31032025-man01-15-01} and two towers of massless integer-spin fields \rf{27032025-man01-41}. We find the following general solution to equations    \rf{30032025-man01-16}-\rf{30032025-man01-25}:
\beq
\label{31032025-man01-15-05} && \hspace{-1cm} p_\smp3^- =  (L_1^u)^{\SSc_1} V(\Qbf_{12}^{u\alpha}\,, Q_{23}^{\alpha\alpha}\,, \Qbf_{31}^{\alpha u};q_{22}^{\alpha\alpha}, q_{33}^{\alpha\alpha})\,,
\\
\label{31032025-man01-15-06} &&  L_1^u =  B_1^u + \frac{\betach_1}{2\beta_1} |m_1| \,,
\nonumber\\
&& Q_{12}^{u\alpha}  =  q_{12}^{u\alpha} + \frac{1}{|m_1|} B_2^\alpha  + \frac{2}{m_1^2} L_1^u B_2^\alpha \,,
\nonumber\\
&& Q_{23}^{\alpha\alpha}  =  q_{23}^{\alpha\alpha}  -  \frac{ 2 }{ m_1^2} B_2^\alpha B_3^\alpha\,,  \hspace{2cm} Q_{31}^{\alpha u}  =  q_{31}^{\alpha u} -  \frac{1}{ |m_1| } B_3^\alpha  + \frac{ 2 }{ m_1^2}  B_3^\alpha L_1^u\,,
\nonumber\\
&& \Qbf_{12}^{u\alpha}  = \frac{Q_{12}^{u\alpha}}{L_1^u}\,, \qquad \Qbf_{31}^{\alpha u}  = \frac{Q_{31}^{\alpha u}}{L_1^u}\,,
\eeq
where, in \rf{31032025-man01-15-05}, we introduce vertex $V$ which depends on five variables  shown explicitly in \rf{31032025-man01-15-05} and defined in \rf{31032025-man01-15-06}, \rf{30032025-man01-10}. The following remarks are in order.

\noinbf{i)} Vertex $V$ \rf{31032025-man01-15-05} is not fixed by equations \rf{30032025-man01-16}-\rf{30032025-man01-25}. The vertex $V$ is a freedom of our solution to cubic vertex $p_\smp3^-$. So, there are many cubic vertices.

\noinbf{ii)} The momentum $\Po^I$ enters the variables $B_a^u$, $B_a^\alpha$ \rf{30032025-man01-10}. From  \rf{31032025-man01-15-05}, \rf{31032025-man01-15-06}, we see that the dependence of vertex $p_\smp3^-$ on the momentum $\Po^I$ is governed by the variables $L_1^u$, $\Qbf_{12}^{u\alpha}$, $Q_{23}^{\alpha\alpha}$, $\Qbf_{31}^{\alpha u}$.

\noinbf{iii)} Vertex $p_\smp3^-$ \rf{31032025-man01-15-05} is non-polynomial in $\Po^I$ and hence non-local. For example, taking into account the  values of $\SSc$ given in \rf{27032025-man01-60} and the pre-factor $(L_1^u)^{\SSc_1}$ in \rf{31032025-man01-15-05}, we see that cubic vertex \rf{31032025-man01-15-05} exhibits $PL$-non-locality defined in \rf{30032025-man01-35}. Note that the $PL$-non-locality is unavoidable feature of the vertex $p_\smp3^-$ as either choice of the vertex $V$ leads to $PL$-non-local $p_\smp3^-$.

\noinbf{iv}) The variables $L_2^\alpha$, $\Qbf_{12}^{u\alpha}$, $\Qbf_{31}^{\alpha u}$ and $q_{22}^{\alpha\alpha}$,  $Q_{23}^{\alpha\alpha}$, $q_{33}^{\alpha\alpha}$ are the respective linear and quadratical forms of the oscillators. Therefore, in order for our solution to be sensible,  vertex $p_\smp3^-$ \rf{31032025-man01-15-05} should be expandable in the just mentioned variables (see definition \rf{30032025-man01-50}).

\noinbf{One massive CSF and two massless integer-spin triplet fields}. Vertex $p_\smp3^-$ for CSF in \rf{31032025-man01-15-01} and two massless spin-$s_2$ and spin-$s_3$ triplet fields is given by \rf{31032025-man01-15-05}, \rf{31032025-man01-15-06}, where $V$ takes the form
{\small
\beq
\label{31032025-man01-15-15} && \hspace{-1.2cm} V = (\Qbf_{12}^{u\alpha})^{n_3} (Q_{23}^{\alpha\alpha})^{n_1} (\Qbf_{31}^{\alpha u})^{n_2} (q_{22}^{\alpha\alpha})^{l_2} (q_{33}^{\alpha\alpha})^{l_3}\,,
\nonumber\\
&& \hspace{-0.3cm}    s_2 = n_1 + n_3 + 2 l_2\,, \quad  s_3 =  n_1 + n_2 + 2 l_3\,,\quad  n_1\,, n_2\,, n_3\,, l_2\,, l_3 \in \No_0\,.
\eeq
}
\!The integers $n_1$, $n_2$, $n_3$, $l_2$, $l_3$ in \rf{31032025-man01-15-15} express the freedom of our solution for the vertex $V$.

\noinbf{One massive CSF and two massless integer-spin fields}.
Cubic vertex $p_\smp3^-$ for fields in \rf{31032025-man01-15-01} is given by \rf{31032025-man01-15-05}, \rf{31032025-man01-15-06}, where $V$ takes the form
{\small
\be
\label{31032025-man01-15-30} V = (\Qbf_{12}^{u\alpha})^{n_3} (Q_{23}^{\alpha\alpha})^{n_1} (\Qbf_{31}^{\alpha u})^{n_2}\big|_\Thsm\,,\quad s_2 =  n_1 + n_3 \,, \quad s_3 =  n_1 + n_2\,,\quad  n_1\,, n_2\,, n_3 \in \No_0\,.
\ee
}
\!In \rf{31032025-man01-15-30}, the integers $n_1$, $n_2$, $n_3$ express the freedom of our solution for the vertex $V$, while the notation $|_\Thsm$ implies that the oscillators $\alpha_2^I$, $\alpha_3^I$ in \rf{31032025-man01-15-30} and \rf{30032025-man01-30} should be replaced as $\alpha_2^I\rightarrow \alpha_{\Thsm\,2}^I$, $\alpha_3^I\rightarrow \alpha_{\Thsm\,3}^I$, where $\alpha_\Thsm^I$ is given in \rf{12042025-man01-05x} in Appendix A. Such replacements are immaterial in \rf{29032025-man01-05} (see our comment at the end of Sec.\,\ref{sec-5-2}).

\vspace{-0.2cm}
\subsection{ One massless CSF and two massive integer-spin fields (non-equal masses)}

Using the shortcuts $(0,\kappa)_\CSFsm$ and $(m,s)$ for the respective  massless CSF and  massive integer-spin field, we consider the cubic vertex that involves the following fields:
\beq
\label{31032025-man01-16-01} && (0,\kappa_1)_\CSFsm\hbox{-}(m_2,s_2) \hbox{-}(m_3,s_3) \,, \qquad m_2^2 > 0\qquad m_3^2>0\,,\qquad m_2^2 \ne m_3^2\,,
\nonumber\\
&&  \hbox{\small one massless CSF and two massive integer-spin fields with non-equal masses.}\quad
\eeq
\noinbf{One massless CSF and two towers of massive triplet fields}. First, we present cubic vertex for CSF shown in \rf{31032025-man01-16-01} and two towers of massive integer-spin fields \rf{27032025-man01-41} having masses $m_2$ and $m_3$, $m_2\ne m_3$. We find the following general solution to equations    \rf{30032025-man01-16}-\rf{30032025-man01-25}:
\beq
\label{31032025-man01-16-05} && \hspace{-1cm} p_\smp3^- =   e^W V(L_2^\alpha,L_3^\alpha,Q_{12}^{u\alpha}\,, Q_{23}^{\alpha\alpha}\,, Q_{31}^{\alpha u};\, Q_{22}^{\alpha\alpha}\,, Q_{33}^{\alpha \alpha})\,,
\\
&& W = -\frac{2\irm \kappa_1 B_1^u}{m_2^2 - m_3^2} \,,
\nonumber\\
\label{31032025-man01-16-06} &&  L_2^\alpha =  B_2^\alpha - \frac{\betach_2}{2\beta_2} m_2\zeta_2 - \frac{m_3^2}{2m_2} \zeta_2\,, \qquad L_3^\alpha =  B_3^\alpha - \frac{\betach_3}{2\beta_3} m_3 \zeta_3 + \frac{m_2^2}{2m_3} \zeta_3\,,
\nonumber\\
&& Q_{12}^{u\alpha}  =  q_{12}^{u\alpha} - \frac{\zeta_2}{m_2} B_1^u  + \frac{2}{m_2^2-m_3^2} B_1^u L_2^\alpha \,,
\nonumber\\
&& Q_{23}^{\alpha\alpha}  =  q_{23}^{\alpha\alpha}  + \frac{\zeta_2}{m_2} L_3^\alpha - \frac{\zeta_3}{m_3} L_2^\alpha  - \frac{m_2^2+m_3^2}{2m_2m_3} \zeta_2 \zeta_3\,,
\nonumber\\
&& Q_{31}^{\alpha u}  =  q_{31}^{\alpha u} + \frac{\zeta_3}{ m_3 } B_1^u  - \frac{ 2 }{ m_2^2 - m_3^2}  L_3^\alpha B_1^u\,,
\nonumber\\
&& Q_{aa}^{\alpha\alpha} = q_{aa}^{\alpha\alpha} + \zeta_a\zeta_a\,, \qquad a=2,3\,,
\eeq
where, in \rf{31032025-man01-16-05}, we introduce vertex $V$ which depends on seven variables  shown explicitly in \rf{31032025-man01-16-05} and defined in \rf{31032025-man01-16-06}, \rf{30032025-man01-10}.

\noinbf{i)} Dependence of vertex $V$ \rf{31032025-man01-16-05} on the seven variables is not fixed by equations \rf{30032025-man01-16}-\rf{30032025-man01-25}. The vertex $V$ is a freedom of our solution to cubic vertex $p_\smp3^-$. So, there are many cubic vertices.

\noinbf{ii)} The momentum $\Po^I$ enters variables $B_a^u$, $B_a^\alpha$ \rf{30032025-man01-10}. From  \rf{31032025-man01-16-05}, \rf{31032025-man01-16-06}, we then see that the dependence of $p_\smp3^-$ on $\Po^I$ is governed by the variables $W$ and  $L_2^\alpha$, $L_3^\alpha$, $Q_{12}^{u\alpha}$, $Q_{23}^{\alpha\alpha}$, $Q_{31}^{\alpha u}$.

\noinbf{iii)} Vertex $p_\smp3^-$ \rf{31032025-man01-16-05} is non-polynomial in $\Po^I$ and hence non-local. For example, taking into account the  values of the spin parameter $\SSc$ given in \rf{27032025-man01-60} and the pre-factor $e^W$ in \rf{31032025-man01-15-05}, we see that  vertex $p_\smp3^-$ \rf{31032025-man01-16-05} exhibits $E$-non-locality defined in \rf{30032025-man01-35}. Note that the $E$-non-locality is unavoidable feature of the vertex $p_\smp3^-$ as either choice of the vertex $V$ leads to $PL$-non-local $p_\smp3^-$.

\noinbf{iv}) The variables $L_2^\alpha$, $L_3^\alpha$, $Q_{12}^{u\alpha}$, $Q_{31}^{\alpha u}$ and $Q_{22}^{\alpha\alpha}$,  $Q_{23}^{\alpha\alpha}$, $Q_{33}^{\alpha\alpha}$ are the respective linear and quadratical forms of the oscillators. Therefore, in order for our solution to be sensible,  vertex $p_\smp3^-$ \rf{31032025-man01-16-05} should be expandable in the just mentioned variables (see \rf{30032025-man01-50}).

\noinbf{One massless CSF and two massive integer-spin triplet fields}. The cubic vertex $p_\smp3^-$ for CSF in \rf{31032025-man01-16-01} and two massive spin-$s_2$ and spin-$s_3$ triplet fields having the respective masses $m_2$ and $m_3$, $m_2\ne m_3$, is given by \rf{31032025-man01-16-05}, \rf{31032025-man01-16-06}, where $V$ takes the form
{\small
\beq
\label{31032025-man01-16-15} && \hspace{-1.2cm} V = (L_2^\alpha)^{k_2} (L_3^\alpha)^{k_3} (Q_{12}^{u\alpha})^{n_3} (Q_{23}^{\alpha\alpha})^{n_1} (Q_{31}^{\alpha u})^{n_2} (Q_{22}^{\alpha\alpha})^{l_2} (Q_{33}^{\alpha\alpha})^{l_3}\,,
\nonumber\\
&& \hspace{-0.3cm}   k_2 = s_2 - n_1 - n_3 - 2 l_2\,, \quad k_3  = s_3 - n_1 - n_2 - 2 l_3\,,\quad k_2\,, k_3\,, n_1\,, n_2\,, n_3\,, l_2\,, l_3 \in \No_0\,,\qquad
\eeq
}
and integers $n_1$, $n_2$, $n_3$, $l_2$, $l_3$ \rf{31032025-man01-16-15} express the freedom of our solution for the vertex $V$.

\noinbf{One massless CSF and two massive integer-spin fields}. Cubic vertex $p_\smp3^-$ for fields in \rf{31032025-man01-16-01} is given by \rf{31032025-man01-16-05}, \rf{31032025-man01-16-06}, where $V$ takes the form
{\small
\beq
\label{31032025-man01-16-30} && \hspace{-1.2cm} V = (L_2^\alpha)^{k_2} (L_3^\alpha)^{k_3} (Q_{12}^{u\alpha})^{n_3} (Q_{23}^{\alpha\alpha})^{n_1} (Q_{31}^{\alpha u})^{n_2}\big|_\Thsm\,,
\nonumber\\
&& \hspace{-0.3cm}   k_2 = s_2 - n_1 - n_3 \,, \quad k_3  = s_3 - n_1 - n_2\,,\quad k_2\,, k_3\,, n_1\,, n_2\,, n_3 \in \No_0\,.
\eeq
}
\!In \rf{31032025-man01-16-30}, the integers $n_1$, $n_2$, $n_3$  express the freedom of our solution for the vertex $V$, while the notation $|_\Thsm$ implies that the oscillators $\alpha_a^I$, $\zeta_a$ in \rf{31032025-man01-16-30} and \rf{30032025-man01-30} should be replaced as
\be \label{31032025-man01-16-35}
\alpha_a^I\rightarrow \alpha_{\Thsm\,a}^I\,, \qquad \zeta_a\rightarrow \zeta_{\Thsm\,a}\,,\qquad a=1,2\,,
\ee
where $\alpha_\Thsm^I$, $\zeta_\Thsm$  are given in \rf{12042025-man01-05} in Appendix A.
Replacements \rf{31032025-man01-16-35} are immaterial in \rf{29032025-man01-05} (see our comment at the end of Sec.\,\ref{sec-5-1}).

\vspace{-0.2cm}
\subsection{ One massless CSF and two massive integer-spin fields (equal masses)} \label{subsec-6-5}

Using the shortcuts $(0,\SSc)_\CSFsm$ and  $(m,s)$ for the respective  massless CSF and massive integer-spin field, we consider the cubic vertex that involves the following fields:
\beq
\label{31032025-man01-17-01} && (0,\kappa_1)_\CSFsm\hbox{-}(m_2,s_2) \hbox{-}(m_3,s_3) \,,\qquad m_2^2=m^2\,, \qquad m_3^2=m^2\,, \qquad m^2 > 0\,,
\nonumber\\
&&  \hbox{\small one massless CSF and two massive integer-spin fields with equal masses.}\quad
\eeq
We state that {\it there are no f-solutions to equations for cubic vertex   \rf{30032025-man01-16}-\rf{30032025-man01-25}.} For proof, see our comment below.

For the  simplest case of two scalar fields in \rf{31032025-man01-17-01}, $s_1=0$, $s_2=0$, we were able to find a distributional solution to cubic vertex which we now discuss.%
\footnote{Finding distributional solutions for integer-spin $s_1$ and $s_2$ fields in \rf{31032025-man01-17-01} is beyond the scope of the present paper.
}

\noinbf{Distributional solution}. We consider the cubic vertex that involves the following fields:
\beq
\label{31032025-man01-17-10} && (0,\kappa_1)_\CSFsm\hbox{-}(m_2,0) \hbox{-}(m_3,0) \,,\qquad m_2^2=m^2\,, \qquad m_3^2=m^2\,, \qquad m^2 > 0\,,
\nonumber\\
&&  \hbox{\small one massless CSF and two massive scalar fields with equal masses.}\quad
\eeq
We find two distributional solutions to cubic vertices for fields in \rf{31032025-man01-17-10} which we denote as $p_\smp3^-(\even)$ and $p_\smp3^-(\odd)$. The explicit form of the solutions is given by
\be \label{31032025-man01-17-15}
2\NN^{-1}\, p_\smp3^-(\even \,,  \odd)  =    e^{- \frac{\kappa_1 \betach_1}{2m \beta_1 } }  c(\frac{\irm \Po}{m\beta_1},u_1)  \pm e^{\frac{\kappa_1 \betach_1}{2 m \beta_1} }  c(-\frac{\irm \Po}{m\beta_1},u_1)\,,
\ee
where $c(U,u)$ stands for a distribution which we refer to as $c$ -- distribution (see Appendix D). For the derivation of vertex \rf{31032025-man01-17-15}, see Appendix F. It remains to understand about whether and not and in which ways such cubic vertex might be interesting and relevant for CSF theory.

\noinbf{Comment on non-sensible f-solution}. All f-solutions to equations \rf{30032025-man01-16}-\rf{30032025-man01-25} for cubic vertices \rf{31032025-man01-17-01} are found to be
\beq
\label{31032025-man01-17-05} && \hspace{-1cm} p_\smp3^- =  e^W   V\big(L_2^\alpha, L_3^\alpha, Q_{23}^{\alpha\alpha}, C_{123}^{u\alpha\alpha};\,  Q_{22}^{\alpha\alpha}\,; Q_{33}^{\alpha \alpha}\big)\,,  \hspace{1cm} \hbox{non-sensible solution};
\\
\label{31032025-man01-17-06} &&  L_2^\alpha =  B_2^\alpha + \frac{\beta_1}{\beta_2} m\zeta_2 \,, \qquad  L_3^\alpha =  B_3^\alpha - \frac{\beta_1}{\beta_3} m \zeta_3 \,,
\nonumber\\
&& Q_{12}^{u\alpha}  =  q_{12}^{u\alpha} - \frac{\zeta_2}{m} B_1^u  \,, \hspace{1cm} Q_{23}^{\alpha\alpha}  =  q_{23}^{\alpha\alpha}  + \frac{\zeta_2}{m} L_3^\alpha - \frac{\zeta_3}{m} L_2^\alpha  -  \zeta_2 \zeta_3\,,
\nonumber\\
&& Q_{31}^{\alpha u}  =  q_{31}^{\alpha u} + \frac{\zeta_3}{ m } B_1^u  \,, \hspace{1cm}  Q_{aa}^{\alpha\alpha} = q_{aa}^{\alpha\alpha} + \zeta_a\zeta_a\,, \qquad a=1,2\,,
\nonumber\\
&& W =  \frac{\irm \kappa_1}{2L_2^\alpha L_3^\alpha} \big( Q_{12}^{u\alpha} L_3^\alpha -  Q_{31}^{\alpha u} L_2^\alpha)\,,\qquad C_{123}^{u\alpha\alpha} = Q_{12}^{u\alpha} L_3^\alpha +  Q_{31}^{\alpha u} L_2^\alpha\,,
\eeq
where, in \rf{31032025-man01-17-05}, we introduce vertex $V$ which depends on six variables  shown explicitly in \rf{31032025-man01-17-05} and defined in \rf{31032025-man01-17-06}, \rf{30032025-man01-10}. In view of negative powers of oscillator variables $L_2^\alpha$, $L_3^\alpha$ appearing in the expression for $W$ given in \rf{31032025-man01-17-06}, we see that vertex $p_\smp3^-$ \rf{31032025-man01-17-05} is indeed non-sensible.

\vspace{-0.2cm}
\subsection{ One massless CSF, one massive integer-spin field and one massless integer-spin field}

Using the shortcuts $(0,\kappa)_\CSFsm$, $(m,s)$, and  $(0,s)$ for the respective  massless CSF,  massive integer-spin field and massless integer-spin field, we consider the cubic vertex for the following fields:
\beq
\label{31032025-man01-18-01} && (0,\kappa_1)_\CSFsm\hbox{-}(m_2,s_2) \hbox{-}(0,s_3) \,,\qquad m_2^2 > 0\,,
\nonumber\\
&&  \hbox{\small one massless CSF, one massive integer-spin field, and one massless integer-spin field.}\quad
\eeq
\noinbf{One massless CSF, one tower of massive triplet fields and one tower of massless triplet fields}. First, we present cubic vertex for CSF shown in \rf{31032025-man01-18-01}, one tower of massive integer-spin triplet fields \rf{27032025-man01-41} having mass $m_2$, and one tower of massless integer-spin triplet fields \rf{27032025-man01-41}. We find the following general solution to equations \rf{30032025-man01-16}-\rf{30032025-man01-25}:
\beq
\label{31032025-man01-18-05} && \hspace{-1cm} p_\smp3^- =   e^W  V(L_2^\alpha\,,Q_{12}^{u\alpha}\,, Q_{23}^{\alpha\alpha}\,, Q_{31}^{\alpha u};\, Q_{22}^{\alpha\alpha}\,, q_{33}^{\alpha \alpha})\,,
\\
\label{31032025-man01-18-06} && W  = -\frac{2\irm \kappa_1 B_1^u}{m_2^2} \,,   \hspace{1cm}L_2^\alpha =  B_2^\alpha - \frac{\betach_2}{2\beta_2} m_2\zeta_2\,,
\nonumber\\
&& Q_{12}^{u\alpha}  =  q_{12}^{u\alpha} - \frac{\zeta_2}{m_2} B_1^u  + \frac{2}{m_2^2} B_1^u L_2^\alpha \,, \hspace{1cm}  Q_{23}^{\alpha\alpha}  =  q_{23}^{\alpha\alpha}  + \frac{\zeta_2}{m_2} B_3^\alpha  + \frac{2}{m_2^2} L_2^\alpha B_3^\alpha\,,
\nonumber\\
&& Q_{31}^{\alpha u}  =  q_{31}^{\alpha u} - \frac{ 2 }{ m_2^2 }  B_3^\alpha B_1^u\,,
\hspace{2.6cm} Q_{22}^{\alpha\alpha} = q_{22}^{\alpha\alpha} + \zeta_2\zeta_2\,,
\eeq
where, in \rf{31032025-man01-18-05}, we introduce vertex $V$ which depends on six variables  shown explicitly in \rf{31032025-man01-18-05} and defined in \rf{31032025-man01-18-06}, \rf{30032025-man01-10}. The following remarks are in order.

\noinbf{i)} Dependence of vertex $V$ \rf{31032025-man01-18-05} on the six variables is not fixed by equations \rf{30032025-man01-16}-\rf{30032025-man01-25}. The vertex $V$ is a freedom of our solution to cubic vertex $p_\smp3^-$. So, there are many cubic vertices.

\noinbf{ii)} The momentum $\Po^I$ enters variables $B_a^u$, $B_a^\alpha$ \rf{30032025-man01-10}. From  \rf{31032025-man01-18-05}, \rf{31032025-man01-18-06}, we then see that the dependence of $p_\smp3^-$ on $\Po^I$ is governed by the variables $W$, and  $L_2^\alpha$, $Q_{12}^{u\alpha}$, $Q_{23}^{\alpha\alpha}$, $Q_{31}^{\alpha u}$.

\noinbf{iii)} Vertex $p_\smp3^-$ \rf{31032025-man01-18-05} is non-polynomial in $\Po^I$ and hence non-local. For example, in view of the pre-factor $e^W$ in \rf{31032025-man01-18-05}, the vertex $p_\smp3^-$ \rf{31032025-man01-18-05} exhibits $E$-non-locality  \rf{30032025-man01-35}. This $E$-non-locality is unavoidable feature of the vertex $p_\smp3^-$ as either choice of the vertex $V$ leads to $E$-non-local $p_\smp3^-$.

\noinbf{iv}) The variables $L_2^\alpha$, $Q_{12}^{u\alpha}$, $Q_{31}^{\alpha u}$ and $Q_{22}^{\alpha\alpha}$,  $Q_{23}^{\alpha\alpha}$, $q_{33}^{\alpha\alpha}$ are the respective linear and quadratical forms of the oscillators. Therefore, in order for our solution to be sensible,  vertex $p_\smp3^-$ \rf{31032025-man01-18-05} should be expandable in the just mentioned variables (see definition \rf{30032025-man01-50}).

\noinbf{One massless CSF, one massive integer-spin triplet field and one massless integer-spin triplet field}. Cubic vertex $p_\smp3^-$ for CSF in \rf{31032025-man01-18-05}, one massive spin-$s_2$ triplet field having mass $m_2$ and one massless spin-$s_3$ triplet field is given by \rf{31032025-man01-18-05}, \rf{31032025-man01-18-06}, where $V$ takes the form
{\small
\beq
\label{31032025-man01-18-15} && \hspace{-1.2cm} V = (L_2^\alpha)^{k_2} (Q_{12}^{u\alpha})^{n_3} (Q_{23}^{\alpha\alpha})^{n_1} (Q_{31}^{\alpha u})^{n_2} (Q_{22}^{\alpha\alpha})^{l_2} (q_{33}^{\alpha\alpha})^{l_3}\,,
\nonumber\\
&& \hspace{-0.3cm}   k_2 = s_2 - n_1 - n_3 - 2 l_2\,, \quad   n_1 +  n_2 + 2 l_3 = s_3\,,\quad k_2\,, n_1\,, n_2\,, n_3\,, l_2\,, l_3 \in \No_0\,.
\eeq
}
\!The integers $n_1$, $n_2$, $n_3$, $l_2$, $l_3$ in \rf{31032025-man01-18-15} express the freedom of our solution for the vertex $V$.

\noinbf{One massless CSF, one massive integer-spin field, and one massless integer-spin field}. Cubic vertex $p_\smp3^-$ for fields in \rf{31032025-man01-18-01} is given by \rf{31032025-man01-18-05}, \rf{31032025-man01-18-06}, where $V$ takes the form
{\small
\beq
\label{31032025-man01-18-30} && \hspace{-1.2cm} V = (L_2^\alpha)^{k_2}  (Q_{12}^{u\alpha})^{n_3} (Q_{23}^{\alpha\alpha})^{n_1} (Q_{31}^{\alpha u})^{n_2}\big|_\Thsm\,,
\nonumber\\
&& \hspace{-0.3cm}   k_2 = s_2 - n_1 - n_3 \,, \quad  n_1 +  n_2 = s_3\,,\quad k_2\,, n_1\,, n_2\,, n_3 \in \No_0\,.
\eeq
}
\!The integers $n_1$, $n_2$, $n_3$  subjected to the conditions in \rf{31032025-man01-18-30} express the freedom of our solution for the vertex $V$, while the notation $|_\Thsm$ implies that, in \rf{31032025-man01-18-30} and \rf{30032025-man01-30}, the oscillators $\alpha_2^I$, $\zeta_2$ and $\alpha_3^I$  should be replaced by the respective new oscillators $\alpha_{\Thsm\,2}^I$,  $\zeta_{\Thsm\,2}$ and $\alpha_{\Thsm\,3}^I$,
\be \label{31032025-man01-18-35}
\alpha_2^I\rightarrow \alpha_{\Thsm\,2}^I\,, \qquad \zeta_2\rightarrow \zeta_{\Thsm\,2}\,, \qquad \alpha_3^I\rightarrow \alpha_{\Thsm\,3}^I\,,
\ee
where we use the notation given in \rf{12042025-man01-05}, \rf{12042025-man01-05x} in Appendix A. Replacements \rf{31032025-man01-18-35} are immaterial in \rf{29032025-man01-05} (see our remarks at the end of Secs. \ref{sec-5-1} and \ref{sec-5-2}).

\vspace{-0.2cm}
\subsection{ One massless CSF and two massless integer-spin fields}

Using the shortcuts $(0,\kappa)_\CSFsm$ and $(0,s)$ for the respective  massless CSF and massless integer-spin field, we consider the cubic vertex that involves the following fields:
\beq
\label{31032025-man01-19-01} && (0,\kappa_1)_\CSFsm\hbox{-}(0,s_2) \hbox{-}(0,s_3) \,,
\nonumber\\
&&  \hbox{\small one massless CSF and two massless integer-spin fields.}\quad
\eeq

\noinbf{Statement}. There are no f-solutions to equations    \rf{30032025-man01-16}-\rf{30032025-man01-25} for cubic vertex \rf{31032025-man01-19-01}, while the particular distributional solution to equations    \rf{30032025-man01-16}-\rf{30032025-man01-25} turns out to be non-sensible.

\noinbf{Proof of Statement}. As the proof of the Statement is similar to the one in Sec.\, \ref{sec-4-5} we skip the details and just present the result. Namely, we verify that the f-solution is not available, while the distributional solution takes formally the following form
\beq
\label{31032025-man01-19-05} &&  \hspace{-1cm} p_\smp3^- =   e^W   V(B_2^\alpha, B_3^\alpha, C_{123}^{u\alpha\alpha},q_{22}^{\alpha\alpha}\,, q_{33}^{\alpha \alpha}) \delta(B_1^u)\,,  \hspace{1cm} \hbox{non-sensible solution};
\\
\label{31032025-man01-19-06}  && W =  \frac{ \irm \kappa_1 }{2B_2^\alpha B_3^\alpha}\big(q_{12}^{u\alpha} B_3^\alpha -  q_{31}^{\alpha u} B_2^\alpha\big) \,,
\qquad C_{123}^{u\alpha\alpha} = q_{12}^{u\alpha} B_3^\alpha +  q_{31}^{u\alpha} B_2^\alpha\,,
\eeq
where, in \rf{31032025-man01-19-05}, we introduce vertex $V$ which depends on five variables  shown explicitly in \rf{31032025-man01-19-05} and defined in \rf{31032025-man01-19-06}, \rf{30032025-man01-10}. In view of negative powers of the oscillator variables $B_2^\alpha$, $B_3^\alpha$ appearing in $W$, we see that vertex $p_\smp3^-$ \rf{31032025-man01-19-05} is indeed non-sensible. We recall that our equations \rf{30032025-man01-16}-\rf{30032025-man01-25}, though applicable for finding some particular distributional solutions, are not applicable for finding all distributional solutions. For this reason, our Statement does not imply that there are no sensible distributional solutions at all. We recall also that our results are adopted to CSF in $\Ro^{d-1,1}$, with $d>4$. For $d=4$, our analysis requires to be carried out separately.

\vspace{-0.2cm}
\newsection{ Conclusions}\label{concl}

In this paper, we developed the light-cone gauge vector formulation of CSFs and applied such formulation for study of cubic interaction vertices of CSFs and integer-spin fields which involve at least one CSF.
For self-interactions of CSF and cross-interactions between CSFs and integer-spin fields, we found all parity-even cubic vertices realized as functions on the vector superspace.%
\footnote{For some particular cases, we found also distributional solutions for cubic vertices. Our study does not provide solution for all distributional cubic vertices and this is beyond the scope of the present paper.
}
In our earlier study in Refs.\cite{Metsaev:2017cuz,Metsaev:2018moa}, we used the light-cone gauge oscillator formulation of CSFs which leads to cubic vertices expressed in terms of special functions (hypergeometric and Bessel functions) and some complicated dressing operators acting on such special functions. The main benefit of the vector formulation, as compared to the oscillator formulation, is that the vector formulation leads to simple expressions for cubic vertices (rational or exponential functions). We believe therefore that such simplification will be helpful in future studies of interacting CSFs. Also, in this paper, we considered the interrelation between the light-cone gauge oscillator and vector formulations of CSFs and their equivalence to covariant formulations. We presented the explicit map (and its inverse) of the light-cone gauge CSF of the vector formulation into the one of the oscillator formulation. Besides this we demonstrated an equivalence of our light-cone gauge formulations and the corresponding Lorentz covariant formulations. We believe that these results will be helpful for study of interrelations between vector and oscillator formulations of interacting CSFs. We expect that our study in this paper might have the following interesting applications and generalizations.

\noinbf{\ibf)} Generalization of our study to 4-point vertices for CSFs seems to be the most interesting direction to go. Various methods for analysis of 4-point light-cone gauge vertices for integer-spin fields may be found, e.g., in Refs. \cite{Bengtsson:2016alt}-\cite{Ivanovskiy:2025kok}.

\noinbf{\iibf)} As we obtained a lot of cubic vertices for CSFs it is tempting to impose some additional restrictions to decrease a number of allowed cubic vertices. In general, $\NN=1$ SUSY might be helpful for this purpose. In $4d$, as shown in Ref.\cite{Metsaev:2019dqt}, all light-cone gauge cubic vertices for the massless integer-spin fields in Ref.\cite{Bengtsson:1983pd} admit $\NN=1$ supersymmetrization and hence, in $4d$, $\NN=1$ SUSY does not seem to be helpful for our purpose. Fortunately, as is well known, for the higher dimensions, $d>4$, the $\NN=1$ SUSY turns out to be more restrictive (see, e.g., Ref.\cite{Buchbinder:2021qrg}). We expect therefore that attempts towards for $\NN=1$ supersymmetrization  of our vertices for $d>4$  will lead to more short list of cubic vertices for CSFs.%
\footnote{Alternative way to a more short list of cubic vertices might be related to extended SUSY. For $\NN=4\No$-extended massless scalar supermultiplet in $4d$, light-cone gauge cubic vertex was obtained in Ref.\cite{Bengtsson:1983pg}, while for {\it arbitrary spin} (integer and half-integer) $\NN=4\No$-extended massless supermultiplets all light-cone gauge cubic vertices were studied Ref.\cite{Metsaev:2019aig}. Recent progress in the study of $\NN=2$ SUSY can be found in Refs.\cite{Buchbinder:2022kzl}.
}

\noinbf{\iiibf)} In Refs.\cite{Metsaev:2017cuz,Metsaev:2018moa}, we obtained the complete list of cubic vertices available in the framework of the oscillator formulation. Unfortunately, these vertices turn out to be much more complicated as compared to the vertices obtained by using vector formulation in this paper. Matching of the vertices in Refs.\cite{Metsaev:2017cuz,Metsaev:2018moa} and the ones obtained in this paper should lead to better understanding of the interacting CSFs. The matching is an open problem at present time.

\noinbf{\ivbf)} Other interesting generalization of our study is related to CSF in AdS. Light-cone gauge formulation of CSF in AdS was discussed in Ref.\cite{Metsaev:2019opn}. Application of the method in this paper and the one in Ref.\cite{Metsaev:2018xip} might  provide new possibilities for the study of interacting CSF in AdS.

\vspace{-0.2cm}
\setcounter{section}{0}\setcounter{subsection}{0}
\appendix{ \large Notation and conventions  }\label{app-notation}

\noinbf{Notation and relations of oscillator formalism}. The vector indices of the $so(d-2)$ algebra $I,J,K$ run over $1,\ldots ,d-2$. Creation operators $\alpha^I$, $\upsilon$, $\zeta$ and the respective annihilation operators $\alphab^I$, $\upsilonb$, $\zetab$ are referred to as oscillators. Our conventions for the commutators, the vacuum $|0\rangle $, and hermitian conjugation rules are summarized as follows
\beq
\label{12042025-man01-01} && [ \alphab^I,\alpha^J] = \delta^{IJ}, \quad [\upsilonb,\upsilon]=1,  \quad [\zetab,\zeta]=1,  \quad \alphab^I |0\rangle = 0\,,\quad \upsilonb |0\rangle = 0\,,   \quad \zetab |0\rangle = 0\,,\qquad
\nonumber\\
&& \alpha^{I \dagger} = \alphab^I\,, \hspace{1.1cm} \upsilon^\dagger = \upsilonb\,, \hspace{1cm} \zeta^\dagger = \zetab\,.
\eeq
Various shortcuts for scalar products of the oscillators are summarized as follows,
\be \label{12042025-man01-02}
\alpha^2 := \alpha^I\alpha^I\,,\qquad \alphab^2 := \alphab^I \alphab^I\,,\qquad N_\alpha  := \alpha^I \alphab^I\,, \qquad N_\zeta  := \zeta \zetab\,,\qquad N_\upsilon  := \upsilon \upsilonb\,.
\ee
Oscillators that respect the traceless conditions are defined as
{\small
\beq
\label{12042025-man01-05} &&  \alpha_\Thsm^I := \alpha^I - (\alpha^2+\zeta^2)\frac{1}{2N_\alpha + 2 N_\zeta + d-1}\alphab^I\,, \hspace{0.8cm}
\nonumber\\
&&  \zeta_\Thsm := \zeta - (\alpha^2+\zeta^2)\frac{1}{2N_\alpha + 2 N_\zeta + d-1}\zetab\,, \hspace{1.3cm} \hbox{for massive integer-spin fields},
\\
\label{12042025-man01-05x}  &&  \alpha_\Thsm^I := \alpha^I - \alpha^2\frac{1}{2N_\alpha + d-2}\alphab^I\,, \hspace{3cm} \hbox{for massless integer-spin fields}.\qquad
\eeq
}
We now note the main property of the two set of oscillators $\alpha_\Thsm^I,\zeta_\Thsm$ and $\alpha_\Thsm^I$. Let $P_\msv(\alpha_\Thsm,\zeta_\Thsm)$ and $P_\msl(\alpha_\Thsm)$ be polynomials of the respective oscillators $\alpha_\Thsm^I$, $\zeta_\Thsm$ and $\alpha_\Thsm^I$ corresponding to massive and massless integer-spin fields. Then one has the following relations:
{\small
\be \label{12042025-man01-06}
(\alphab^2 + \zetab^2) P_\msv(\alpha_\Thsm^I,\zeta_\Thsm)|0\rangle =  0\,, \quad \hbox{and}  \quad \alphab^2 P_\msl(\alpha_\Thsm^I)|0\rangle = 0\,.
\ee
}
The starred fields and inner product for any quantities $A(\alpha)$ and $B(\alpha)$ depending, among other things, on the oscillators are defined by the relation
{\small
\beq
\label{12042025-man01-08} &&   \phi^*(p,\alpha): =  (\phi(p,\alpha))^\dagger \,,\hspace{2cm} \phi^*(p,\alpha) \cdot \phi(p,\alpha) : = \langle 0 | (\phi(p,\alpha))^\dagger \phi(p,\alpha)|0\rangle\,,
\nonumber\\
&&  A(\alpha) \cdot B(\alpha) : = \langle A(\alpha)| B(\alpha)\rangle\,, \qquad \langle A(\alpha)| B(\alpha)\rangle : = \langle 0 | (A(\alpha))^\dagger B(\alpha)|0\rangle\,.\qquad
\eeq
}
\noinbf{ Notation and relations of embedding space method}.  Let $\PP^I$ be a tangential derivative with respect to unit vector $u^I$, $u^Iu^I=1$. We note the following well-known commutators and relations:
\beq
\label{17042025-man01-appa-01} && [\PP^I, u^J] =  \delta^{IJ} - u^I u^J\,,
\qquad  [\PP^I,\PP^J] =   u^I\PP^J - u^J \PP^I \,,
\nonumber\\
&& u^I \PP^I = 0 \,,\qquad \PP^I u^I = N-1\,, \qquad I,J = 1,\dots, N\,.
\eeq
For arbitrary rank tensor fields denoted symbolically as  $f_1$, $f_2$, and $f$, rules for integration by parts and for a total derivative take the form
{\small
\beq
\label{17042025-man01-appa-05} && \int du\,\, f_1 \PP^I f_2  =    \int du\,\,  \big( -f_2 \PP^I f_1  + (N-1)  u^I f_1 f_2 \big) \,,
\nonumber\\
&& \int du\,\, \PP^I f  =   (N-1)  \int du\,\, u^I f \,, \qquad \int du= S_{N-1}\,, \qquad du := d^N u\,\, \delta(u^I u^I- 1)\,.\qquad
\eeq
}
\!The inner product of fields depending on unit vector $u^I$ \rf{27032025-man01-10} required for the building of generators \rf{27032025-man01-75}, \rf{29032025-man01-05}  is defined by the relation
{\small
\be \label{12042025-man01-08-u}
\phi^*(p,u) \cdot \varphi(p,u) = \int du\, \phi^*(p,u) \varphi(p,u)\,,
\ee
}
\!where the definition of the starred fields depending on the unit vector $u^I$ is given in \rf{17042025-appk-05}.

\noinbf{Explicit expressions for the operators $G_a^\alpha$, $G_a^u$, $G_\beta$}. Derivative of  a quantity $X$ is denoted as
\be  \label{12042025-man01-10}
\partial_X = \partial/\partial X\,.
\ee
Using notation in \rf{30032025-man01-10}, \rf{12042025-man01-10}, we present explicit form of operators in \rf{30032025-man01-16},
\beq
&& G_a^\alpha =   (B_{a+2}^\alpha - \frac{\beta_a}{\beta_{a+2}} m_{a+2} \zeta_{a+2} )\partial_{q_{a+2a}^{\alpha\alpha}} - ( B_{a+1}^\alpha + \frac{\beta_a}{\beta_{a+1}} m_{a+1} \zeta_{a+1} ) \partial_{q_{aa+1}^{\alpha\alpha}}
\nonumber\\
&&  \hspace{0.7cm} +\,\,   (B_{a+2}^u + \frac{\beta_a}{\beta_{a+2}} |m_{a+2}| )\partial_{q_{a+2 a}^{u\alpha}} - ( B_{a+1}^u -  \frac{\beta_a}{\beta_{a+1}} |m_{a+1}| ) \partial_{q_{aa+1}^{\alpha u}}
\nonumber\\
&& \hspace{0.7cm} +\,\, \half \Bigl( \frac{\betach_a}{ \beta_a} m_a^2 +  m_{a+1}^2 - m_{a+2}^2  \Bigr) \partial_{B_a^\alpha} + m_a\partial_{\zeta_a}\,,
\nonumber\\
&& G_a^u =  (B_{a+2}^\alpha - \frac{\beta_a}{\beta_{a+2}} m_{a+2} \zeta_{a+2} )\partial_{ q_{a+2 a}^{\alpha u} } - ( B_{a+1}^\alpha  +   \frac{\beta_a}{\beta_{a+1}} m_{a+1} \zeta_{a+1} ) \partial_{ q_{aa+1}^{u\alpha} }
\nonumber\\
&&  \hspace{0.7cm} +\,\, (B_{a+2}^u + \frac{\beta_a}{\beta_{a+2}} |m_{a+2}|)\partial_{ q_{a+2a}^{uu} } - ( B_{a+1}^u - \frac{\beta_a}{\beta_{a+1}} |m_{a+1}|  ) \partial_{ q_{aa+1}^{uu} }
\nonumber\\
&& \hspace{0.5cm} + \,\, \half \Bigl( \frac{\betach_a}{ \beta_a} m_a^2 +  m_{a+1}^2 - m_{a+2}^2  \Bigr) \partial_{ B_a^u }
\nonumber\\
&& \hspace{0.7cm} - \,\,  |m_a| \big( q_{aa+1}^{uu}\partial_{ q_{aa+1}^{uu} } +  q_{a+2a}^{uu} \partial_{ q_{a+2a}^{uu} } + q_{aa+1}^{u\alpha} \partial_{ q_{aa+1}^{u\alpha} } +  q_{a+2a}^{\alpha u} \partial_{ q_{a+2a}^{\alpha u} } +  B_a^u \partial_{ B_a^u } - \SSc_a \big)  + \irm \kappa_a\,,\qquad
\nonumber\\
&&  G_\beta =  - \frac{1}{\beta}  \No_\beta - \sum_{a=1,2,3} \frac{1}{\beta_a^2} \big( m_a \zeta_a \partial_{B_a^\alpha} - |m_a|\partial_{B_a^u})\,.
\eeq

\vspace{-0.2cm}
\appendix{ \large Derivation of operator $M^I$ for massive CSF}\label{app-bb}

Discussion of the original form of the BB-constraints and their interrelation with the constraints we use may be found at the end of this Appendix. We consider the field $\Phi= \Phi(p,\xi)$ depending on momentum $p^\mu$ and Lorentz $so(d-1,1)$ algebra vector $\xi^\mu$. The momentum $p^\mu$ and the vector $\xi^\mu$ are constrained to the surface defined by the relations
\be \label{12042025-04-bb-app-01}
\ibf) \ \  p^\mu p^\mu + m^2 = 0 \,,\qquad  \iibf) \ \ p^\mu \xi^\mu   = 0 \,,\qquad \iiibf) \ \ \xi^\mu \xi^\mu   = 0 \,,
\ee
$m^2<0$, while the field $\Phi$ satisfies the differential constraint,
\be \label{22012021-15-a1}
\big( \xi^\mu \partial_{\xi^\mu} - \SSc \big) \Phi = 0 \,.
\ee
Generators of the Poincar\'e algebra $iso(d-1,1)$ are realized on the field $\Phi(x,\xi)$ as
\be \label{22012021-04-a1}
J^{\mu\nu} = p^\mu\partial_{p^\nu} - p^\nu\partial_{p^\mu} + \xi^\mu\partial_{\xi^\nu} -  \xi^\nu\partial_{\xi^\mu}\,, \qquad   P^\mu = p^\mu\,.
\ee
We now derive the operator $M^I$ for massive CSF \rf{27032025-man01-55} in the following 3 steps.

\noinbf{Step 1}. From \ibf), \iibf), \iiibf) in \rf{12042025-04-bb-app-01}, we find the respective relations,
{\small
\be \label{22012021-05}
\ibf) \ \ p^- = - \frac{p^Ip^I+m^2}{2\beta}\,, \quad \iibf) \ \ \xi^- = -\frac{p^-}{\beta} \xi^+    -\frac{p^I}{\beta} \xi^I \,,\qquad \iiibf) \ \ \xi^- = -\frac{\xi^I \xi^I}{2\xi^+}\,.
\ee
}
Solution for $p^-$ in \ibf) \rf{22012021-05} and two solutions for $\xi^-$ in \iibf), \iiibf)  \rf{22012021-05} lead to the relation
{\small
\be \label{22012021-09}
\eta^I \eta^I  = - m^2\,, \qquad  \eta^I := \frac{\beta}{\xi^+}\xi^I - p^I\,.
\ee
}
\noinbf{Step 2}. Solution for $p^-$ and $\xi^-$ in \rf{22012021-05} motivates us to introduce field $\Phi^{(1)}$ independent of $p^-$, $\xi^-$, while the definition of $\eta^I$ \rf{22012021-09} motivates us to introduce field $\Phi^{(2)}$ depending on $\eta^I$ in place of $\xi^I$,
{\small
\be
\Phi(p^-,\beta,p^I,\xi^-,\xi^+, \xi^I) =  \beta \Phi^{(1)}(\beta,p^I,\xi^+, \xi^I)\,,\qquad \Phi^{(1)}(\beta,p^I,\xi^+, \xi^I) =  \Phi^{(2)}(\beta,p^I,\xi^+, \eta^I)\,.
\ee
}
Using \rf{22012021-04-a1}, we find the realization of the generators $J^{+-}$, $J^{\pm I}$ on $\Phi^{(2)}$ given by
{\small
\beq
\label{22012021-22} && \hspace{-1.7cm} J^{+-} = \partial_{\beta}\beta + N_{\xi^+}\,, \qquad J^{+I} = \partial_{p^I}\beta  \,, \qquad J^{-I} = l^{-I} +   M^{IJ}\frac{p^J}{\beta}   + \frac{1}{\beta} M^I - \frac{p^I}{\beta} N_{\xi^+}\,,
\nonumber\\
&& \hspace{-0.5cm} M^I =   m^2\partial_{\eta^I}  + \eta^I (N_\eta - N_{\xi^+}) \,, \quad \ \ M^{IJ} = \eta^I \partial_{\eta^J} -  \eta^J \partial_{\eta^I}\,,\qquad l^{-I} = \partial_{p^I}p^- -\partial_\beta p^I \,,
\eeq
}
where $N_{\xi^+} : = \xi^+ \partial_{\xi^+}$, $N_\eta : = \eta^I \partial_{\eta^I}$. Also we note that the constraint \rf{22012021-15-a1} takes the form
\be \label{22012021-25-a1}
(\xi^+ \partial_{\xi^+}  - \SSc)\Phi^{(2)}(\beta,p^I,\xi^+, \eta^I) = 0\,.
\ee

\noinbf{Step 3}. Constraint \rf{22012021-25-a1} and our desire to get simple form of $J^{+-}$ suggest to introduce field $\Phi^{(3)}$,
\be \label{22012021-26}
\Phi^{(2)}(\beta,p^I,\xi^+, \eta^I) = \Big(\frac{\xi^+}{\beta} \Big)^\SSc \Phi^{(3)}(\beta,p^I,\eta^I)\,.
\ee
Realization of generators \rf{22012021-22} on the field $\Phi^{(3)}$ takes the form
{\small
\beq
\label{22012021-27} && J^{+-} = \partial_{\beta}\beta \,,
\qquad J^{+I} = \partial_{p^I}\beta  \,,\qquad J^{-I} = l^{-I} +   M^{IJ}\frac{p^J}{\beta}   + \frac{1}{\beta} M^I\,,
\nonumber\\
&& M^I =   m^2 \partial_{\eta^I}  + \eta^I (N_\eta - \SSc)\,, \qquad \ \ \ M^{IJ} = \eta^I \partial_{\eta^J} -  \eta^J \partial_{\eta^I}\,.
\eeq
}
Introducing $\eta^I  =  |m| u^I$, we see that expression for $M^I$ \rf{22012021-27} takes the same form as the one for massive CSF in \rf{27032025-man01-55}.

\noinbf{BB constraints}. In Ref.\cite{Bekaert:2006py}, the BB-constraints are formulated for field $\Phi_\smBB = \Phi_\smBB(p,\xi_{_\smBB})$ depending on momentum $p^\mu$ and Lorentz $so(d-1,1)$ algebra vector $\xi_\smBB^\mu$. The momentum $p^\mu$ is considered to be on-shell,
\be
\label{05072021-01-x}   p^\mu p^\mu + m^2 = 0 \,,
\ee
while the field $\Phi_\smBB$ obeys the constraints
\be \label{05072021-02-x}
p^\mu \partial_{\xi_\smBB^\mu}\Phi_\smBB   = 0 \,,\quad   \partial_{\xi_\smBB^\mu} \partial_{\xi_\smBB^\mu}\Phi_\smBB   = 0 \,,\qquad (\xi_\smBB^\mu \partial_{\xi_\smBB^\mu} - s_\smBB)\Phi_\smBB = 0 \,,
\ee
where $s_\smBB = \frac{3-d}{2} + \irm \sigma$. The field $\Phi_\smBB(p,\xi)$ in \rf{05072021-02-x} is expressed in terms of our field $\Phi$ in \rf{22012021-15-a1} as
{\small
\be \label{05072021-06-x}
\Phi_\smBB(p,\xi_\smBB) = \int d^d\xi\,\, e^{\irm \xi_\smBB^\mu \xi^\mu}\,\, \widetilde\Phi(p,\xi)\,, \qquad   \widetilde\Phi(p,\xi) : = \delta(\xi^2)\delta(p\xi) \Phi(p,\xi)\,,
\ee
}
while the parameter $\SSc$ in \rf{22012021-15-a1} and the parameter $s_\smBB$ in \rf{05072021-02-x} are related as $\SSc  = 3 -d - s_\smBB$.

We prefer to use constraints \rf{12042025-04-bb-app-01}, \rf{22012021-15-a1} for two reasons: \ibf) derivation of the operator $M^I$ turns out to be simpler; \iibf) constraints \rf{12042025-04-bb-app-01} are similar to the Wigner constraints \rf{02012021-01} for massless CSF and hence derivations of the operator $M^I$ for massive and massless CSFs turn out to be  similar.

In Ref.\cite{Bekaert:2006py}, constraints \rf{05072021-02-x} were proposed for the principal series. Our considerations show that these remarkable constraints can also be used for the derivation of the operator $M^I$ corresponding to the complementary and discrete series. The relevant scalar products \rf{17042025-appk-01} and values $n_\minrm$ \rf{27032025-man01-08} are  then obtained by requiring the operator $M^I$ to be anti-hermitian.

\vspace{-0.2cm}
\appendix{ \large Derivation of operator $M^I$ for massless CSF}\label{app-wigner}

We consider field $\Phi= \Phi(p,\xi)$ depending on momentum $p^\mu$ and Lorentz $so(d-1,1)$ algebra vector $\xi^\mu$. In the framework of Wigner approach, the momentum $p^\mu$ and the vector $\xi^\mu$ are constrained to the surface defined by the relations%
\footnote{ Wigner's CSF $\Phi_\Wsm$ is expressed in terms of our CSF $\Phi$ by the relation  $\Phi_\Wsm = \delta(\xi^2-\kappa^2)\delta(p\xi)\Phi$ supplemented by the replacement $\xi^\mu\rightarrow \kappa\xi^\mu$. Various modified forms of Wigner equations are discussed in Refs.\cite{Najafizadeh:2017tin}.
}
\be \label{02012021-01}
\ibf) \ \ p^\mu p^\mu = 0 \,, \qquad \iibf) \ \ p^\mu \xi^\mu = 0 \,, \qquad
\iiibf) \ \  \xi^\mu\xi^\mu  - \kappa^2 = 0 \,,
\ee
$\kappa^2>0$, while the field $\Phi$ satisfies the differential constraint
\be \label{02012021-03}
\big(p^\mu\partial_{\xi^\mu} -\irm\big)\Phi  = 0 \,.
\ee
Generators of the Poincar\'e algebra $iso(d-1,1)$ are realized on $\Phi(x,\xi)$ as in \rf{22012021-04-a1}. We now derive the operator $M^I$ for massless CSF \rf{27032025-man01-55} in the following 3 steps.

\noinbf{Step 1}. From \ibf) and \iiibf) in \rf{02012021-01}, we find
\be \label{02012021-05-a1}
p^- = - \frac{p^Ip^I}{2\beta}\,, \qquad  \xi^- =  \frac{\kappa^2 - \xi^I\xi^I}{2\xi^+}\,.
\ee
Plugging $p^-$ and $\xi^-$ \rf{02012021-05-a1} into \iibf) \rf{02012021-01}, we get the constraint
\be \label{02012021-05-a3}
\eta^I \eta^I = \kappa^2 \,, \qquad \eta^I := \xi^I - \frac{p^I}{\beta} \xi^+\,.
\ee

\noinbf{Step 2}. Solution for $p^-$ and $\xi^-$ \rf{02012021-05-a1} motivates us to introduce field $\Phi^{(1)}$ independent of $p^-$, $\xi^-$, while the definition of $\eta^I$ \rf{02012021-05-a3} motivates us to introduce field $\Phi^{(2)}$ depending on $\eta^I$ in place of $\xi^I$,
{\small
\be
\Phi(p^-,\beta,p^I,\xi^-,\xi^+, \xi^I) =  \beta \Phi^{(1)}(\beta,p^I,\xi^+, \xi^I)\,,\qquad \Phi^{(1)}(\beta,p^I,\xi^+, \xi^I) =  \Phi^{(2)}(\beta,p^I,\xi^+, \eta^I)\,.
\ee
Using \rf{22012021-04-a1}, we find the realization of the generators $J^{+-}$, $J^{\pm I}$ on $\Phi^{(2)}$ given by
{\small
\be
\label{02012021-28-a} J^{+-} = \partial_{\beta}\beta + N_{\xi^+} \,, \qquad  J^{+I} = \partial_{p^I}\beta \,, \qquad J^{-I} = l^{-I} + M^{IJ} \frac{p^J}{\beta} - \eta^I \partial_{\xi^+} - \frac{p^I}{\beta} N_{\xi^+} \,,
\ee
}
\!where $N_{\xi^+} : = \xi^+ \partial_{\xi^+}$, while $l^{-I}$ and $M^{IJ}$ take the same form as in \rf{22012021-22}.
Also we note that, on space of $\Phi^{(2)}$, constraint \rf{02012021-03} takes the form
\be \label{02012021-21}
\big(\beta\partial_{\xi^+}    - \irm \big)\Phi^{(2)}= 0\,.
\ee

\noinbf{Step 3}. Constraint \rf{02012021-21} is solved as
\be \label{02012021-23}
\Phi^{(2)}(\beta,p^I,\xi^+, \eta^I) =  \exp\big(   \frac{\irm }{\beta} \xi^+\big) \Phi^{(3)}(\beta,p^I,\eta^I)\,.
\ee
Realization of generators \rf{02012021-28-a} on $\Phi^{(3)}$ takes then the form
{\small
\be \label{02012021-28-23-xx1}
J^{+-} = \partial_{\beta}\beta \,, \qquad  J^{+I} = \partial_{p^I}\beta \,, \qquad J^{-I} = l^{-I} + M^{IJ} \frac{p^J}{\beta} + \frac{1}{\beta} M^I \,, \qquad M^I = -\irm \eta^I \,,
\ee
}
\!where $M^{IJ}$ is given in \rf{22012021-27}. Using constraint \rf{02012021-05-a3}, we introduce a unit vector $u^I$ by the relation $\eta^I  =  \kappa u^I$, and note then that the $M^I$ in \rf{02012021-28-23-xx1} takes the same form as the $M^I$ for massless CSF in \rf{27032025-man01-55}.

\vspace{-0.2cm}
\appendix{ \large $c$ -- distribution  }\label{app-cdis}

\noinbf{$c$ -- distribution}. Let $U_1^I$, $U_2^I$ be two vectors in $\Eo^N$. We define a distribution $c(U_1,U_2)$ by the relations%
{\small
\beq
\label{18042025-m1-apd-01}  && c(U_1,U_2) := \sum_{n=0}^\infty c_n(U_1,U_2)\,,
\nonumber\\
&& c_n(U_1,U_2) := \frac{2}{S_{N-1}(N-2)} \big( n + \frac{N-2}{2} \big) |U_1|^n|U_2|^n C_n^{\frac{N-2}{2}}(u_1u_2)\,,\qquad u_a^I = \frac{U_a^I}{|U_a|}\,, \qquad
\eeq
}
\!where $C_n^\alpha(t)$ is the Gegenbauer polynomial, while $S_{N-1}$ is given in \rf{17042025-appk-03}.%
\footnote{For  the Gegenbauer polynomial $C_n^\alpha(t)$, we use conventions in the handbook \cite{NIST}. For $N=2$, the $c_n$ \rf{18042025-m1-apd-01} is expressed in terms of the  Chebyshev polynomial (see, e.g., Ref.\cite{Ponomarev:2016jqk}).}
We note the harmonicity property, $\partial_{U_1^I} \partial_{U_1^I}c(U_1,U_2)=0$, while the remaining properties of the distribution $c(U_1,U_2)$ are as follows.

\noinbf{i)} We accept expressions \rf{18042025-m1-apd-01} as starting point for the study of $c(U_1,U_2)$. If $|U_1||U_2|=1$, then the $c$ -- distribution is realized as the standard $(N-1)$-dimensional Dirac delta function on $S^{N-1}$,
\be \label{18042025-m1-apd-05}
c(U_1,U_2) = \delta^{N-1}(u_1,u_2)\qquad \hbox{for}  \quad  |U_1||U_2| = 1\,.
\ee
It is relation \rf{18042025-m1-apd-05} that motivates us refer $c(U_1,U_2)$ \rf{18042025-m1-apd-01} to as $c$-distribution.

\noinbf{ii)} For $|U_1||U_2| < 1$, the $c(U_1,U_2)$ is well defined. Using relations 18.12.4 and 18.12.5 in Ref.\cite{NIST}, we find
\be \label{18042025-m1-apd-06}
c(U_1,U_2) =  \frac{ 1 - U_1^2 U_2^2 }{ S_{N-1} (1-2 U_1U_2 + U_1^2 U_2^2)^{N/2} }\,, \qquad \hbox{for}  \quad  |U_1||U_2| <  1\,.
\ee
For $|U_1||U_2| > 1$, the $c(U_1,U_2)$ can then be defined by analytical continuation of the expression in \rf{18042025-m1-apd-06}.

\noinbf{iii}) Considering $U_1^I=X^I$, $|X| \ne 1$, and $U_2^I = u^I$, $u^Iu^I=1$, we note the following relations for $c(X,u)$:
{\small
\beq
\label{18042025-m1-apd-09} && u^I c (X,u) = \big( \frac{1}{2N_X+N}\partial_{X^I}  + X_\Thsm^I \big) c(X,u)\,,
\nonumber\\
&& \PP^I c(X,u) = \big( -\frac{N_X}{2N_X+N} \partial_{X^I}  + (N+N_X-2) X_\Thsm^I\big)  c(X,u)\,,
\nonumber\\
&& \hspace{2cm} X_\Thsm^I := X^I - X^2 \frac{1}{2N_X+N}\partial_{X^I}\,, \qquad N_X := X^I\partial_{X^I}\,,
\eeq
}
\!where $\PP^I$ is defined in \rf{17042025-man01-appa-01}. For the derivation of \rf{18042025-m1-apd-09}, we use \rf{18042025-m1-apd-01} and the following relations for the functions $c_n(X,u)$:
{\small
\beq
\label{18042025-m1-apd-10} && u^I c_n(X,u) = \frac{1}{2n+N}\partial_{X^I} c_{n+1}(X,u) + X_\Thsm^I c_{n-1}(X,u)\,,
\nonumber\\
&& \PP^I c_n(X,u) = -\frac{n}{2n+N} \partial_{X^I} c_{n+1}(X,u) + (N+n-2) X_\Thsm^I c_{n-1}(X,u)\,.
\eeq
}
\!We note also the useful relations for the functions $c_n(U_1,U_2)$ \rf{18042025-m1-apd-01},
{\small
\beq
\label{17042025-m1-apd-35} && \int du\,\, c_n(U_1,u) c_n(U_2,u) = c_n(U_1,U_2)\,,
\\
\label{17042025-m1-apd-36} && \langle c_n(\alpha,U_1)| c_n(\alpha,U_2)\rangle = \frac{1}{\tau_n S_{N-1}} c_n(U_1,U_2)\,,
\\
\label{17042025-m1-apd-37} && \langle c_n(\alpha,U_1)|(\alpha U_2)^n\rangle = n! c_n(U_1,U_2)\,, \qquad \alpha U : = \alpha^I U^I\,,
\eeq
}
\!where $\alpha$ stands for the oscillator $\alpha^I$. The scalar product $\langle A|B\rangle $ is defined in \rf{12042025-man01-08}, while $\tau_n$ and $S_{N-1}$ in \rf{17042025-appk-03}.

\appendix{ \large Oscillator and vector formulations}\label{app-osc-vec}

In this Appendix, to simplify our formulas, we hide momentum arguments of fields. For example, in place of fields $\phi(p,u)$ and $\phi(p,\alpha,\upsilon)$ of the vector and oscillator formulations we use $\phi(u)$ and $\phi(\alpha,\upsilon)$ respectively.

\noinbf{Vector formulation}. Field content of the vector formulation is presented in \rf{27032025-man01-08}, \rf{27032025-man01-10}. Scalar product of field $\phi(u)$ is defined to be
{\small
\beq
\label{17042025-appk-01} && \hspace{-1.4cm} (\phi,\phi) = \sum_{n=n_\minrm}^\infty \mu_n \int du (\phi_n(u))^\dagger \phi_n(u)\,,
\nonumber\\
&& \mu_n=1\,, \hspace{5.5cm} \hbox{for massive CSF of principal series};\qquad
\nonumber\\
&& \mu_n  = \frac{ \Gamma(\frac{N-1}{2} + n  + q) \Gamma(\frac{N-1}{2} - q) }{ \Gamma(\frac{N-1}{2} + n  - q)\Gamma(\frac{N-1}{2} + q)}\,, \hspace{1cm} \hbox{for massive CSF of complementary series};\qquad
\nonumber\\
&& \mu_n= \frac{\Gamma(n-s)\Gamma(2s+N)}{\Gamma( n + s +N-1)}\,, \hspace{2.3cm}  \hbox{for massive CSF of discrete series};
\nonumber\\
&& \mu_n=1\,, \hspace{5.4cm} \hbox{for massless CSF};\qquad
\eeq
}
\!where $n_\minrm$ is defined in \rf{27032025-man01-08}. To complete formulas in \rf{27032025-man01-10}, we present the expressions for $\tau_n$ and $S_{N-1}$,
{\small
\be \label{17042025-appk-03}
\tau_n = \frac{ \Gamma(\frac{N}{2}) }{ 2^n\Gamma(\frac{N}{2}+n) }\,,\qquad S_{N-1} =  \frac{ 2\pi^{N/2} }{ \Gamma(\frac{N}{2}) }\,, \qquad N:= d-2\,,
\ee
}
\!where the $S_{N-1}$ is a surface area of the unit $(N-1)$-sphere of radius 1
embedded in $\Eo^N$. The hermicity properties of operators in \rf{27032025-man01-49} are valid with respect to norm defined in \rf{17042025-appk-01}. For the complementary and discrete series, the $\mu_n$ are normalized by the condition $\mu_{n_\minrm}=1$.%
\footnote{In mathematical handbook \cite{vilenkin}, $\mu_n$ for the three series  \rf{17042025-appk-01} are derived by using a representation of the spin operators in terms of the  Euler angles parametrization of $S^{N-1}$. We derived $\mu_n$  by using the embedding space representation of spin operators \rf{27032025-man01-55}. Up to overall normalization factor, our $\mu_n$ coincides with the one in Ref.\cite{vilenkin}.}
Using the expressions for $\mu_n$ \rf{17042025-appk-01}, starred fields entering  generators \rf{29032025-man01-05} are defined by the relation
{\small
\be \label{17042025-appk-05}
\phi^*(u) : = \sum_{n=n_\minrm}^\infty  \mu_n (\phi_n(u))^\dagger\,, \qquad (\phi,\phi) = \int du \phi^*(u) \phi(u)\,,
\ee
}
\!where we represent norm \rf{17042025-appk-01} in terms of $\phi^*(u)$ and $\phi(u)$. Relations \rf{17042025-appk-05} can alternatively be represented as
{\small
\beq \label{17042025-appk-05-add}
&& \phi^*(u) : = \int du_1 (\phi(u_1))^\dagger \mu(u_1,u)\,, \qquad \qquad (\phi,\phi) = \int du_1 du_2 (\phi(u_1))^\dagger \mu(u_1,u_2) \phi(u_2)\,,
\nonumber\\
\label{17042025-appk-06-add} && \hspace{3cm} \mu(u_1,u_2) : = \sum_{n=n_\minrm} \mu_n c_n(u_1,u_2)\,,
\eeq
\!where $\mu(u_1,u_2)$ in \rf{17042025-appk-06-add} is defined by using $c_n(u_1,u_2)$ in \rf{18042025-m1-apd-01} and $\mu_n$ in \rf{17042025-appk-01}. We note the helpful relation
{\small
\be
\phi_n(u_1) = \int du_2\,\, c_n(u_1,u_2)\phi_n(u_2) \,.
\ee
}
\noinbf{Oscillator formulation}. Field content of the oscillator formulation is the same as in \rf{27032025-man01-08}. However, now we collect fields \rf{27032025-man01-08} into ket-vectors defined by the relations
{\small
\be \label{17042025-appk-07}
\phi(\alpha,\upsilon) = \sum_{n=n_\minrm}^\infty \phi_n(\alpha,\upsilon) \,, \qquad \phi_n(\alpha,\upsilon) : = \frac{\upsilon^n}{n!\sqrt{n!}} \alpha^{I_1} \ldots \alpha^{I_n} \phi^{I_1\ldots I_n}\,.
\ee
}
Using definition of the scalar product $\langle A|B\rangle$ in \rf{12042025-man01-08}, we define scalar product of the ket-vector as
{\small
\be \label{17042025-appk-09}
(\phi,\phi) : = \langle \phi(\alpha,\upsilon)|\phi(\alpha,\upsilon)\rangle \,, \qquad (\phi,\phi) = \sum_{n=n_\minrm}^\infty  \frac{1}{n!}  \phi^{I_1\ldots I_n\, \dagger} \, \phi^{I_1\ldots I_n}\,.
\ee
}
The normalization is chosen so that the scalar product in \rf{17042025-appk-09} is equal to scalar product in \rf{17042025-appk-01}.
Realization of the spin operators $M^{IJ}$, $M^I$ \rf{27032025-man01-49} is given by
{\small
\beq
 \label{17042025-appk-15} && \hspace{-1.1cm} M^{IJ} = \alpha^I \alphab^J - \alpha^J \alphab^I\,,
\nonumber\\
&& \hspace{-1.1cm} M^I = m_\upsilon \alphab^I \upsilonb - \alpha_\Thsm^I v m_\upsilon\,, \qquad m_\upsilon := \NN_\upsilon \sqrt{F_\upsilon}\,, \qquad \NN_\upsilon = \big( (N_\upsilon+1)(2N_\upsilon+N) \big)^{-\half}\,, \qquad
\nonumber\\
&& F_\upsilon =  \kappa^2 - m^2N_\upsilon(N_\upsilon + N-1)\,, \hspace{2.1cm} \hbox{for massive/massless CSF},
\eeq
}
\!where the operator $\alpha_\Thsm^I$ appearing in \rf{17042025-appk-15} takes the same form as the one for massless integer-spin field in \rf{12042025-man01-05}. We recall the inequality $m^2 < 0$ for massive CSF. The remaining parameter $\kappa$ satisfies the constraints
{\small
\beq
\label{17042025-appk-17} && \kappa^2 > - \frac{1}{4} (N-1)^2 m^2\,, \hspace{1.6cm}  \hbox{for CSF of principal series};
\nonumber\\
&&  0 < \kappa^2 < - \frac{1}{4} (N-1)^2 m^2\,,  \qquad \ \hbox{for CSF of complementary series};
\nonumber\\
&&   \kappa^2  = m^2 s(s+N-1)\,,  \hspace{1.4cm} \hbox{for CSF of discrete series}.
\eeq
}
Note that, in view of $m^2 < 0$, we get $\Re \kappa=0$ for CSF of discrete series.

In Ref.\cite{Metsaev:2019opn}, for massive CSF of principal series, we found alternative representation for operator $M^I$,
{\small
\beq
\label{17042025-appk-20} && \hspace{-1.2cm}  M^I = m_\upsilon \alphab^I \upsilonb - \alpha_\Thsm^I v \mb_\upsilon\,, \hspace{2.1cm} m_\upsilon := \NN_\upsilon f_\upsilon \,, \qquad \mb_\upsilon := \NN_\upsilon \fb_\upsilon\,,
\nonumber\\
&& f_\upsilon = |m|\big(  N_\upsilon + \frac{N-1}{2}\big)  + \sigma \,, \quad \fb_\upsilon = |m|\big( N_\upsilon + \frac{N-1}{2} \big) - \sigma\,,  \qquad
\nonumber\\
&& \sigma = \irm \epsilon \Big( \kappa^2 + \frac{1}{4} (N-1)^2 m^2 \Big)^\half \,, \quad \epsilon^2 = 1\,, \hspace{0.3cm} \hbox{for massive CSF of principal series},
\eeq
}
\!where $\NN_\upsilon$ is given in  \rf{17042025-appk-15}. So, for the massive CSF of principal series, we have two equivalent representations for the operator $M^I$ given in \rf{17042025-appk-15} and \rf{17042025-appk-20}.

\noinbf{Interrelation between vector and oscillator formulations}. Fields of the vector and oscillator formulations defined in \rf{27032025-man01-10} and \rf{17042025-appk-07} are related as%
\footnote{Relation \rf{17042025-appk-30} provides the explicit map between the oscillator and vector light-cone formulations. For massless integer-spin fields, a map between oscillator and vector Lorentz covariant formulations was discussed in Ref.\cite{Rivelles:2014fsa}, where it was noted that explicit map between two such formulations remains to be fixed.}
{\small
\be \label{17042025-appk-30}
\phi(\alpha,\upsilon) = \int du\,  \Pi_\pi(\alpha,\upsilon;u) \phi(u)\,,\qquad \phi(u) = \langle \Pi_\pi(\alpha,\upsilon;u) | \phi(\alpha,\upsilon)\rangle\,,
\ee
}
\!where we introduce the intertwine operator $\Pi_\pi(\alpha,\upsilon;u)$,
{\small
\beq
\label{17042025-appk-25} && \hspace{-0.5cm} \Pi_\pi(\alpha,\upsilon;u) =  \sum_{n=n_\minrm}^\infty \pi_n \upsilon^n\, c_n(\alpha,u) \,,
\nonumber\\
&& \hspace{-0.5cm} \pi_n   = (-)^n \Big( \frac{ t_n \tau_n S_{N-1}  }{ n!
} \Big)^\half\,,\quad  t_n  = \frac{ \Gamma(\frac{N-1}{2} + n  + q) \Gamma(\frac{N-1}{2} + n_\minrm - q) }{ \Gamma(\frac{N-1}{2} + n  - q)\Gamma(\frac{N-1}{2} + n_\minrm + q)}\,, \hspace{0.3cm} \hbox{for massive CSF}\,, \qquad
\nonumber\\
&& \hspace{-0.5cm} \pi_n = \irm^{-n} \Big(\frac{\tau_nS_{N-1} }{n!}\Big)^{1/2}   \,, \hspace{8cm} \hbox{for massless CSF}\,, \qquad
\eeq
}
while $n_\minrm$ and $\tau_n$ are defined in \rf{27032025-man01-08} \rf{17042025-appk-03}. Restrictions on $q$ are given in \rf{27032025-man01-60}.
Using \rf{17042025-appk-25}, we verify that: \ibf) the expressions for $M^I$, $M^{IJ}$  \rf{27032025-man01-55} amount to the expressions for $M^I$, $M^{IJ}$  \rf{17042025-appk-15}; \iibf) the scalar product of the vector formulation in \rf{17042025-appk-01} is equal to the one for the oscillator formulation in \rf{17042025-appk-15}; \iiibf) The parameter $\kappa$ and $\SSc$ entering the respective oscillator and vector formulations of massive CSF are related as
\be
\kappa^2 = m^2  \SSc(\SSc+N-1)\,,  \qquad \hbox{or in terms of $q$}: \quad q^2 =  \frac{\kappa^2}{m^2} + \frac{1}{4}(N-1)^2\,.
\ee

For the massive CSF of principal series, the $M^I$ has alternative oscillator formulation \rf{17042025-appk-20}. The field of vector formulation in \rf{27032025-man01-10} and a field associated with the alternative oscillator formulation in \rf{17042025-appk-20} are related as in \rf{17042025-appk-30}, \rf{17042025-appk-25}, where we should use
$t_n = 1$, while the corresponding parameters $q$ and $\sigma$ are related as $\sigma = |m| q $.

\vspace{-0.2cm}
\appendix{ \large Derivation and some properties of vertex \rf{31032025-man01-17-15} }\label{app-der-dis}

\noinbf{Derivation of vertex \rf{31032025-man01-17-15}}.
We are interested in cubic vertex for fields shown in \rf{31032025-man01-17-10}.
Let us use the notation $p_{\smp3\, \vecrm}^-$ for cubic vertex of the vector formulation and the notation $p_{\smp3\, \osc}^-$ for the corresponding cubic vertex of the oscillator formulation. The cubic vertex $p_{\smp3\, \osc}^-$ was obtained in Ref.\cite{Metsaev:2017cuz} (see relations (4.33)-(4.35) in Ref.\cite{Metsaev:2017cuz}). We obtained two solutions for $p_{\smp3\, \osc}^-$ which we denote as $p_{\smp3\,\osc}^-(\even)$ and $p_{\smp3\,\osc}^-(\odd)$. The solutions are given by
\beq
\label{17042025-man01-appf-01} && p_{\smp3\,\osc}^-(\even) = U \cosh\big(  \frac{ \upsilon_1 B_1}{m}  - \frac{\kappa_1 \betach_1}{2m\beta_1 } \big)\,, \qquad p_{\smp3,\osc}^-(\odd) = U \sinh\big(  \frac{ \upsilon_1 B_1}{m}  - \frac{\kappa_1 \betach_1}{2m\beta_1 } \big)\,, \qquad
\nonumber\\
&& \hspace{4cm} U = \Big(\frac{  2^{N_{B_1} } \Gamma(N_{B_1} + \frac{N}{2}) }{ \Gamma( N_{B_1} + 1 ) } \Big)^{1/2}\,.
\eeq
Using the vertices in \rf{17042025-man01-appf-01} we find the corresponding $p_{\smp3\,\vecrm}^-(\even)$ and $p_{\smp3\,\vecrm}^-(\odd)$ by the 3 steps.

\noinbf{Step 1}. Relation for fields \rf{17042025-appk-30} and definition of the scalar product $\langle A|B\rangle$ in \rf{12042025-man01-08} imply the following relation for the corresponding cubic vertices:
\be \label{17042025-man01-appf-05}
p_{\smp3\, \vecrm}^- = \langle \Pi_\pi(\alpha_1,\upsilon_1;u_1)|p_{\smp3\,\osc}^-\rangle\,.
\ee

\noinbf{Step 2}. Keeping in mind the notation in \rf{17042025-appk-03}, we note the relation
\be
U \frac{1}{n!} \Big( \frac{ \upsilon_1 B_1}{m}\Big)^n |0\rangle =  \frac{1}{n!} \Big(\frac{ \Gamma(\frac{N}{2}) }{ n!\tau_n  } \Big)^{1/2}  \Big( \frac{ \upsilon_1 B_1}{m}\Big)^n |0\rangle\,.
\ee

\noinbf{Step 3}. Using $\Pi_\pi$ corresponding to the massless CSF in \rf{17042025-appk-25} and relation \rf{17042025-m1-apd-37}, we find
\be
\langle \Pi_\pi(\alpha_1,\upsilon_1;u_1)| U \frac{1}{n!} \Big(\frac{ \upsilon_1 B_1}{m} \Big)^n \rangle = \NN  c_n(\frac{\irm\, \Po}{m\beta_1},u_1)\,, \qquad \NN := \big( \Gamma(\frac{N}{2}) S_{N-1} \big)^{1/2}\,,
\ee
which, in view of definition of the $c$ -- distribution \rf{18042025-m1-apd-01} leads immediately to the relation
\be \label{17042025-man01-appf-30}
\langle \Pi_\pi(\alpha_1,\upsilon_1;u_1)| U e^{\frac{ \upsilon_1 B_1}{m}}\rangle =  \NN c(\frac{\irm\, \Po}{m\beta_1},u_1)\,.
\ee
Finally, using \rf{17042025-man01-appf-30} in \rf{17042025-man01-appf-05} and taking into account expressions \rf{17042025-man01-appf-01}, we find the expressions for vertices $p_{\smp3\, \vecrm}^-(\even/\odd)$ given in \rf{31032025-man01-17-15}.

\noinbf{Direct check of solution \rf{31032025-man01-17-15}}. Using \rf{18042025-m1-apd-01}, we verify that vertex \rf{31032025-man01-17-15} satisfies the  harmonic condition
\be \label{31032025-man01-17-20}
\partial_{\Po^I}^{\vphantom{5pt}} \partial_{\Po^I}^{\vphantom{5pt}} p_\smp3^- =  0 \,,
\ee
and is referred therefore to as harmonic cubic vertex.
In Sec.\,\ref{sec-3-1}, we derived equations \rf{30032025-man01-16}-\rf{30032025-man01-25} for cubic vertices independent of $\Po^I\Po^I$-terms, while
the harmonic cubic vertex depends on $\Po^I\Po^I$-terms and hence equations \rf{30032025-man01-16}-\rf{30032025-man01-25} cannot be used to validate the harmonic cubic vertex \rf{31032025-man01-17-15}.
Explicit equations for the harmonic cubic vertices were presented in (4.26), (4.27), and (4.32) in Ref.\cite{Metsaev:2005ar}. We now confirm that vertex \rf{31032025-man01-17-15} satisfies the just mentioned equations in Ref.\cite{Metsaev:2005ar}. The above given 1st relation in \rf{18042025-m1-apd-09}, Appendix D, in this paper turns out to be helpful to this end.

\noinbf{Properties of vertex \rf{31032025-man01-17-15}}. {\bf i}) Using the definition of $\Pbf^-$ \rf{29032025-man01-30}, we find for fields in \rf{31032025-man01-17-10} the following relations:
\be \label{17042025-man01-appf-35}
\Pbf^- = \frac{1}{2\beta}(\Po^I\Po^I + m^2 \beta_1^2)\,, \qquad \Pbf^-\big|_{\Po^I=\Po_\crt^I}=0\,, \qquad \Po_\crt^I := - \irm m \beta_1 u_\crt^I\,, \qquad u_\crt^I u_\crt^I=1\,. \qquad
\ee
Using \rf{18042025-m1-apd-05}, we get $c(\irm \Po_\crt/m\beta_1,u_1) = \delta^{N-1}(u,u_1)$, which is Dirac delta function on $S^{N-1}$.
So we see that, for $\Po^I=\Po_\crt^I$ (unphysical sheet), cubic vertex \rf{31032025-man01-17-15} is realized as the distribution given by,
\be \label{17042025-man01-appf-38-00}
2\NN^{-1}\, p_\smp3^-(\even,\odd)|_{\Po^I = \Po_\crt^I}  =    e^{- \frac{\kappa_1 \betach_1}{2m \beta_1 } }  \delta^{N-1}(u_1,u_\crt)  \pm e^{\frac{\kappa_1 \betach_1}{2 m \beta_1} }  \delta^{N-1}(u_1,-u_\crt)\,.
\ee

\noinbf{ii)} Plugging $U_1 = u_1^I$, $U_2 = \frac{\irm \Po^I}{m\beta_1}$ in \rf{18042025-m1-apd-06}, we get the following representation for the $c$-distribution:
\be \label{18042025-m1-apd-07}
c(\frac{\irm \Po}{m\beta_1},u)   = \Pbf^- V(\Po)\,, \qquad  V(\Po): = \frac{2\beta}{  m^2 \beta_1^2S_{N-1}}   \Big| u  - \frac{\irm \Po}{m \beta_1} \Big|^{-N}\,,
\ee
while, plugging $c(\frac{\irm \Po}{m\beta_1},u) $ \rf{18042025-m1-apd-07}, into \rf{31032025-man01-17-15} we get the following representation for cubic vertex \rf{31032025-man01-17-15}:
\be \label{17042025-man01-appf-38}
p_\smp3^-({\small\even,\odd})  = \Pbf^- V(\even,\odd)\,, \qquad 2\NN^{-1}V(\even,\odd) :=   e^{- \frac{\kappa_1 \betach_1}{2m \beta_1 } }  V(\Po)  \pm e^{\frac{\kappa_1 \betach_1}{2 m \beta_1} }  V(-\Po)\,.
\ee
We now see that vertex $p_\smp3^-$ \rf{17042025-man01-appf-38} does not satisfy requirement for $p_\smp3^-$ in \rf{29032025-man01-24}. This provokes us to remove the vertex $p_\smp3^-$ from our game by using field redefinitions. Instead we inspect the vertex $V$ by using  requirement for $V_\frd$ in \rf{29032025-man01-24}. As seen from \rf{17042025-man01-appf-35}, the equation $\Pbf^-=0$ implies  $\Po^I = \Po_\crt^I$. Plugging $\Po_\crt^I$ \rf{17042025-man01-appf-35} into $V$ \rf{18042025-m1-apd-07}, we get
\be  \label{18042025-m1-apd-08}
V(\Po)\big|_{\Pbf^- = 0}  = \frac{2\beta}{  m^2 \beta_1^2S_{N-1}}   \big| u  - u_\crt \big|^{-N}\,.
\ee
From \rf{18042025-m1-apd-08}, we see that $V$ is not well defined when $\Pbf^-=0$ and hence requirement for $V_\frd$ in \rf{29032025-man01-24} is not satisfied.
For this reason, we do not eliminate cubic vertex $p_\smp3^-$ \rf{31032025-man01-17-15} from our game.

\vspace{-0.2cm}


\small
\begin{thebibliography}{100}

\parskip=-3.3pt








\bibitem{Bekaert:2006py}
  X.~Bekaert and N.~Boulanger,
 ``The Unitary representations of the Poincare group in any spacetime dimension,''
   in 2nd Modave Summer School in Theoretical Physics Modave, Belgium, August 6-12, 2006, 2006.
  hep-th/0611263.



\bibitem{Bekaert:2017khg}
  X.Bekaert and E.D.~Skvortsov,
  Int.J.Mod.Phys.\ A {\bf 32}, no.23n24, 1730019 (2017)
  [arXiv:1708.01030].


\bibitem{Brink:2002zx}
  L.~Brink, A.~M.~Khan, P.~Ramond and X.~z.~Xiong,
  J.\ Math.\ Phys.\  {\bf 43}, 6279 (2002)
  [hep-th/0205145].


\bibitem{Schuster:2013pta}
P.~Schuster and N.~Toro,
JHEP \textbf{10} (2013), 061
[arXiv:1302.3225 [hep-th]].
%
\\
%
  P.~Schuster and N.~Toro,
  Phys.\ Rev.\ D {\bf 91}, 025023 (2015)
  [arXiv:1404.0675 [hep-th]].



\bibitem{Najafizadeh:2015uxa}
  X.Bekaert, M.Najafizadeh, M.R.Setare,
  Phys.\ Lett.\ B {\bf 760}, 320 (2016)
  [arXiv:1506.00973 [hep-th]].






\bibitem{Metsaev:2017cuz}
  R.~R.~Metsaev,
  JHEP {\bf 1711}, 197 (2017)
  [arXiv:1709.08596 [hep-th]].


\bibitem{Bekaert:2017xin}
  X.~Bekaert, J.~Mourad and M.~Najafizadeh,
  JHEP {\bf 1711}, 113 (2017)
  [arXiv:1710.05788 [hep-th]].


\bibitem{Rivelles:2018tpt}
V.~O.~Rivelles,
``A Gauge Field Theory for Continuous Spin Tachyons,''
arXiv:1807.01812 [hep-th].



\bibitem{Metsaev:2018moa}
R.~R.~Metsaev,
JHEP \textbf{12} (2018), 055
[arXiv:1809.09075 [hep-th]].


\bibitem{Rivelles:2023hzo}
V.~O.~Rivelles,
``Virtual Exchange of Continuous Spin Particles,''
[arXiv:2303.06490 [hep-th]].










\bibitem{Schuster:2023xqa}
P.~Schuster, N.~Toro and K.~Zhou,
JHEP \textbf{04} (2023), 010
[arXiv:2303.04816 [hep-th]].
%
\\
%
P.~Schuster and N.~Toro,
Phys. Rev. D \textbf{109} (2024) no.9, 096008
[arXiv:2308.16218 [hep-th]].
%
\\
%
S.~Kundu, P.~Schuster and N.~Toro,
[arXiv:2503.03817 [gr-qc]].






\bibitem{Schuster:2013pxj}
P.~Schuster and N.~Toro,
JHEP \textbf{09} (2013), 104
[arXiv:1302.1198 [hep-th]].



\bibitem{Bellazzini:2024dco}
B.~Bellazzini, S.~De Angelis and M.~Romano,
[arXiv:2406.17017 [hep-th]].





\bibitem{Schuster:2024wjc}
P.~Schuster, G.~Sundaresan and N.~Toro,
Phys. Rev. D \textbf{111} (2025) no.5, 056019
[arXiv:2406.14616]








\bibitem{Najafizadeh:2019mun}
M.~Najafizadeh,
JHEP \textbf{03} (2020), 027
[arXiv:1912.12310 [hep-th]].
%
\\
%
M.~Najafizadeh,
JHEP \textbf{02} (2022), 038
[arXiv:2112.10178 [hep-th]].


\bibitem{Buchbinder:2019kuh}
I.~L.~Buchbinder, M.~V.~Khabarov, T.~V.~Snegirev, Y.~M.~Zinoviev,
Nucl. Phys. B \textbf{946} (2019), 114717


\bibitem{Buchbinder:2022msd}
I.Buchbinder, S.Fedoruk, A.Isaev, V.Krykhtin,
Phys.Lett.B\textbf{829} (2022), 137139
[arXiv:2203.12904]






\bibitem{Metsaev:2016lhs}
  R.~R.~Metsaev,
  Phys.\ Lett.\ B {\bf 767}, 458 (2017)
  [arXiv:1610.00657 [hep-th]].


\bibitem{Metsaev:2017ytk}
  R.~R.~Metsaev,
  Phys.\ Lett.\ B {\bf 773}, 135 (2017)
  [arXiv:1703.05780 [hep-th]].


\bibitem{Zinoviev:2017rnj}
  Y.~M.~Zinoviev,
  Universe {\bf 3}, no. 3, 63 (2017)
  [arXiv:1707.08832 [hep-th]].


\bibitem{Khabarov:2017lth}
M.~V.~Khabarov and Y.~M.~Zinoviev,
Nucl. Phys. B \textbf{928} (2018), 182-216
[arXiv:1711.08223 [hep-th]].


\bibitem{Metsaev:2017myp}
R.~R.~Metsaev,
J. Phys. A \textbf{51} (2018) no.21, 215401
[arXiv:1711.11007 [hep-th]].


\bibitem{Metsaev:2019opn}
R.~R.~Metsaev,
Phys. Lett. B \textbf{793} (2019), 134-140
[arXiv:1903.10495 [hep-th]].



\bibitem{Metsaev:2021zdg}
R.~R.~Metsaev,
Phys. Lett. B \textbf{820} (2021), 136497
[arXiv:2105.11281 [hep-th]].


\bibitem{Buchbinder:2024hea}
I.Buchbinder, S.Fedoruk, A.Isaev, V.Krykhtin,
Phys.Lett.B\textbf{853} (2024), 138689
[arXiv:2402.13879]




\bibitem{Basile:2023vyg}
T.~Basile, E.~Joung and T.~Oh,
JHEP \textbf{01} (2024), 018
[arXiv:2307.13644 [hep-th]].







\bibitem{Alkalaev:2017hvj}
  K.~B.~Alkalaev and M.~A.~Grigoriev,
  JHEP {\bf 1803}, 030 (2018)
  [arXiv:1712.02317 [hep-th]].
%
\\
%
  K.~Alkalaev, A.~Chekmenev and M.~Grigoriev,
  arXiv:1808.09385 [hep-th].





\bibitem{Ponomarev:2010st}
D.~S.~Ponomarev and M.~A.~Vasiliev,
Nucl. Phys. B \textbf{839} (2010), 466-498
[arXiv:1001.0062 [hep-th]].


\bibitem{Khabarov:2020glf}
M.~V.~Khabarov and Y.~M.~Zinoviev,
Nucl. Phys. B \textbf{953} (2020), 114959
[arXiv:2001.07903 [hep-th]].











\bibitem{Najafizadeh:2017tin}
  M.~Najafizadeh,
  Phys.\ Rev.\ D {\bf 97}, no. 6, 065009 (2018)
  [arXiv:1708.00827 [hep-th]].
%
\\
%
H.~Takata,
Nucl. Phys. B \textbf{1005} (2024), 116599
[arXiv:2404.14118 [hep-th]].







\bibitem{Bengtsson:1983pd}
  A.~K.~H.~Bengtsson, I.~Bengtsson and L.~Brink,
  Nucl.\ Phys.\ B {\bf 227}, 31 (1983).
%
\\
%
  A.~K.~H.~Bengtsson, I.~Bengtsson and N.~Linden,
  Class.\ Quant.\ Grav.\  {\bf 4}, 1333 (1987).



\bibitem{Metsaev:2005ar}
  R.~R.~Metsaev,
  Nucl.\ Phys.\ B {\bf 759}, 147 (2006)
  [hep-th/0512342].



\bibitem{Metsaev:2007rn}
  R.~R.~Metsaev,
  Nucl.\ Phys.\ B {\bf 859}, 13 (2012)
  [arXiv:0712.3526 [hep-th]].


\bibitem{Metsaev:2022yvb}
R.~R.~Metsaev,
Nucl. Phys. B \textbf{984} (2022), 115978
[arXiv:2206.13268 [hep-th]].

\bibitem{Manvelyan:2010jr}
R.Manvelyan, K.Mkrtchyan and W.Ruhl,
Nucl.\ Phys.\ B {\bf 836}, 204 (2010)
[arXiv:1003.2877 [hep-th]].
%
\\
%
  R.Manvelyan, K.Mkrtchyan and W.Ruehl,
  Phys.\ Lett.\  B {\bf 696}, 410 (2011)
  [arXiv:1009.1054 [hep-th]].



\bibitem{Sagnotti:2010at}
A.~Sagnotti and M.~Taronna,
Nucl. Phys. B \textbf{842} (2011), 299-361
[arXiv:1006.5242 [hep-th]].








\bibitem{Metsaev:2020gmb}
R.~R.~Metsaev,
J. Phys. A \textbf{53} (2020) no.44, 445401
[arXiv:2005.12224 [hep-th]];
%
\\
R.~R.~Metsaev,
JHEP \textbf{12} (2021), 069
[arXiv:2110.02696 [hep-th]].


\bibitem{Mkrtchyan:2017ixk}
K.~Mkrtchyan,
Phys. Rev. Lett. \textbf{120} (2018) no.22, 221601
[arXiv:1712.10003 [hep-th]].

\bibitem{Zinoviev:2021cmi}
Y.~M.~Zinoviev,
JHEP \textbf{11} (2021), 022
[arXiv:2109.08480];
%
JHEP \textbf{03} (2023), 058
[arXiv:2211.09405]


\bibitem{Sharapov:2024rud}
A.Sharapov, D.Shcherbatov and E.~Skvortsov,
Eur. Phys. J. C \textbf{85} (2025) no.3, 252
[arXiv:2412.11052]


\bibitem{Delplanque:2024enh}
W.~Delplanque and E.~Skvortsov,
Class. Quant. Grav. \textbf{41} (2024) no.24, 245018
[arXiv:2405.13706]
%
\\
%
W.~Delplanque and E.~Skvortsov,
JHEP \textbf{08} (2024), 173
[arXiv:2406.14148 [hep-th]].
%
\\
%
W.~Delplanque,
Phys. Rev. D \textbf{111} (2025) no.4, 4
[arXiv:2411.03463 [hep-th]].








\bibitem{Dirac:1949cp}
  P.~A.~M.~Dirac,
  Rev.\ Mod.\ Phys.\  {\bf 21}, 392 (1949).





\bibitem{Vasiliev:1990en}
M.~A.~Vasiliev,
Phys. Lett. B \textbf{243} (1990), 378-382
%
\\
%
M.~A.~Vasiliev,
Phys. Lett. B \textbf{567} (2003), 139-151
[arXiv:hep-th/0304049 [hep-th]].



\bibitem{Didenko:2021vdb}
V.~E.~Didenko and A.~V.~Korybut,
JHEP \textbf{01} (2022), 125
[arXiv:2110.02256]; \
%
%
Phys. Rev. D \textbf{110} (2024) no.2, 026007
[arXiv:2312.11096]; \
%
%
Phys. Rev. D \textbf{108} (2023) no.8, 086031
[arXiv:2304.08850]

\bibitem{Tatarenko:2024csa}
Y.~A.~Tatarenko and M.~A.~Vasiliev,
JHEP \textbf{07} (2024), 246
[arXiv:2405.02452 [hep-th]].


\bibitem{Didenko:2019xzz}
V.Didenko, O.Gelfond, A.Korybut, M.~A.~Vasiliev,
JHEP \textbf{12} (2019), 086
[arXiv:1909.04876];
JHEP \textbf{12} (2020), 184
[arXiv:2009.02811];
%
J. Phys. A \textbf{51} (2018) no.46, 465202
[arXiv:1807.00001].

\bibitem{DeFilippi:2019jqq}
D.~De Filippi, C.~Iazeolla and P.~Sundell,
JHEP \textbf{10} (2019), 215
[arXiv:1905.06325 [hep-th]];
JHEP \textbf{07} (2022), 003
[arXiv:2111.09288 [hep-th]].





\bibitem{Bengtsson:1986ys}
A.~K.~H.~Bengtsson,
Phys. Lett. B \textbf{182} (1986), 321-325

\bibitem{Pashnev:1989gm}
A.~I.~Pashnev,
Theor. Math. Phys. \textbf{78} (1989), 272-277


\bibitem{Francia:2002pt}
D.~Francia and A.~Sagnotti,
Class. Quant. Grav. \textbf{20} (2003), S473-S486
[arXiv:hep-th/0212185].


\bibitem{Sagnotti:2003qa}
A.~Sagnotti and M.~Tsulaia,
Nucl. Phys. B \textbf{682} (2004), 83-116
[arXiv:hep-th/0311257 [hep-th]].


\bibitem{Sorokin:2008tf}
D.~P.~Sorokin and M.~A.~Vasiliev,
Nucl. Phys. B \textbf{809} (2009), 110-157
[arXiv:0807.0206 [hep-th]].


\bibitem{Sorokin:2018djm}
D.~Sorokin and M.~Tsulaia,
Nucl. Phys. B \textbf{929} (2018), 216-242
[arXiv:1801.04615 [hep-th]].


\bibitem{Campoleoni:2012th}
A.~Campoleoni and D.~Francia,
JHEP \textbf{03} (2013), 168
[arXiv:1206.5877 [hep-th]].
%
\\
%
D.~Francia, G.~L.~Monaco and K.~Mkrtchyan,
JHEP \textbf{04} (2017), 068
[arXiv:1611.00292 [hep-th]].







\bibitem{Konstein:1989ij}
S.~E.~Konstein and M.~A.~Vasiliev,
Nucl. Phys. B \textbf{331} (1990), 475-499



\bibitem{Metsaev:1991nb}
  R.~R.~Metsaev,
  Mod.\ Phys.\ Lett.\ A {\bf 6}, 2411 (1991).


\bibitem{Skvortsov:2020wtf}
E.~Skvortsov, T.~Tran and M.~Tsulaia,
Phys. Rev. D \textbf{101} (2020) no.10, 106001
[arXiv:2002.08487].
%
\\
%
  E.~D.~Skvortsov, T.~Tran, M.~Tsulaia,
  Phys.\ Rev.\ Lett.\  {\bf 121}, no. 3, 031601 (2018)
  [arXiv:1805.00048].
%
\\
%
M.~G\"unaydin, E.~D.~Skvortsov and T.~Tran,
JHEP \textbf{11} (2016), 168
[arXiv:1608.07582 [hep-th]].






\bibitem{Bekaert:2005in}
  X.~Bekaert and J.~Mourad,
  JHEP {\bf 0601}, 115 (2006)
  [hep-th/0509092].



\bibitem{Khan:2004nj}
A.~M.~Khan and P.~Ramond,
J. Math. Phys. \textbf{46} (2005), 053515
[arXiv:hep-th/0410107]
%
\\
%
A.~M.~Khan,
J. Math. Phys. \textbf{62} (2021) no.3, 032305
[arXiv:2102.08932 [hep-th]].











\bibitem{Kessel:2018ugi}
  P.~Kessel and K.~Mkrtchyan,
  Phys.\ Rev.\ D {\bf 97}, no. 10, 106021 (2018)
  [arXiv:1803.02737 [hep-th]].
%
\\
%
S.Fredenhagen, F.Lausch, K.Mkrtchyan,
Phys. Rev. D \textbf{110} (2024) no.8, L081702
[arXiv:2404.00497]







\bibitem{Bengtsson:1987jt}
A.~K.~H.~Bengtsson,
Class. Quant. Grav. \textbf{5} (1988), 437

\bibitem{Bekaert:2005jf}
X.~Bekaert, N.~Boulanger and S.~Cnockaert,
JHEP \textbf{01} (2006), 052
[arXiv:hep-th/0508048 [hep-th]].

\bibitem{Fotopoulos:2010ay}
  A.~Fotopoulos and M.~Tsulaia,
  JHEP {\bf 1011}, 086 (2010)
  [arXiv:1009.0727 [hep-th]].


\bibitem{Metsaev:2012uy}
  R.~R.~Metsaev,
  Phys.\ Lett.\ B {\bf 720}, 237 (2013)
  [arXiv:1205.3131 [hep-th]].


\bibitem{Grigoriev:2020lzu}
M.Grigoriev, K.Mkrtchyan, E.Skvortsov,
Phys. Rev. D \textbf{102} (2020) no.6, 066003
[arXiv:2005.05931]



\bibitem{Buchbinder:2021xbk}
I.~L.~Buchbinder and A.~A.~Reshetnyak,
Phys. Lett. B \textbf{820} (2021), 136470
[arXiv:2105.12030].



\bibitem{Vasiliev:2025hfh}
M.~A.~Vasiliev,
JHEP \textbf{07} (2025), 110
[arXiv:2503.10967 [hep-th]].






\bibitem{Bengtsson:2013vra}
A.~K.~H.~Bengtsson,
JHEP \textbf{10} (2013), 108
[arXiv:1303.3799 [hep-th]].



\bibitem{Metsaev:2018lth}
R.~R.~Metsaev,
Phys. Lett. B \textbf{781} (2018), 568-573
[arXiv:1803.08421 [hep-th]].



\bibitem{Buchbinder:2018yoo}
  I.~L.~Buchbinder, V.~A.~Krykhtin and H.~Takata,
  Phys.\ Lett.\ B {\bf 785}, 315 (2018)
  [arXiv:1806.01640]



\bibitem{Buchbinder:2020nxn}
I.L.Buchbinder, S.Fedoruk, A.Isaev, V.Krykhtin,
Nucl.Phys.B\textbf{958}(2020), 115114,
arXiv:2005.07085


\bibitem{Burdik:2020ror}
\v{C}.~Burd\'\i{}k and A.~A.~Reshetnyak,
Nucl. Phys. B \textbf{965} (2021), 115357
[arXiv:2010.15741 [hep-th]].








\bibitem{Bengtsson:2016alt}
A.~K.~H.~Bengtsson,
``Quartic amplitudes for Minkowski higher spin,''
[arXiv:1605.02608 [hep-th]].



\bibitem{Metsaev:1991mt}
R.~R.~Metsaev,
Mod. Phys. Lett. A \textbf{6} (1991), 359-367


\bibitem{Ponomarev:2016lrm}
D.~Ponomarev and E.~D.~Skvortsov,
J. Phys. A \textbf{50} (2017) no.9, 095401
[arXiv:1609.04655 [hep-th]].


\bibitem{Ivanovskiy:2025kok}
V.~Ivanovskiy and D.~Ponomarev,
[arXiv:2503.11546 [hep-th]].





\bibitem{Metsaev:2019dqt}
R.~R.~Metsaev,
JHEP \textbf{08} (2019), 130
[arXiv:1905.11357 [hep-th]].





\bibitem{Buchbinder:2021qrg}
I.~L.~Buchbinder, V.~A.~Krykhtin, M.~Tsulaia and D.~Weissman,
Nucl. Phys. B \textbf{967} (2021), 115427








\bibitem{Bengtsson:1983pg}
A.~K.~H.~Bengtsson, I.~Bengtsson and L.~Brink,
Nucl. Phys. B \textbf{227} (1983), 41-49



\bibitem{Metsaev:2019aig}
R.~R.~Metsaev,
JHEP \textbf{11} (2019), 084
[arXiv:1909.05241 [hep-th]].






\bibitem{Buchbinder:2022kzl}
I.~Buchbinder, E.~Ivanov and N.~Zaigraev,
JHEP \textbf{05} (2022), 104
[arXiv:2202.08196 [hep-th]].
%
\\
%
I.~Buchbinder, E.~Ivanov and N.~Zaigraev,
JHEP \textbf{12} (2021), 016
[arXiv:2109.07639 [hep-th]].









\bibitem{Metsaev:2018xip}
R.~R.~Metsaev,
Nucl. Phys. B \textbf{936} (2018), 320-351
[arXiv:1807.07542 [hep-th]].









\bibitem{NIST}
Olver F. W. J. (ed.). NIST handbook of mathematical functions hardback and CD-ROM. – Cambridge University Press, 2010









\bibitem{vilenkin}
N. Ja. Vilenkin, Special Functions and Theory of Group Representations,
Moscow, 1965, Transl. Math. Monographs, Vol. 22, Amer. Math. Soc, Providence, R. I.,1968.








\bibitem{Ponomarev:2016jqk}
D.~Ponomarev and A.~A.~Tseytlin,
JHEP \textbf{05} (2016), 184
[arXiv:1603.06273 [hep-th]].






\bibitem{Rivelles:2014fsa}
V.~O.~Rivelles,
Phys. Rev. D \textbf{91} (2015) no.12, 125035
[arXiv:1408.3576 [hep-th]].



\end{thebibliography}
\end{document}